\documentclass[aps,letterpaper,twocolumn,nofootinbib]{revtex4}

\pdfoutput=1

\usepackage[latin2]{inputenc}
\usepackage{graphicx}
\usepackage{epsfig}
\usepackage{amssymb}
\usepackage{amsfonts}
\usepackage{amsmath,euscript,array,mathrsfs}
\usepackage{epsf}
\usepackage{slashed}
\usepackage{comment}
\usepackage{colortbl}
\usepackage{parskip}
\usepackage{mathtools}
\usepackage{upgreek}

\usepackage[colorlinks=true,linktocpage=true,linkcolor=blue,citecolor=blue]{hyperref}

\usepackage{tikz}
\usetikzlibrary{decorations.markings,decorations.pathmorphing}

\usepackage{float}

\newcommand{\be}{\begin{equation}}
\newcommand{\ee}{\end{equation}}
\newcommand{\beqs}{\begin{eqnarray}}
\newcommand{\eeqs}{\end{eqnarray}}

\numberwithin{equation}{section} 
\allowdisplaybreaks

\begin{document}

\title{Cauchy Evolution of Asymptotically Global AdS Spacetimes with No Symmetries}
\author{Hans Bantilan}
\email{h.bantilan@qmul.ac.uk}
\author{Pau Figueras}
\email{p.figueras@qmul.ac.uk}
\author{Lorenzo Rossi}
\email{l.rossi@qmul.ac.uk}
\affiliation{School of Mathematical Sciences, Queen Mary University of London, Mile End Road, London E1 4NS, United Kingdom}
		
\begin{abstract}
We present the first proof-of-principle Cauchy evolutions of asymptotically global AdS spacetimes with no imposed symmetries, employing a numerical scheme based on the generalized harmonic form of the Einstein equations.
In this scheme, the main difficulty in removing all symmetry assumptions can be phrased in terms of finding a set of generalized harmonic source functions that are consistent with AdS boundary conditions.
In four spacetime dimensions, we detail an explicit set of source functions that achieves evolution in full generality.
A similar prescription should also lead to stable evolution in higher spacetime dimensions, various couplings with matter fields, and on the Poincar\'{e} patch.
We apply this scheme to obtain the first long-time stable 3+1 simulations of four dimensional spacetimes with a negative cosmological constant, using initial data sourced by a massless scalar field.
We present preliminary results of gravitational collapse with no symmetry assumptions, and the subsequent quasi-normal mode ringdown to a static black hole in the bulk, which corresponds to evolution towards a homogeneous state on the boundary.
\end{abstract}

\maketitle

\section{Introduction}

In recent years anti-de Sitter (AdS) space has proven to be a particularly exciting theoretical laboratory for studying the strong-field regime of General Relativity (GR).
AdS with reflective boundary conditions plays the role of a box that naturally keeps propagating waves confined to its interior, where they are perpetually interacting.
Thus, even the smallest perturbations in AdS can enter the strong-field regime, where qualitatively new gravitational phenomena emerge.
One of the most important of these is gravitational collapse -- the growth of curvatures that eventually leads to the formation of a singularity in spacetime potentially associated with a black hole.
Obtaining the details of this fundamental process in full generality in AdS is still an open problem. 
In asymptotically flat spacetimes, although it has not yet been proven rigorously, this process of gravitational collapse is expected to generically end in a rotating black hole that is characterized by two conserved numbers: total mass and total angular momentum. 
In asymptotically AdS spacetimes, the endpoint is less clear.
Small, rapidly rotating black holes are unstable due to a process known as superradiance -- the amplification of waves that scatter off a rotating object. Along with the box-like nature of AdS, this amplification leads to a runaway process whose endpoint is unknown. 

In an unprecedented way, the simulation of asymptotically AdS spacetimes has also opened up the field of numerical relativity to the study of phenomena in areas beyond the traditional astrophysical setting.
At the heart of this push to understand AdS is a deep connection between gravity in AdS to certain conformal field theories (CFT), now known as the AdS/CFT correspondence~\cite{Maldacena:1997re,Gubser:1998bc,Witten:1998qj}. 
Through this connection, the study of AdS spacetimes has become immediately relevant to fundamental questions in many areas in physics, such as fluid dynamics~\cite{Baier:2007ix,Bhattacharyya:2010owp,Hubeny:2011hd}, relativistic heavy ion collisions~\cite{Chesler:2010bi,Casalderrey-Solana:2013aba,Casalderrey-Solana:2013sxa,Chesler:2015wra}, and superconductivity~\cite{Gubser:2008px,Hartnoll:2008kx,Hartnoll:2008vx}.
See, for example, \cite{CasalderreySolana:2011us,Chesler:2015lsa,Zaanen:2015oix,Hartnoll:2016apf} for excellent reviews. 
The reason why the study of AdS is crucial for our understanding of these phenomena is that AdS/CFT provides an important -- and in most cases the only -- window into the real-time dynamics of strongly interacting quantum field theories far from equilibrium. 
The dynamical far-from-equilibrium strongly interacting regime is precisely the one that is least explored and understood, and the one that has the best chance of making contact with certain experiments.

Our current understanding of gravity in AdS remains limited for several reasons.
First, evolution in AdS is notoriously hard, in part because it is an initial-boundary value problem whose systematic study is still in its infancy. 
Cauchy evolution in AdS requires data to be prescribed not only at an initial spacelike hypersurface, but also at spatial and null infinity which constitute the timelike boundary of an asymptotically AdS spacetime.
Second, the most interesting phenomena involve spacetimes that have very little or no symmetry, making these evolutions beyond the reach of most numerical codes. 
Third, for many of these phenomena, there is a variety of physical scales that must be adequately resolved to correctly capture the relevant physics.

The main purpose of this article is to present the first proof-of-principle Cauchy evolution of asymptotically AdS spacetimes that has been achieved with no symmetry assumptions, and to describe the framework that makes Cauchy evolution in AdS possible in full generality.
The results presented here are based on a code with adaptive mesh refinement (AMR) capabilities that solves the Einstein equations in generalized harmonic form for asymptotically AdS spacetimes, subject to reflective (i.e., Dirichlet) boundary conditions. We couple gravity to a massless scalar field, 
but the latter does not play any fundamental role in our scheme; we introduce it as a convenient mechanism to arrange for initial data whose future Cauchy development contain trapped surfaces.

Ingoing characteristic (e.g., Eddington-Finkelstein) coordinates have been successfully used to simulate dynamical spacetimes containing black branes in asymptotically AdS spacetimes in Poincar\'{e} coordinates in full generality, i.e., no symmetry assumptions.\footnote{The same coordinates were used to successfully evolve single black holes in asymptotically flat spacetimes \cite{Gomez:1998uj} and in global AdS \cite{Chesler:2018txn}.} This method has been applied to a variety of settings and by now the literature on the subject is vast and we will not review it here. We refer the reader to \cite{Chesler:2013lia} for a detailed review. This approach, however, will fail if the ingoing radial null geodesics form caustics within the numerical domain, which can happen whenever there is a strong localized perturbation of the background spacetime. For instance, the dynamical formation of localized black holes in the background of the AdS soliton spacetime \cite{Bantilan:2020pay} or even a localized black hole falling through the Poincar\'{e} horizon of AdS are just two possible examples where the ingoing coordinates of \cite{Chesler:2013lia} are likely to become singular due to the formation of caustics.\footnote{In asymptotically flat spaces, it has not been possible thus far to simulate all stages of a black hole binary with characteristic coordinates precisely because of the formation of caustics outside the black holes \cite{Bishop:1997ik,Lehner:2001wq}.} 
On the other hand, Cauchy evolution in conjunction with generalized harmonic coordinates is well-known to successfully handle strong, highly dynamical and localized gravitational fields, such as those produced by the individual black holes in a binary. 
Whilst it is possible that many problems that have been solved using ingoing coordinates in the Poincar\'{e} patch of AdS can also be solved with Cauchy evolution, the latter can be applied to situations where ingoing coordinates will almost certainly fail. Furthermore, the use of Cauchy evolution benefits from the infrastructure developed over many years to numerically solve the black hole binary problem in general relativity \cite{Pretorius:2005gq,Campanelli:2005dd,Baker:2005vv}. 
In particular, the code described in the present work has built-in AMR and is designed to run in large supercomputing clusters; both of these features will likely turn out to be crucial in solving certain key open problems in AdS. 

A key requirement for obtaining stable evolution in AdS is a gauge choice that is consistent with the conditions imposed at the AdS boundary (see, for example, \cite{Bantilan:2012vu}).
In most cases, a gauge choice leading to stable numerical evolution is typically found in spacetimes with a certain degree of symmetry.
In the present work, we detail a gauge choice in $D=4$ spacetime dimensions that leads to stable evolution in an asymptotically global AdS setting with no symmetry assumptions.
This work is a direct precursor to fully general studies of gravitational collapse and black hole formation in AdS.
In this context, Cartesian coordinates are suitable as they are regular everywhere, do not contain coordinate singularities, and do not have the well-known limitation suffered by spherical coordinates in the form of severely shorter time steps imposed by the Courant-Friedrichs-Lewy (CFL) condition. In addition, most AMR infrastructures are designed for this type of coordinates. 
Similar coordinates were used in \cite{Bantilan:2017kok} to study the non-spherically symmetric collapse of a massless scalar field in global AdS$_5$ with SO(3) symmetry. 
In anticipation of fully general studies, we choose to write our prescription in terms of global Cartesian coordinates, using second order finite difference derivative stencils to discretize the initial constraint equations and the evolution equations. 
The framework we present here straightforwardly generalizes to other settings and other discretization schemes.

The rest of this article is organized as follows.
In Section~\ref{sec:setup} we describe the setup, starting with a short review of anti-de Sitter spacetime, and two complementary characterizations of asymptotically AdS boundary conditions.
In Section~\ref{sec:pre_sta} we detail our prescription for obtaining stable Cauchy evolution with no symmetries in Cartesian coordinates. 
The crucial ingredients for this perscription are reflective Dirichlet boundary conditions imposed on appropriate evolution variables, and a specific choice of generalized harmonic source functions. In Section~\ref{sec:bouset2} we define boundary quantities whose evolution describes the physics at the AdS boundary.
In Section~\ref{sec:numerical_scheme} we outline the generalized harmonic scheme that we use in our simulations. 
Section~\ref{sec:results} contains preliminary results of simulations of gravitational collapse with no symmetry assumptions.
We conclude with a discussion in Section~\ref{sec:Discussion}. We have relegated some technical details to several appendices. In Appendix \ref{sec:GHfor} we write down the Einstein equations in harmonic coordinates. In Appendix~\ref{sec:sphevvarboucon}, we follow our prescription for the interesting case of global AdS in spherical coordinates and we obtain the corresponding stable gauge. In Appendix~\ref{sec:poincare}, we do the same for the Poincar\'e patch. Appendix \ref{sec:initdata} contains a description of our construction of initial data for the class of spacetimes considered in the paper, while in Appendix \ref{sec:GCbulk} we provide the details of our complete gauge choice, including the bulk. In Appendix \ref{sec:extrapconvbdy} we explain how we carry out the extrapolation to read off the boundary quantities. Some convergence tests are presented in Appendix \ref{sec:convbulk}.
Throughout, we use geometric units where Newton's constant is set to $G=1$ and the speed of light is set to $c=1$.

\section{Setup}\label{sec:setup}

\subsection{Anti-de Sitter Spacetime}\label{subsec:AdS}

The dynamics of gravity with a cosmological constant $\Lambda$ in four dimensions coupled to a real massless scalar field $\varphi$ can be described by the following action:
\begin{equation}\label{eqn:action}
S = \int d^4 x \sqrt{-g} \left( \frac{1}{16\pi} \left( R - 2\Lambda \right) - g^{\alpha\beta} \partial_\alpha \varphi \partial_\beta \varphi \right),
\end{equation}
where $R$ is the Ricci scalar of the metric $g_{\alpha\beta}$ with determinant $g$.
The variation of the action \eqref{eqn:action} with respect to $g_{\alpha\beta}$ and $\varphi$ gives the equations of motion
\begin{eqnarray}
\label{eqn:eoms1}
R_{\alpha\beta}- \frac{1}{2} R g_{\alpha\beta}+ \Lambda g_{\alpha\beta}&=&\nonumber\\
&&\hspace{-2.8cm} 8\pi \left( \partial_\alpha \varphi \partial_\beta \varphi - g_{\alpha\beta} \frac{1}{2} g^{\gamma\delta} \partial_{\gamma} \varphi \partial_{\delta} \varphi \right), \\
\label{eqn:eoms2}
g^{\alpha\beta} \nabla_{\alpha} \nabla_{\beta} \varphi &=& 0\,.
\end{eqnarray}

We then recast \eqref{eqn:eoms1} into generalized harmonic form.
See Appendix~\ref{sec:GHfor} for the explicit form of the resulting equations that we evolve, and \cite{Pretorius:2004jg} for more details about the theoretical aspects of the formulation.
The numerical solution we obtain is given 
in terms of the spacetime metric $g_{\alpha\beta}$, the scalar field $\varphi$ and a choice of gauge source functions $H_\alpha$. 

The metric of AdS$_4$ is the maximally symmetric vacuum (i.e., $\varphi=0$) solution of \eqref{eqn:eoms1} and \eqref{eqn:eoms2} in four dimensions.
In terms of global coordinates that cover the whole spacetime, given by $(t,r,\theta,\phi)\in(-\infty,+\infty)\times(0,+\infty)\times[0,\pi]\times[0,2\pi)$, this metric can be expressed as
\begin{equation}\label{eqn:ads4}
\hat{g}= -\left(1+\frac{r^2}{L^2}\right) dt^2 + \left(1+\frac{r^2}{L^2}\right)^{-1} dr^2 +r^2 d{\Omega_2}^2 \,,
\end{equation}
with a characteristic length scale $L$, also called AdS radius, that is related to the cosmological constant by $\Lambda = - 3/L^2$, and where $d{\Omega_2}^2 = d\theta^2 + \sin^2\theta d\phi^2$ is the metric of the round unit 2-sphere. 
A crucial feature of this spacetime is the presence of a \emph{timelike} boundary at $r \rightarrow +\infty$, which makes stable evolution of initial data possible only if boundary conditions are imposed on the evolved fields. 
In other words, any Cauchy problem in this setting is an initial-boundary value problem.

To proceed further, we first compactify $r=2\rho/(1-\rho^2/\ell^2)$ so that the AdS boundary at $r \rightarrow +\infty$ is at a finite value of the new radial coordinate, $\rho=\ell$.\footnote{We emphasize that the arbitrary compactification scale $\ell$ is completely independent of the AdS length scale $L$.}
We hereafter set $\ell=1$ without loss of generality, so that the AdS boundary is at $\rho=1$. In this way, we obtain (compactified) spherical coordinates $x^\alpha=(t,\rho,\theta,\phi)$.
Defining a convenient function $\hat{f}(\rho) = (1-\rho^2)^2+4\rho^2/L^2$, the metric of AdS$_4$ in this set of coordinates reads 
\begin{equation}\label{eqn:ads4_compact}
\hat{g} = \frac{1}{(1-\rho^2)^2} \left( -\hat{f}(\rho) dt^2 + \frac{4(1+\rho^2)^2}{\hat{f}(\rho)} d\rho^2 + 4\rho^2 d{\Omega_2}^2 \right).
\end{equation}

Second, we make use of Cartesian coordinates $x^\mu=(t,x,y,z)$ defined by $x=\rho\cos\theta$, $y=\rho\sin\theta\cos\phi$, $z=\rho\sin\theta\sin\phi$.
This allows us to bypass the severe restriction that would be imposed on the time step size near $\rho=0$ on a grid in spherical coordinates.
The metric of AdS$_4$ in Cartesian coordinates reads
\begin{align}
\label{eqn:ads4_final}
\hat{g} = \frac{1}{\left(1-\rho^2\right)^2 }\Big[ &-\hat{f}(\rho)dt^2\nonumber\\
&+\frac{4 \left(1+\rho^2\right)^2}{\rho^2 \hat{f}(\rho)} (x dx + y dy + z dz)^2\nonumber\\
&\hspace{-2.0cm}+\frac{4}{\rho^2} \Big(\left(y^2+z^2\right) dx^2 + \left(x^2+z^2\right) dy^2 + \left(x^2+y^2\right) dz^2 \nonumber\\
&\hspace{-2.0cm}-2 x y\, dx dy - 2 y z\, dy dz - 2 x z\,dx dz\Big)\Big]\,,
\end{align}
where $\rho=\rho(x,y,z)\equiv\sqrt{x^2+y^2+z^2}$.
Without loss of generality, in the rest of this work we set the AdS length scale to $L=1$. With this choice, the metric \eqref{eqn:ads4_final} takes the diagonal form
\begin{equation}
\label{eq:ads4_diag}
\hat{g} =-\left(\frac{1+\rho^2}{1-\rho^2}\right)^2dt^2+\frac{4}{(1-\rho^2)^2} \left(dx^2 + dy^2+dz^2\right).
\end{equation}

\subsection{Asymptotically anti-de Sitter Spacetimes}\label{subsec:asyAdS}

We will be interested in the Cauchy evolution of asymptotically AdS spacetimes. 
In this section we present a review of two different characterizations of such spacetimes and the relation between them, specializing to the case of $D=4$ spacetime dimensions for concreteness. 
In doing so, we will also be able to write down the boundary conditions for asymptotically AdS spacetimes in terms of these two different characterizations.

Let us start from the original arguments presented in \cite{Henneaux:1985tv}. 
The authors implicitly considered spacetimes $(M,g)$ that admit a conformal compactification, and thus a definition of conformal boundary $\partial M$.
Then they define asymptotically AdS spacetimes by requiring that the spacetime asymptotically approaches the pure AdS solution.
More precisely, for any set of global coordinates $x^\alpha$, the authors required the deviation of the full metric $g_{\alpha\beta}$ from the pure AdS metric $\hat{g}_{\alpha\beta}$, given by $h_{\alpha\beta}=g_{\alpha\beta}-\hat{g}_{\alpha\beta}$, to satisfy three conditions:

\begin{enumerate}
\item[(i)] It is consistent with the asymptotic decay of the Kerr-AdS metric near $\partial M$ in this set of coordinates.
\item[(ii)] Its fall-off near $\partial M$ is invariant under the global AdS symmetry group $O(3,2)$, i.e.,
\begin{equation}\label{eqn:asyKilleq}
(\mathcal{L}_X h)_{\alpha\beta}=\mathcal{O}(h_{\alpha\beta})\,,
\end{equation}
near the boundary $\partial M$ for any generator $X$ of $O(3,2)$.
\item[(iii)] The surface integral charges associated with the generators of $O(3,2)$ are finite.
\end{enumerate}
In addition, for the purposes of this article, we restrict this definition to spacetimes that satisfy the Einstein equations \eqref{eqn:eoms1}.

It is important to recognize that conditions (i), (ii) and (iii) can be condensed into one. Ref. \cite{Henneaux:1985tv} already shows that the explicit fall-off satisfying (i) and (ii) automatically implies (iii). 
Furthermore, requiring (ii) is sufficient to obtain the fall-off near the boundary that satisfies also (i) and (iii). 
This can be seen from the results of \cite{Henneaux:2006hk}, in which \eqref{eqn:asyKilleq} is solved in any spacetime dimension and the 4-dimensional case coincides with the fall-offs in \cite{Henneaux:1985tv}.

The condition (ii) amounts to a full spacetime metric $g_{\alpha\beta}$ that approaches the pure AdS metric $\hat{g}_{\alpha\beta}$ near $\partial M$.
This has two consequences for the terminology commonly used in the literature, as well as in this work.
First, we can refer to $\partial M$ as the AdS boundary because it has the same conformal structure as the boundary of pure AdS, i.e, $\mathbb{R}\times S^2$ topology, and metric given by that of the Einstein Static Universe. 
Second, we can define certain classes of coordinates in terms of the corresponding fall-offs of the metric components near the boundary as follows.
Given a set of coordinates $x^\alpha$ in which the pure AdS metric components are $\hat{g}_{\alpha\beta}$, we denote by $x^\alpha$ all sets of coordinates in which the full metric components $g_{\alpha\beta}$ approach the pure AdS metric components in the form $\hat{g}_{\alpha\beta}$. For example, we will denote any set of coordinates in which the metric $g$ asymptotes to $\hat{g}$ in the form \eqref{eqn:ads4_compact} by $(t,\rho,\theta,\phi)$ and we will refer to them as spherical coordinates. Similarly, we will denote any set of coordinates in which $g$ asymptotes to $\hat{g}$ in the form \eqref{eq:ads4_diag} by $(t,x,y,z)$ and we will refer to them as Cartesian coordinates.\footnote{Note that these coordinates should only be regarded as asymptotically spherical and Cartesian coordinates respectively, since they are only completely specified near the boundary $\partial M$.}

The fall-offs for the metric obtained by \cite{Henneaux:1985tv} can thus be written in the form
\begin{eqnarray}
\label{eq:sphbounconh}
h_{\rho\alpha}&=&f_{\rho\alpha}(t,\theta,\phi)(1-\rho)^2+\mathcal{O}((1-\rho)^3), \, \textrm{ if $\alpha\neq\rho$}, \nonumber \\ 
h_{\alpha\beta}&=&f_{\alpha\beta}(t,\theta,\phi)(1-\rho)+\mathcal{O}((1-\rho)^{2}), \, \; \textrm{ otherwise}, \nonumber \\
\end{eqnarray}
for arbitrary functions $f_{\alpha\beta}(t,\theta,\phi)$. 
These are supplemented by the fall-offs for the scalar field, given in \cite{Henneaux:2006hk}.
Here we restrict the discussion to a massless scalar field $\varphi$ with a fast fall-off that preserves the asymptotics \eqref{eq:sphbounconh}, for which
\begin{equation}\label{eq:sphbounconphi}
\varphi=f(t,\theta,\phi)(1-\rho)^3+\mathcal{O}((1-\rho)^4)
\end{equation}
for arbitrary $f(t,\theta,\phi)$. In Cartesian coordinates, these fall-offs read
\begin{eqnarray}
\label{eq:carbouncondh}
h_{\mu\nu}&=&f_{\mu\nu}(t,x,y,z)(1-\rho)+\mathcal{O}((1-\rho)^{2}), \\
\label{eq:carbouncondphi}
\varphi&=&f(t,x,y,z)(1-\rho)^3+\mathcal{O}((1-\rho)^{4}), 
\end{eqnarray}
for arbitrary $f_{\mu\nu}$ and $f$, and where $\rho = \rho(x,y,z)$.

The fall-offs of the source functions, involved in the generalized harmonic formulation employed in this study, can be deduced from \eqref{eq:sphbounconh} through the definition 
\begin{align}
\label{eq:defsoufunsph}
H^\alpha\equiv \Box x^\alpha&=\frac{1}{\sqrt{-g}}\partial_\beta (\sqrt{-g}g^{\beta\gamma}x^\alpha_{\;\;,\gamma})\nonumber\\
&=\frac{1}{\sqrt{-g}}\partial_\beta(\sqrt{-g}g^{\beta\alpha}).
\end{align}
In spherical coordinates, denoting the pure AdS values by $\hat{H}_\alpha$, \eqref{eq:sphbounconh} and \eqref{eq:defsoufunsph} imply
\begin{eqnarray}\label{eq:sphbouncondsoufunc}
H_\alpha&=&\hat{H}_\alpha+f_\alpha(t,\theta,\phi)(1-\rho)^3+\mathcal{O}((1-\rho)^4), \, \textrm{ if $\alpha\neq\rho$,} \nonumber \\
H_\rho&=&\hat{H}_\rho+f_\rho(t,\theta,\phi)(1-\rho)^2+\mathcal{O}((1-\rho)^3),
\end{eqnarray}
for arbitrary $f_\alpha$.
In Cartesian coordinates, denoting the pure AdS values by $\hat{H}_\mu$, \eqref{eq:carbouncondh} and \eqref{eq:defsoufunsph} imply
\begin{equation}\label{eq:carbouncondsoufun}
H_\mu=\hat{H}_\mu+f_\mu(t,x,y,z)|_{\rho=1}(1-\rho)^2+\mathcal{O}((1-\rho)^3)
\end{equation}
for arbitrary $f_\mu$ and $\rho=\rho(x,y,z)$.

A different characterization of asymptotically AdS spacetimes can be given in terms of the well-known Fefferman-Graham (FG) expansion \cite{AST_1985__S131__95_0}.
In this approach, one starts with the definition of a locally asymptotically AdS spacetime $(M,g)$ as a spacetime that admits a conformal compactification, thus allowing the definition of a conformal boundary $\partial M$, and that satisfies the Einstein equations \eqref{eqn:eoms1}. No assumption is made at this stage on the topology of the boundary.
The FG theorem states that one can always find a coordinate system $x^{\bar{\alpha}}=(\bar{t},\bar{z},\bar{\theta},\bar{\phi})$ in a neighbourhood of the boundary for which the boundary is at $\bar{z}=0$ and the metric can be written in the form
\begin{equation}
\label{eqn:FGmetric}
g=\frac{1}{\bar{z}^2}(d\bar{z}^2+g_{\bar{a}\bar{b}}dx^{\bar{a}}dx^{\bar{b}}),
\end{equation}
where 
\begin{equation}
\label{eqn:FGbdymetric}
g_{\bar{a}\bar{b}}(\bar{t},\bar{z},\bar{\theta},\bar{\phi})=g_{(0)\bar{a}\bar{b}}(\bar{t},\bar{\theta},\bar{\phi})+g_{(2)\bar{a}\bar{b}}(\bar{t},\bar{\theta},\bar{\phi})\bar{z}^2+\mathcal{O}(\bar{z}^3).
\end{equation}
Then, the near-boundary (i.e., about $\bar{z}=0$) expansion of the Einstein equations completely determines the coefficient $g_{(2)\bar{a}\bar{b}}$ in terms of $g_{(0)\bar{a}\bar{b}}$.
Therefore the dynamics that makes this spacetime differ from pure AdS appears at order $\bar{z}^3$ in the expansion of $g_{\bar{a}\bar{b}}$. If we make the further requirement that the topology of the boundary is the same as in the pure AdS case, i.e., $\mathbb{R}\times S^2$, the spacetime becomes globally asymptotically AdS and this characterization becomes equivalent to the one obtained from the original arguments in \cite{Henneaux:1985tv}. 
The FG form \eqref{eqn:FGmetric} of the metric immediately provides the near-boundary behaviour and shows that coordinates can be defined so that the $\bar{z}\bar{z}$ component of any asymptotically AdS metric goes as $1/\bar{z}^2$, and the $\bar{z}\bar{a}$ components vanish in a neighbourhood of the AdS boundary.

We conclude this section by showing an explicit example of how FG coordinates can be found in the case of general asymptotically AdS$_4$ spacetimes. 
We start from the general form for the asymptotically AdS metric in spherical coordinates $x^\alpha$, given by $g_{\alpha\beta}=\hat{g}_{\alpha\beta}+h_{\alpha\beta}$.
The deviations $h_{\alpha\beta}$ from the pure AdS metric $\hat{g}_{\alpha\beta}$ have fall-offs that are given by the asymptotically AdS boundary conditions \eqref{eq:sphbounconh}.
Defining $z=2(1-\rho)/(1+\rho)$, we can bring the pure AdS metric \eqref{eqn:ads4_compact} into the FG form.
Since $z$ asymptotes to $1-\rho$ near the AdS boundary $\rho=1$, we can use \eqref{eq:sphbounconh} to immediately write down the metric fall-offs in terms of our new coordinate $z$.
The metric in these coordinates reads
\begin{eqnarray}
\label{eq:metnewcoords}
g
&=\frac{1}{z^2}\biggl[& 
- \left(1+z^2/2+f_{tt}z^3+\mathcal{O}(z^4)\right)dt^2 \nonumber \\
&&\hspace{0.0cm} + \left(1+f_{\rho\rho}z^3+\mathcal{O}(z^4)\right)dz^2 \nonumber \\
&&\hspace{0.0cm} + \left(1-z^2/2+f_{\theta\theta}z^3 +\mathcal{O}(z^4) \right)d\theta^2 \nonumber \\
&&\hspace{0.0cm} + \sin^2\theta \left(1-z^2/2+\frac{f_{\phi\phi}}{\sin^2\theta} z^3 +\mathcal{O}(z^4) \right)d\phi^2 \nonumber \\ 
&&\hspace{-0.8cm}+ 2 \left( f_{t\theta}z^3 +\mathcal{O}(z^4)\right)dt d\theta + 2 \left( f_{t\phi} z^3+\mathcal{O}(z^4)\right)dt d\phi \nonumber \\
&&\hspace{-0.8cm}+ 2 \left(f_{\theta\phi} z^3 +\mathcal{O}(z^4)\right)d\theta d\phi - 2\left(f_{t\rho}z^4 +\mathcal{O}(z^5)\right)dtdz \nonumber \\
&&\hspace{-0.8cm}-2\left(f_{\rho\theta}z^4+\mathcal{O}(z^5)\right)dzd\theta \nonumber \\
&&\hspace{-0.8cm}-2\left(f_{\rho\phi}z^4 +\mathcal{O}(z^5)\right)dzd\phi\biggr] ,
\end{eqnarray}
where the coefficients $f_{\alpha\beta}$ in the expansion above are functions of $(t,\theta,\phi)$. 
Notice that the metric in \eqref{eq:metnewcoords} is not in the FG form yet because the $zz$ component is not $1/z^2$ up to the desired order in $z$, and the $tz$, $z\theta$, $z\phi$ components do not vanish up to $\mathcal{O}(z^2)$.
Defining 
\begin{eqnarray}
\label{eqn:FGcoords}
\bar{t}&=&t+\frac{1}{10 } \left(2 f_{t\rho}(t,\theta,\phi)+\frac{1}{3}f_{\rho\rho,t}(t,\theta,\phi)\right)z^5+\mathcal{O}(z^6),\nonumber \\
\bar{z}&=&z+\frac{1}{6}f_{\rho\rho}(t,\theta,\phi)z^4+\mathcal{O}(z^5),\nonumber \\
\bar{\theta}&=&\theta-\frac{1}{10} \left(2 f_{\rho\theta}(t,\theta,\phi)+\frac{1}{3}f_{\rho\rho,\theta}(t,\theta,\phi)\right)z^5+\mathcal{O}(z^6),\nonumber \\
\bar{\phi}&=&\phi-\frac{1}{10} \left(2 f_{\rho\phi}(t,\theta,\phi)+\frac{1}{3}f_{\rho\rho,\phi}(t,\theta,\phi)\right)z^5+\mathcal{O}(z^6),\nonumber \\
\end{eqnarray}
which can be inverted near the boundary as
\begin{eqnarray}
\label{eqn:invertFGcoords}
t&=&\bar{t}-\frac{1}{10} \left(2 f_{t\rho}(\bar{t},\bar{\theta},\bar{\phi})+\frac{1}{3}f_{\rho\rho,\bar{t}}(\bar{t},\bar{\theta},\bar{\phi})\right)\bar{z}^5+\mathcal{O}(\bar{z}^6),\nonumber \\
z&=&\bar{z}-\frac{1}{6}f_{\rho\rho}(\bar{t},\bar{\theta},\bar{\phi})\bar{z}^4+\mathcal{O}(\bar{z}^5),\nonumber \\
\theta&=&\bar{\theta}+\frac{1}{10} \left(2 f_{\rho\theta}(\bar{t},\bar{\theta},\bar{\phi})+\frac{1}{3}f_{\rho\rho,\bar{\theta}}(\bar{t},\bar{\theta},\bar{\phi})\right)\bar{z}^5+\mathcal{O}(\bar{z}^6),\nonumber \\
\phi&=&\bar{\phi}+\frac{1}{10} \left(2 f_{\rho\phi}(\bar{t},\bar{\theta},\bar{\phi})+\frac{1}{3}f_{\rho\rho,\bar{\phi}}(\bar{t},\bar{\theta},\bar{\phi})\right)\bar{z}^5+\mathcal{O}(\bar{z}^6),\nonumber \\
\end{eqnarray}
we finally obtain the metric in FG form:
\begin{align}
g=\frac{1}{\bar{z}^2}\biggl[ 
&d\bar{z}^2 \nonumber \\
&-\left(1+\frac{\bar{z}^2}{2}+\left( \textstyle{f_{tt}-\frac{1}{3}f_{\rho\rho}}\right)\bar{z}^3+\mathcal{O}(\bar{z}^4) \right)d\bar{t}^2 \nonumber \\
&+ \left(1-\frac{\bar{z}^2}{2}+\left(\textstyle{f_{\theta\theta}+\frac{1}{3}f_{\rho\rho}} \right)\bar{z}^3 +\mathcal{O}(\bar{z}^4)\right)d\bar{\theta}^2\nonumber\\
&\hspace{-0.8cm} +  \sin^2\bar{\theta} \left(1-\frac{\bar{z}^2}{2}+\left(\textstyle{\frac{f_{\phi\phi}}{ \sin^2\bar{\theta}}+\frac{1}{3}f_{\rho\rho}}\right)\bar{z}^3 +\mathcal{O}(\bar{z}^4)\right)d\bar{\phi}^2\nonumber\\
&\hspace{-0.8cm}+ 2 \left(f_{t\theta}\bar{z}^3 +\mathcal{O}(\bar{z}^4) \right) d\bar{t} d\bar{\theta} + 2 \left( f_{t\phi}\bar{z}^3 +\mathcal{O}(\bar{z}^4)\right)d\bar{t} d\bar{\phi} \nonumber \\
&\hspace{-0.8cm}+ 2\left( f_{\theta\phi} \bar{z}^3+\mathcal{O}(\bar{z}^4)\right)d\bar{\theta} d\bar{\phi} 
\nonumber\\
&\hspace{-0.8cm}+\mathcal{O}(\bar{z}^5)d\bar{t}d\bar{z}+\mathcal{O}(\bar{z}^5)d\bar{z}d\bar{\theta}+\mathcal{O}(\bar{z}^5)d\bar{z}d\bar{\phi}
\biggr]\,,
\label{eq:asyFG}
\end{align}
where now the coefficients $f_{\alpha\beta}$ are functions of $(\bar t, \bar\theta, \bar\phi)$. 
Notice that $f_{\rho\rho}$ has been reabsorbed in $g_{\bar{t}\bar{t}}, g_{\bar{\theta}\bar{\theta}}, g_{\bar{\phi}\bar{\phi}}$. From the form of the metric in \eqref{eq:asyFG}, we could use the holographic renormalization prescription of \cite{deHaro:2000vlm} to read off the boundary CFT stress tensor. See Appendix \ref{sec:HoloRen} for more details.

\section{Boundary Prescription}\label{sec:pre_sta}

In this section, we present our prescription to obtain a choice of generalized harmonic gauge source functions that achieves stable evolution.
We choose to do so using Cartesian coordinates, as they provide a suitable chart to evolve points near the centre of the grid, which is necessary when analyzing gravitational collapse and black hole formation. 
This procedure generalizes in a straightforward manner to other asymptotically AdS spacetimes in $D\geq 4$ spacetime dimensions, different coupling with matter fields, and coordinates on global AdS or on the Poincar\'{e} patch. We consider the application to spherical coordinates in Appendix~\ref{sec:sphevvarboucon}, and to the Poincar\'e patch in Appendix~\ref{sec:poincare}.
We impose asymptotically AdS boundary conditions \eqref{eq:carbouncondh}, \eqref{eq:carbouncondphi}, \eqref{eq:carbouncondsoufun} as reflective Dirichlet boundary conditions on appropriate evolution variables, as explained in the next section.
For a discussion in a simpler context with more symmetry, see~\cite{Bantilan:2012vu}.

\subsection{Evolution Variables and Boundary Conditions}\label{subsec:cartevvarboucon}

The boundary conditions on asymptotically AdS spacetimes, discussed in Section~\ref{subsec:asyAdS}, can be imposed as Dirichlet boundary conditions at the AdS boundary.
This requires appropriately defining and evolving a new set of variables, from which the full solution $(g_{\mu\nu},\varphi,H_\mu)$ can be subsequently reconstructed. 
Here, we define evolution variables in the Cartesian coordinates employed by our numerical scheme.
Later, in Section~\ref{sec:bouset2} we will show expressions for quantities at the AdS boundary in spherical coordinates.
In Appendix~\ref{sec:sphevvarboucon}, we explicitly show how these spherical variables relate to our Cartesian evolution variables.

The Cartesian metric evolution variables, $\bar{g}_{\mu\nu}$, are defined by first considering the deviation from pure AdS in Cartesian coordinates, $h_{\mu\nu}=g_{\mu\nu}-\hat{g}_{\mu\nu}$, then stripping $h_{\mu\nu}$ of as many factors of $(1-\rho^2)$ as needed so that each component falls off linearly in $(1-\rho)$ near the AdS boundary at $\rho=1$.\footnote{Looking at the boundary conditions \eqref{eq:carbouncondh}, it seems natural to factor out $(1-\rho)$ rather than $(1-\rho^2)$. However, the latter is preferred since it preserves the even/odd character in the $\rho$ variable.}
We see from \eqref{eq:carbouncondh} that in four dimensions, the metric evolution variables $\bar{g}_{\mu\nu}$ that satisfy these requirements are simply 
\begin{equation}\label{eq:gbarcart}
\bar{g}_{\mu\nu}=h_{\mu\nu}\,.
\end{equation}
Similarly, the Cartesian boundary condition on the scalar field \eqref{eq:carbouncondphi} suggests that we use the evolution variable
\begin{equation}
\label{eq:phibarcart}
\bar{\varphi}=\frac{\varphi }{(1-\rho^2)^2}\,.
\end{equation}
Finally, the boundary conditions \eqref{eq:carbouncondsoufun} on $H_\mu$ suggest the use of
\begin{equation}\label{eq:soufunb}
\bar{H}_\mu=\frac{H_\mu-\hat{H}_\mu}{1-\rho^2 }\,.
\end{equation}

For evolved variables defined in this way, the boundary conditions \eqref{eq:carbouncondh}, \eqref{eq:carbouncondphi}, \eqref{eq:carbouncondsoufun} can be easily imposed as Dirichlet boundary conditions at the AdS boundary:
\begin{equation}
\label{eq:dirbc}
 \bar{g}_{\mu\nu}\big|_{\rho=1}=0\,,\quad \bar{\varphi}\big|_{\rho=1}=0\,,\quad \bar{H}_\mu\big|_{\rho=1}=0\,.
 \end{equation}

\subsection{Gauge Choice for Stability}\label{sec:gauge_choice}

Coordinates over the entire spacetime are fully determined only once we choose the gauge source functions $H_\mu$. 
In Cartesian coordinates, as can be seen from \eqref{eq:carbouncondsoufun}, $H_\mu$ are fixed up to order $1-\rho$ by its pure AdS values $\hat{H}_\mu$ in an expansion near the AdS boundary.
As we shall see, the choice of $H_\mu$ at the next order in this expansion, $(1-\rho)^2$, cannot be completely arbitrary if we wish to achieve stable evolution.
A specification of generalized harmonic source functions at order $(1-\rho)^2$ that provides stable Cauchy evolution can be obtained following the procedure detailed in this section.

The first step involves expanding the evolved variables, $\bar{g}_{\mu \nu}$, $\bar{H}_{\mu}$ and $\bar{\varphi}$, in a power series about $(1-\rho) \equiv q = 0$. 
By construction, these evolved variables are linear in $q$ at leading order:
\begin{eqnarray}\label{eqn:qexp}
&\bar{g}_{\mu \nu} = \bar{g}_{(1) \mu \nu} q + \bar{g}_{(2) \mu \nu} q^2 + \bar{g}_{(3) \mu \nu} q^3 + \mathcal{O}(q^4), \label{eqn:qexpg} \\
&\bar{H}_{\mu} = \bar{H}_{(1) \mu} q + \bar{H}_{(2) \mu} q^2 + \bar{H}_{(3) \mu} q^3 + \mathcal{O}(q^4) ,\label{eqn:qexpH}\\
&\hspace{-0.5cm}\bar{\varphi} = \bar{\varphi}_{(1)} q + \bar{\varphi}_{(2)} q^2 + \bar{\varphi}_{(3)} q^3 + \mathcal{O}(q^4), \label{eqn:qexpphi}
\end{eqnarray}
where all the coefficients are functions of the coordinates $(t,x,y,z)$ on the boundary $\rho(x,y,z)\equiv\sqrt{x^2+y^2+z^2}=1$ (or $q(x,y,z)=0$).
We now substitute these variables into the evolution equations \eqref{eqn:efe_gh_modified}, and we expand each component in powers of $q$. The three lowest orders, $q^{-2}$, $q^{-1}$, $q^0$, are fixed by the pure AdS metric $\hat{g}$ which itself is a solution of \eqref{eqn:efe_gh_modified}, so these terms vanish trivially. The remaining orders vanish only if $\bar{g}_{\mu \nu}$, $\bar{H}_{\mu}$, $\bar{\varphi}$ are a solution of \eqref{eqn:efe_gh_modified}.

We are now interested in identifying the order of $q$ at which the second derivatives of $ \bar{g}_{(1) \mu \nu}$ with respect to $(t,x,y,z)$ appear.
For each component, we denote their combination by $\tilde{\Box}\bar{g}_{(1)\mu\nu}$, i.e., 
\begin{equation}
\label{eq:tildeboxbarg}
\tilde{\Box}\bar{g}_{(1)\mu\nu}\equiv\biggl(c^t_{\mu\nu}\frac{\partial^2}{\partial t^2}+c^x_{\mu\nu}\frac{\partial^2}{\partial x^2}+c^y_{\mu\nu}\frac{\partial^2}{\partial y^2}+c^z_{\mu\nu}\frac{\partial^2}{\partial z^2}\biggr)\bar{g}_{(1)\mu\nu}\,,
\end{equation}
for some functions $c^t_{\mu\nu}$, $c^x_{\mu\nu}$, $c^y_{\mu\nu}$, $c^z_{\mu\nu}$ of $(t,x,y,z)$ at $\rho(x,y,z)=1$.\footnote{None of these coefficients are tensors, despite the notation, and there is no sum over repeated indices.}
These derivative terms are included in the first piece of \eqref{eqn:efe_gh_modified}, namely in $-\frac{1}{2}g^{\rho \sigma} \bar{g}_{\mu \nu, \rho \sigma}$. From this, we can easily find their order of $q$ by recalling that the leading order of the inverse metric is given by its purely AdS piece, $g^{\mu\nu}=\mathcal{O}(\hat{g}^{\mu\nu})=\mathcal{O}(q^{2})$, and $\bar{g}_{(1)\mu\nu}$ is multiplied by $q$ in the near-boundary expression of $\bar{g}_{\mu\nu}$ (see eq. \eqref{eqn:qexpg}). Thus, $\tilde{\Box}\bar{g}_{(1)\mu\nu}$ must appear in the coefficient of order $q^{3}$ for every component of \eqref{eqn:efe_gh_modified}.\footnote{$\mathcal{O}(\hat{g}^{\mu\nu})=\mathcal{O}(q^{2})$ is true in any number of dimensions but only for Cartesian coordinates. For an arbitrary set of coordinates, the leading power in $\hat{g}^{\mu\nu}$, and hence the order at which the operator \eqref{eq:tildeboxbarg} appears, depends on the specific component under consideration.
See Appendix~\ref{sec:sphevvarboucon} and \cite{Bantilan:2012vu} for examples in spherical coordinates in 4 and 5 dimensions, respectively.}
In other words, each component of the expansion of \eqref{eqn:efe_gh_modified} near $q=0$ can be written in the schematic form:
\begin{eqnarray}\label{eq:efefullexp}
0 
&=& A_{(1)\mu\nu}q+A_{(2)\mu\nu}q^2+A_{(3)\mu\nu}q^3+A_{(4)\mu\nu}q^4+\mathcal{O}(q^5) \nonumber \\
&=& A_{(1)\mu\nu}q+A_{(2)\mu\nu}q^2+(\tilde{\Box}\bar{g}_{(1)\mu\nu}+B_{(3)\mu\nu})q^3\nonumber \\
&&+A_{(4)\mu\nu}q^4  +\mathcal{O}(q^5)
\end{eqnarray}
or, rearranging the terms in order to obtain wave-like equations,
\begin{equation}
\label{eq:waveEFE}
\tilde{\Box}\bar{g}_{(1)\mu\nu}=-A_{(1)\mu\nu}\frac{1}{q^2}-A_{(2)\mu\nu}\frac{1}{q}-B_{(3)\mu\nu}-A_{(4)\mu\nu}q+\mathcal{O}(q^2).
\end{equation}
Similar arguments show that the terms involving the scalar field, with the fast fall-off that we have chosen in \eqref{eq:sphbounconphi}, appear in $A_{(4)\mu\nu}$ and higher order coefficients of \eqref{eq:efefullexp}. A similar result holds in any number of dimensions and any set of coordinates $x^\alpha$: the terms involving fastly-decaying matter fields appear at the next order with respect to the order of $\tilde{\Box}\bar{g}_{(1)\alpha\beta}$ in the near-boundary expansion of the Einstein equations. This implies that the details of the matter sector, e.g., the value of the mass of a matter field, do not affect the results of the prescription presented here, since only the lowest order coefficients in the expansion of the Einstein equations are relevant.

We now explicitly write the lowest order terms of the Einstein equations in the wave-like form \eqref{eq:waveEFE}.
The near-boundary expansion is most easily obtained by first writing the Cartesian coordinates $(x,y,z)$ in terms of the boundary-adapted spherical coordinates $(q,\theta,\phi)$, and then expanding near $q=0$. We find
\begin{widetext}
\begin{eqnarray}\label{eqn:efett}
\tilde{\Box}\bar{g}_{(1)tt}&=&-(\cos \theta (3 \cos \theta \bar{g}_{(1)xx}-2 \bar{H}_{(1)x})+\sin\theta (3 \sin \theta \cos^2\phi \bar{g}_{(1) yy}+3
  \sin \theta \sin \phi (2 \cos \phi \bar{g}_{(1) yz} \nonumber \\
&&+\sin\phi
  \bar{g}_{(1) zz})-2 (\cos \phi \bar{H}_{(1) y}+\sin\phi
  \bar{H}_{(1) z}))+3 \sin 2 \theta \cos \phi \bar{g}_{(1) xy} \nonumber \\
&&+3
  \sin 2 \theta \sin \phi \bar{g}_{(1) xz})q^{-2} +\mathcal{O}(q^{-1}),\\
\label{eqn:efetx}
\tilde{\Box}\bar{g}_{(1)tx}&=&-2 \cos \theta (3 \cos\theta \bar{g}_{(1) tx}+3 \sin \theta
  (\cos \phi \bar{g}_{(1) ty}+\sin \phi \bar{g}_{(1)tz})-2
  \bar{H}_{(1) t})  q^{-2}+\mathcal{O}(q^{-1}),\\
\label{eqn:efety}
\tilde{\Box}\bar{g}_{(1)ty}&=&-2 \cos \phi \sin\theta (3 \cos\theta \bar{g}_{(1) tx}+3 \sin \theta
  (\cos \phi \bar{g}_{(1) ty}+\sin \phi \bar{g}_{(1)tz})-2
  \bar{H}_{(1) t})  q^{-2}+\mathcal{O}(q^{-1}),\\
\label{eqn:efetz}
\tilde{\Box}\bar{g}_{(1)tz}&=&-2 \sin \theta \sin\phi (3 \cos\theta \bar{g}_{(1) tx}+3 \sin \theta
  (\cos \phi \bar{g}_{(1) ty}+\sin \phi \bar{g}_{(1)tz})-2
  \bar{H}_{(1) t})  q^{-2}+\mathcal{O}(q^{-1}),\\
\label{eqn:efexx}
\tilde{\Box}\bar{g}_{(1)xx}&=&\frac{1}{4} (3 (-4 \cos ^2\theta (\bar{g}_{(1) tt}+2 \bar{g}_{(1)
  xx})+(\cos 2 \theta +3) (\bar{g}_{(1) yy}+\bar{g}_{(1)
zz}) +8 \cos \theta \bar{H}_{(1) x})\nonumber \\
&&-8 \sin \theta \cos \phi 
  (3 \cos \theta \bar{g}_{(1)xy}+\bar{H}_{(1) y}) -8 \sin\theta \sin\phi (3 \cos\theta \bar{g}_{(1) xz}+\bar{H}_{(1) z}) \nonumber \\
  &&+6 \sin^2\theta \cos 2 \phi (\bar{g}_{(1)yy}-\bar{g}_{(1)zz})+12
  \sin^2\theta \sin 2 \phi \bar{g}_{(1) yz})  q^{-2} +\mathcal{O}(q^{-1}),\\
\label{eqn:efexy}
\tilde{\Box}\bar{g}_{(1)xy}&=&-\frac{1}{2} (2 \sin \theta \cos \phi (3 \cos \theta (\bar{g}_{(1)tt}+\bar{g}_{(1)xx}+\bar{g}_{(1)yy}-\bar{g}_{(1)zz})-4 \bar{H}_{(1) x}) \nonumber \\
&&+3 \bar{g}_{(1)xy} (2 \cos 2\theta \sin ^2\phi +\cos 2 \phi +3)+6 \sin ^2\theta \sin 2 \phi 
  \bar{g}_{(1)xz} \nonumber \\
  &&+6 \sin 2 \theta \sin \phi \bar{g}_{(1)yz}-8 \cos
  \theta  \bar{H}_{(1) y})    q^{-2} +\mathcal{O}(q^{-1}),\\
\label{eqn:efexz}
\tilde{\Box}\bar{g}_{(1)xz}&=&- (\sin \theta  \sin \phi  (3 \cos \theta  (\bar{g}_{(1)tt}+\bar{g}_{(1)xx}-\bar{g}_{(1)yy}+\bar{g}_{(1)zz})-4
   \bar{H}_{(1) x}) \nonumber \\
   &&+3 \sin ^2\theta \sin 2 \phi  \bar{g}_{(1)xy}-3 \sin
   ^2\theta  \cos 2 \phi  \bar{g}_{(1)xz}+\frac{3}{2} (\cos 2 \theta +3)
   \bar{g}_{(1)xz} \nonumber \\
   &&+3 \sin 2 \theta  \cos \phi  \bar{g}_{(1)yz}-4 \cos
   \theta  \bar{H}_{(1)z})    q^{-2} +\mathcal{O}(q^{-1}),\\
\label{eqn:efeyy}
\tilde{\Box}\bar{g}_{(1)yy}&=&-( (\sin \theta (3 \sin \theta  (2 \cos ^2\phi  \bar{g}_{(1)yy}+\sin 2 \phi  \bar{g}_{(1)yz}-\bar{g}_{(1)zz}) -6 \cos\phi \bar{H}_{(1) y}+2 \sin \phi  \bar{H}_{(1) z}) \nonumber \\
&&+6 \sin \theta \cos
   \theta \cos \phi  \bar{g}_{(1)xy}-6 \sin \theta  \cos \theta  \sin   \phi  \bar{g}_{(1)xz}+2 \cos \theta  \bar{H}_{(1) x}) \nonumber \\
   &&-3 \sin ^2\theta \cos ^2\phi  \bar{g}_{(1)tt}+\frac{3}{4} \bar{g}_{(1)xx} (2 \sin ^2\theta  \cos 2 \phi +\cos 2 \theta +3)  )  q^{-2} +\mathcal{O}(q^{-1}),\\
\label{eqn:efeyz}
\tilde{\Box}\bar{g}_{(1)yz}&=&-\frac{1}{2} \sin \theta (4 \sin \phi  (3 \cos \theta  \bar{g}_{(1)xy}-2 \bar{H}_{(1) y})+4 \cos \phi  (3 \cos \theta  \bar{g}_{(1)xz}-2 \bar{H}_{(1) z}) \nonumber \\
&&+3 \sin \theta  \sin 2 \phi  (\bar{g}_{(1)tt}-\bar{g}_{(1)xx}+\bar{g}_{(1)yy}+\bar{g}_{(1)zz})+12 \sin \theta  \bar{g}_{(1)yz})  q^{-2} +\mathcal{O}(q^{-1}),\\
\label{eqn:efezz}
\tilde{\Box}\bar{g}_{(1)zz}&=&(-2 \cos \theta  (3 \sin \theta  \sin \phi  \bar{g}_{(1)xz}+\bar{H}_{(1)x}) + \sin \theta  (3 \sin \theta  \bar{g}_{(1)yy}\nonumber \\
&&-6 \sin\theta  \sin \phi  (\cos \phi  \bar{g}_{(1)yz}+\sin \phi 
   \bar{g}_{(1)zz}) -2 \cos \phi  \bar{H}_{(1) y} +6 \sin \phi  \bar{H}_{(1)z}) -3  \sin ^2\theta  \sin ^2\phi  \bar{g}_{(1)tt} \nonumber \\
   &&+\frac{3}{4}  \bar{g}_{(1)xx} (-2 \sin ^2\theta  \cos 2 \phi +\cos 2 \theta
   +3) +3 \sin 2 \theta  \cos \phi  \bar{g}_{(1)xy})  q^{-2} +\mathcal{O}(q^{-1}), 
\end{eqnarray}
\end{widetext}

where the coordinates $(q,\theta,\phi)$ should be understood as functions of $(x,y,z)$. 
All that remains is to write down the generalized harmonic constraints $C_\mu \equiv H_\mu-\Box x_\mu = 0$ at leading order in the same near-boundary expansion. We get

\begin{widetext}
\begin{eqnarray}
\label{eqn:ct}
C_t&=&q^2 (-3 \cos \theta \bar{g}_{(1)tx}-3 \sin \theta \cos \phi \bar{g}_{(1)ty}-3 \sin \theta \sin \phi \bar{g}_{(1)tz}+2
  \bar{H}_{(1) t})+\mathcal{O}(q^3),\\
\label{eqn:cx}
C_x&=&\frac{1}{2} q^2 (-3 \cos \theta  \bar{g}_{(1)tt}-3 \cos \theta  \bar{g}_{(1)xx}-6 \sin \theta  \cos \phi  \bar{g}_{(1)xy}-6 \sin
   \theta  \sin \phi  \bar{g}_{(1)xz} \nonumber \\
  &&+3 \cos \theta  \bar{g}_{(1)yy}+3
   \cos \theta  \bar{g}_{(1)zz}+4 \bar{H}_{(1) x})+\mathcal{O}(q^3),\\
\label{eqn:cy}
C_y&=&\frac{1}{2} q^2 (-3 \sin \theta  \cos \phi  \bar{g}_{(1)tt}+3 \sin
   \theta  \cos \phi  \bar{g}_{(1)xx}-6 \cos \theta  \bar{g}_{(1)xy} \nonumber \\
   &&-3
   \sin \theta  \cos \phi  \bar{g}_{(1) yy}-6 \sin \theta  \sin \phi
   \bar{g}_{(1)yz}+3 \sin \theta  \cos \phi  \bar{g}_{(1)zz}+4
   \bar{H}_{(1) y}) +\mathcal{O}(q^3),\\
\label{eqn:cz}
C_z&=&\frac{1}{2} q^2 (-3 \sin \theta \sin \phi  \bar{g}_{(1)tt}+3 \sin
   \theta \sin \phi \bar{g}_{(1)xx}-6 \cos \theta  \bar{g}_{(1)xz} \nonumber \\
   &&+3
   \sin \theta \sin \phi  \bar{g}_{(1)yy}-6 \sin \theta  \cos \phi    \bar{g}_{(1)yz}-3 \sin \theta  \sin \phi  \bar{g}_{(1)zz}+4
   \bar{H}_{(1)z})+\mathcal{O}(q^3).
\end{eqnarray}
\end{widetext}

In the generalized harmonic formulation, choosing a gauge amounts to choosing a set of generalized harmonic source functions $\bar{H}_\mu$ for the entire evolution. Although we expect that many gauge choices are allowed,~\cite{Bantilan:2012vu} mentions a few that do not give rise to stable evolutions. We now present a procedure that provides the stable gauge in our Cartesian simulations. We believe that our prescription provides a stable gauge in a variety of settings of physical interest, such as higher spacetime dimensions, various couplings to matter fields, different types of global coordinates or Poincar\'{e} coordinates. Thus, it enables numerical Cauchy evolution in AdS in full generality, that is, with no symmetry assumptions.
The steps that lead to our stable gauge, in a form that can be easily applied to all previously mentioned cases, are the following.
\begin{enumerate}
\item Solve the leading order of the near-boundary generalized harmonic constraints for $\bar{H}_{(1)\mu}$.
For example, in the Cartesian case, the leading orders of \eqref{eqn:ct}--\eqref{eqn:cz} vanish for:
\begin{eqnarray}\label{eqn:target_gauge_txyz_step1}
\bar{H}_{(1)t}&=&\frac{3}{2\sqrt{x^2+y^2+z^2}}(x \bar{g}_{(1)tx}+y\bar{g}_{(1)ty}+z\bar{g}_{(1)tz}), \nonumber\\
\bar{H}_{(1)x}&=&\frac{3}{4\sqrt{x^2+y^2+z^2}}(2y \bar{g}_{(1)xy}+2z \bar{g}_{(1)xz}\nonumber \\
&&\hspace{1.1cm}+x(\bar{g}_{(1)tt}+ \bar{g}_{(1)xx}-\bar{g}_{(1)yy}-\bar{g}_{(1)zz})), \nonumber \\
\bar{H}_{(1)y}&=&\frac{3}{4\sqrt{x^2+y^2+z^2}}(2x \bar{g}_{(1)xy}+2z \bar{g}_{(1)yz}\nonumber \\
&&\hspace{1.1cm}+y(\bar{g}_{(1)tt}+ \bar{g}_{(1)xx}-\bar{g}_{(1)yy}-\bar{g}_{(1)zz})), \nonumber \\
\bar{H}_{(1)z}&=&\frac{3}{4\sqrt{x^2+y^2+z^2}}(2x \bar{g}_{(1)xz}+2y \bar{g}_{(1)yz}\nonumber \\
&&\hspace{1.1cm}+z(\bar{g}_{(1)tt}+ \bar{g}_{(1)xx}-\bar{g}_{(1)yy}-\bar{g}_{(1)zz})). \nonumber\\
\end{eqnarray}

\item Let $N_{(1)}$ be the lowest order in $q$ appearing in the near-boundary expansions of all the $\tilde{\Box}\bar{g}_{(1)\mu\nu}$. Plug the source functions obtained in step 1 into the $q^{N_{(1)}}$ terms of the near-boundary expansions $\tilde{\Box}\bar{g}_{(1)\mu\nu}$.
This gives a number of independent equations that, together with their derivatives, ensure tracelessness and conservation of the boundary stress-energy tensor (see Section~\ref{sec:bouset2}).\footnote{We show this in Appendix~\ref{sec:sphevvarboucon} using spherical coordinates, since they are adapted to the AdS boundary and make the proof less unwieldy.}
Solve these equations for an equal number of metric coefficients $\bar g_{(1)\mu\nu}$ and their derivatives.
In the Cartesian case, $N_{(1)}=-2$ and there is only one independent equation given by
\begin{equation}
\label{eq:cart_tracelessness}
\bar{g}_{(1)tt}-\bar{g}_{(1)xx}-\bar{g}_{(1)yy}-\bar{g}_{(1)zz}=0,
\end{equation}
which we can solve, for instance, in terms of $\bar g_{(1)tt}$.

\item Plug the solutions to the equations in step 2 into the gauge obtained in step 1.
In Cartesian coordinates, using \eqref{eq:cart_tracelessness} to eliminate $\bar{g}_{(1)tt}$ from \eqref{eqn:target_gauge_txyz_step1}, we have
\begin{eqnarray}\label{eqn:target_gauge_txyz}
\bar{H}_{(1)t}&=&\frac{3}{2\sqrt{x^2+y^2+z^2}}(x \bar{g}_{(1)tx}+y\bar{g}_{(1)ty}+z\bar{g}_{(1)tz}),\nonumber\\
\bar{H}_{(1)x}&=&\frac{3}{2\sqrt{x^2+y^2+z^2}}(x \bar{g}_{(1)xx}+y\bar{g}_{(1)xy}+z\bar{g}_{(1)xz}), \nonumber \\
\bar{H}_{(1)y}&=&\frac{3}{2\sqrt{x^2+y^2+z^2}}(x \bar{g}_{(1)xy}+y\bar{g}_{(1)yy}+z\bar{g}_{(1)yz}), \nonumber \\
\bar{H}_{(1)z}&=&\frac{3}{2\sqrt{x^2+y^2+z^2}}(x \bar{g}_{(1)xz}+y\bar{g}_{(1)yz}+z\bar{g}_{(1)zz}). \nonumber \\
\end{eqnarray}
\end{enumerate}
This is the asymptotic gauge condition that we have empirically verified leads to stable 3+1 evolution of asymptotically AdS$_4$ spacetimes in Cartesian coordinates.
Other choices of asymptotic source functions may enjoy similar stability properties.
The choice of $\bar{H}_\mu$ in the bulk is still completely arbitrary and the functional form that we implement in our simulations is detailed explicitly in Appendix~\ref{sec:GCbulk}.

The rationale for this procedure is as follows.
Recall that if $C_\mu=0$ and $\partial_t C_\mu=0$ are satisfied at $t=0$\footnote{This condition is satisfied by our initial data, see Appendix~\ref{sec:initdata}.}, and the boundary conditions are consistent with $C_\mu=0$ being satisfied at the boundary for all time, then, at the analytical level, the generalized harmonic constraint $C_\mu=0$ remains satisfied in the interior for all time. 
The addition of constraint damping terms to the Einstein equations, eq. \eqref{eqn:efe_gh_modified}, helps to ensure that deviations at the level of the discretized equations remain under control. 
Thus, in solving the expanded system of equations \eqref{eqn:efe_gh_modified}, we are assured that only the subset of solutions that are also solutions of the Einstein equations are being considered.
With this in mind, the near-boundary form of \eqref{eqn:efe_gh_modified}, given by \eqref{eq:efefullexp}, implies that our task in obtaining a solution is to satisfy $A_{(i)\mu\nu}=0$ for all $i$, and for some choice of source function variables $\bar{H}_\mu$. 
This task is significantly eased by picking a gauge, through a suitable choice of $\bar{H}_\mu$, that eliminates $A_{(1)\mu\nu}$, i.e., the lowest order of the expansion of the Einstein equations near the AdS boundary.
This is precisely what the above set of steps is designed to do, and it is why we did not stop at the gauge obtained in step 1, \eqref{eqn:target_gauge_txyz_step1}, which would have resulted in a gauge that does not explicitly set $A_{(1)\mu\nu}=0$. 

Finally, it is also important to develop an understanding of the reason why the choice of $\bar{H}_\mu$ is not completely free. 
Although identifying every cause for the instability of a simulation is usually very complicated, one practical reason is clear and can be understood with the following example in Cartesian coordinates.
Suppose we choose a gauge in which, after some time $t>t_0$, $\bar{H}_{(1)t}$ takes the value
\begin{equation} 
\bar{H}_{(1)t}(t>t_0)=\frac{3}{2\sqrt{x^2+y^2+z^2}}(x \bar{g}_{(1)tx}+y\bar{g}_{(1)ty}+z b_t),
\end{equation} 
where $b_t\in \mathbb{R}$ is a possibly vanishing constant.
According to \eqref{eqn:target_gauge_txyz_step1}, the requirement that $C_t=0$ now implies $\bar{g}_{(1)tz}=b_t$. 
Even though this condition does not violate any of the requirements above, it is an additional Dirichlet boundary condition that must be imposed for $t>t_0$ if we hope to find a solution for this example.\footnote{The Dirichlet boundary condition $\bar{g}_{tz}|_{\rho=1}=0$ clearly does not restrict $\bar{g}_{(1)tz}$.} Although imposing boundary conditions that change with time is of interest in certain studies motivated by the AdS/CFT correspondence, for simplicity we do not consider such cases in this article. It should be straightforward to generalize our prescription for time-dependent boundary conditions.

\section{Boundary Stress Tensor}
\label{sec:bouset2}

In the simulations we output the holographic stress-energy tensor of the dual CFT.
In this section, we obtain the analytic expression for this object in spherical coordinates $x^\alpha=(t,\rho,\theta,\phi)$, as they are adapted to the metric of the AdS boundary in global coordinates. Thus, in order to obtain their numerical values, we will have to convert the evolution variables in Cartesian coordinates $\bar{g}_{\mu\nu}$ provided by our numerical scheme into their counterparts $\bar{g}_{\alpha\beta}$ in spherical coordinates. We do this in Appendix~\ref{sec:sphevvarboucon}, through the transformation \eqref{eq:cartosph}.

Let us denote by $x^a=(t,\theta,\phi)$ the coordinates on timelike hypersurfaces $\partial M_q$ at fixed $\rho$ (or $q$).
To compute the holographic stress-energy tensor of the boundary CFT, $\langle T_{ab}\rangle_{CFT}$, we first compute the quasi-local stress-energy tensor $^{(q)}T_{\alpha\beta}$ at $\partial M_q$ as prescribed in \cite{Balasubramanian:1999re}. 
We have
\begin{equation}
\label{eq:qslocset}
^{(q)}T_{\alpha\beta}=\frac{1}{8\pi}\biggl(\;  \Theta_{\alpha\beta}-\Theta \;\omega_{\alpha\beta}-2\omega_{\alpha\beta}+ G_{\alpha\beta} \biggr),
\end{equation}
where $\Theta_{\alpha\beta}=-\omega^\gamma_{\alpha}\omega^\delta_\beta\nabla_{\gamma}S_{\delta}$ is the extrinsic curvature of $\partial M_q$, $\omega_{\alpha\beta}=g_{\alpha\beta}-S_\alpha S_\beta$ is the induced metric on $\partial M_q$ (in four-dimensional form), $S^\alpha$ is the spacelike, outward pointing timelike unit vector normal to $\partial M_q$ and $G_{\alpha\beta}$ is the Einstein tensor of $\partial M_q$.\footnote{Notice the different sign in the last term of \eqref{eq:qslocset} with respect to \cite{Balasubramanian:1999re}. When comparing the two results, recall that in our expressions we set $L=1$.}$^{,}$\footnote{All these tensors, although defined on the tangent space of the spacetime manifold $M$, are invariant under projection $\omega^\alpha_{\beta}=\delta^\alpha_{\beta}-S^\alpha S_\beta$ onto $\partial M_q$. Therefore, they can be identified, under a natural (i.e., basis-independent) isomorphism, with tensors defined on the tangent space of $\partial M_q$. The components of tensors on $\partial M_q$ in coordinates $x^a$ is simply given by taking the components of tensors on $M$ in coordinates $x^\alpha$ and disregarding every combination of indices that includes an index $\rho$. See \cite{Hawking:1973uf} for more details on this correspondence.} We will be interested in the value of $^{(q)}T_{\alpha\beta}$ for $q$ close to 0, i.e., near the AdS boundary.
Restricting the indices corresponding to the coordinates $x^a$, we can compute the boundary stress-energy tensor as
\begin{equation}
\langle T_{ab}\rangle_{CFT}=\lim_{q\to0}\frac{1}{q} \;^{(q)}T_{ab}\,.
\end{equation}

From $^{(q)}T_{\alpha\beta}$ we also compute the total AdS mass as follows \cite{Balasubramanian:1999re}. At each time $t$ of evolution, we take a spacelike two-dimensional surface $\mathcal{S}$ in $\partial M_q$, with induced metric $\sigma_{ab}=\omega_{ab}+u_a u_b$, where $u_a=-N(dt)_a$ is the future pointing unit 1-form normal to $\mathcal{S}$ in $\partial M_q$, lapse $N$ and shift $N^a$. The total AdS mass is then given by
\begin{equation}
M=\lim_{q\to0}\int_{\mathcal{S}} d\theta d\phi \sqrt{\sigma} N ( ^{(q)}T_{ab} u^a u^b)\,.
\end{equation}

The holographic stress-energy tensor can be expressed in terms of the leading order coefficients of the near-boundary expansion of $\bar{g}_{\alpha\beta}$. We find:\footnote{The expressions~\eqref{eq:set_explicit} have a factor of $1/G$ that corresponds to the large-$N$ scaling of the expectation value of the stress tensor in the boundary 2+1-dimensional CFT. When quoting numerical results, we keep convention of working in geometric units with $G=1$.}
\begin{eqnarray}
\label{eq:set_explicit}
\langle T_{tt}\rangle_{CFT}&\hspace{-0.15cm}=&\hspace{-0.15cm}\frac{1}{16\pi} \biggl(2\bar{g}_{(1)\rho\rho}+3\bar{g}_{(1)\theta\theta}+3\frac{\bar{g}_{(1)\phi\phi}}{\sin^2\theta}\biggr), \nonumber \\
\langle T_{t\theta}\rangle_{CFT}&\hspace{-0.15cm}=&\hspace{-0.15cm}\frac{3}{16\pi}\bar{g}_{(1)t\theta}, \nonumber \\
\langle T_{t\phi}\rangle_{CFT}&\hspace{-0.15cm}=&\hspace{-0.15cm}\frac{3}{16\pi}\bar{g}_{(1)t\phi}, \nonumber \\
\langle T_{\theta\theta}\rangle_{CFT}&\hspace{-0.15cm}=&\hspace{-0.15cm}\frac{1}{16\pi} \biggl(3\bar{g}_{(1)tt}-2\bar{g}_{(1)\rho\rho}-3\frac{\bar{g}_{(1)\phi\phi}}{\sin^2\theta}\biggr), \nonumber \\
\langle T_{\theta\phi}\rangle_{CFT}&\hspace{-0.15cm}=&\hspace{-0.15cm}\frac{3}{16\pi}\bar{g}_{(1)\theta\phi}, \nonumber \\
\langle T_{\phi\phi}\rangle_{CFT}&\hspace{-0.15cm}=&\hspace{-0.15cm}\frac{\sin^2\theta}{16\pi} (3\bar{g}_{(1)tt}-2\bar{g}_{(1)\rho\rho}-3\bar{g}_{(1)\theta\theta}).
\end{eqnarray}
Similarly, for the total mass in AdS we find
\begin{equation}
\label{eq:AdSmasscalc}
M=\int_0^\pi d\theta \int_0^{2\pi}d\phi\frac{\sin\theta}{16\pi} \biggl(2\bar{g}_{(1)\rho\rho}+3\bar{g}_{(1)\theta\theta}+3\frac{\bar{g}_{(1)\phi\phi}}{\sin^2\theta}\biggr).
\end{equation}

We can now use the metric of the AdS boundary, $\lambda_{ab}dx^a dx^b=-dt^2+d\theta^2+\sin^2\theta d\phi^2$, to raise one index of $\langle T_{ab}\rangle_{CFT}$ and solve the eigenvalue problem $\langle {T^a}_{b}\rangle_{CFT} v^b=\Lambda_v v^a$ at each point along the AdS boundary. In this way, assuming that $\langle T_{ab}\rangle_{CFT}$ satisfies the weak energy condition,\footnote{If $\pm \langle T_{ab}\rangle_{CFT}$ fail to satisfy the weak energy condition, the $L^2$-norm of $\langle T_{ab}\rangle_{CFT}$, $||\langle T_{ab}\rangle_{CFT}||_2$, can have complex conjugate pairs of eigenvalues and no real timelike eigenvector, as pointed out in footnote 9 of \cite{Chesler:2013lia}.} we obtain the energy density of the boundary CFT, $\epsilon$, as minus the eigenvalue associated to the unique (up to rescaling) timelike eigenvector. Similarly, the boundary anisotropy is given by $\Delta p\equiv|p_1-p_2|$, where $p_1$ and $p_2$ are the eigenvalues associated with, respectively, the remaining two spacelike eigenvectors.

One useful quantity to compute is the trace of the stress-energy tensor, $\langle \text{tr}T\rangle_{CFT}=\lambda^{ab} \langle T_{ab}\rangle_{CFT}$. We obtain:
\begin{equation}
\label{eq:tracecalc}
\langle \text{tr}T\rangle_{CFT}=\frac{3}{8\pi}\biggl(\bar{g}_{(1)tt}-\bar{g}_{(1)\rho\rho}-\bar{g}_{(1)\theta\theta}-\frac{\bar{g}_{(1)\phi\phi}}{\sin^2\theta}\biggr).
\end{equation}
If we convert the spherical quantities into their Cartesian counterparts 
we see that $\langle \text{tr}T\rangle_{CFT}$ depends only on the factor $\bar{g}_{(1)tt}-\bar{g}_{(1)xx}-\bar{g}_{(1)yy}-\bar{g}_{(1)zz}$. 
We saw in \eqref{eq:cart_tracelessness} that this factor vanishes.
This is an important sanity check: we see that tracelessness of the stress-energy tensor, expected for a CFT in 2+1 dimensions, is ensured by the lowest order in the near boundary expansion of the Einstein equations, provided that the generalized harmonic constraints are satisfied.
In other words, tracelessness of the boundary stress tensor is, in our scheme, directly tied to how close our numerical solution is to a solution of the Einstein field equations. 
We check that we are indeed converging to such a solution in Appendix~\ref{sec:convbulk}.
In practice, we monitor $\langle \text{tr}T\rangle_{CFT}$ to estimate truncation error.
Another important check that we performed is the conservation of the analytic form of $\langle T_{ab}\rangle_{CFT}$. 
The simplest way to prove this is by using the near-boundary expansion of the Einstein equations in spherical coordinates, as done in Appendix~\ref{sec:sphevvarboucon}.

\section{Numerical Scheme}\label{sec:numerical_scheme}

In this section we consider the core elements of the numerical scheme used in this study.
We start by discussing the numerical features on which this scheme relies for solving the initial-boundary value problem in AdS.
We then describe our apparent horizon finder and the method with which we excise trapped regions.

\subsection{Numerics of the Initial-Boundary Value Problem}
\label{sec:numcauprob}
We solve the Einstein equations in generalized harmonic form \eqref{eqn:efe_gh_modified} with constraint damping terms, coupled with the massless Klein-Gordon equation \eqref{eqn:eoms2cart}.
We obtain asymptotically AdS spacetimes in Cartesian coordinates $x^\mu=(t,x,y,z)$. 
The solution is determined in terms of the metric, scalar field and source function variables $(\bar{g}_{\mu\nu},\bar{\varphi},\bar{H}_\mu)$ defined in Section~\ref{subsec:cartevvarboucon}. We substitute the definitions of these variables, \eqref{eq:gbarcart}--\eqref{eq:soufunb}, in the equations of motion and analytically remove all the purely AdS terms.
The resulting partial differential equations (PDEs) are discretized with second order finite difference derivative stencils, and then integrated in time using an iterative Newton-Gauss-Seidel relaxation procedure with a three time level hierarchy.
The source function variables $\bar{H}_\mu$ near the AdS boundary are set as we have prescribed in \eqref{eqn:target_gauge_txyz}, whilst deep in the bulk they are set to zero. 
In between, we use smooth transition functions to interpolate between the near boundary and the bulk regions, see Appendix~\ref{sec:GCbulk} for the details of our full implementation. 

We use the PAMR/AMRD libraries \cite{PAMR} for running these simulations in parallel on Linux computing clusters.
Although these libraries have adaptive mesh refinement capabilities, numerical evolution is performed on a grid with fixed refinement.
The numerical grid is in $(t,x,y,z)$ with $t \in [0,t_{max}]$, $x \in [-1,1]$, $y \in [-1,1]$, $z \in [-1,1]$.
The typical grid resolution uses $N_x=N_y=N_z=325$ points in each of the Cartesian directions, with equal grid spacings $\Delta x = \Delta y = \Delta z\equiv \Delta$. 

The time step of evolution is determined by $\Delta t=\lambda \Delta$. 
Although we do not perform a detailed analysis of the stability of our finite difference scheme, the Courant-Friedrichs-Lewy (CFL) condition for stability is expected to be satisfied as long as the CFL factor $\lambda$ is set to a value well below 1. Thus, we use $\lambda=0.3$. Notice that the most remarkable advantage of using Cartesian coordinates is that the CFL condition does not severely restrict the CFL factor as it would in spherical coordinates, hence allowing simulations to reach large evolution times with modest computational resources.
In contrast, spherical coordinates $(t,\rho,\theta,\phi)$ with fixed resolution $\Delta\rho,\Delta\theta,\Delta\phi$ would necessitate $\Delta t = \lambda \min(\Delta\rho, \rho_{min} \Delta\theta, \rho_{min} \Delta\theta \Delta\phi)$.
At points next to the origin, which must be evolved in studies of gravitational collapse and black hole formation, $\rho$ takes its smallest value $\rho_{min}=\Delta\rho$. 
Hence, in spherical coordinates, $\Delta t$ would become prohibitively small for higher resolutions, i.e., for smaller $\Delta\rho$, $\Delta\theta$, $\Delta\phi$.

\begin{figure*}[t!]
        \centering
        \includegraphics[width=6.0in,clip=true]{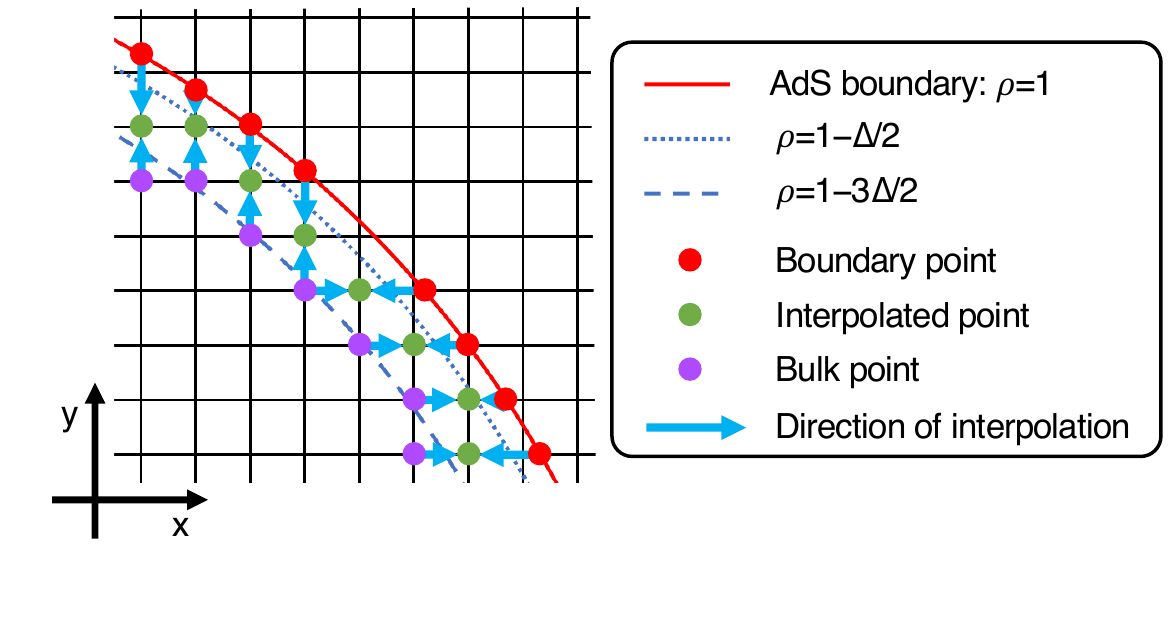}
\parbox{5.0in}{\caption{Visual description of the implementation of Dirichlet boundary conditions through first order interpolation in a portion of a $z=$const. surface for a grid with spatial grid spacing $\Delta$.
        }\label{fig:lego_circle_dirbc}}
\end{figure*}

The following components play a fundamental role in the numerical implementation of the initial-boundary value problem.
Reflective Dirichlet boundary conditions \eqref{eq:dirbc} are imposed at the AdS boundary $\rho=1$.
In general the AdS boundary does not lie on Cartesian grid points, so we set boundary conditions at points at most one grid point away from the boundary via interpolation. Referring to Figure~\ref{fig:lego_circle_dirbc}, for any given evolution variable, we set its value at grid points with $\rho<1-\Delta/2$ (i.e., the green dots inside the blue dotted line in this figure) by first order interpolation between the Dirichlet value at boundary points (red dots) and the value at the adjacent point further into the interior $\rho<1$ (purple dots). To identify the latter, we move along the Cartesian direction corresponding to the coordinate of the green dot with the largest absolute value. This direction is represented by light blue arrows. Notice that points with $\rho\geq1-\Delta/2$ are excised to avoid issues with quantities that would diverge 
at $\rho=1$. Finally, to obtain the values of quantities at the boundary, needed to extract the holographic observables, we use third order extrapolation from their bulk point values. 
The details of the implementation in our numerical simulations can be found in Appendix~\ref{sec:extrapconvbdy}.

Last but not least, time-symmetric initial data, sourced by a massless real scalar field, are obtained by solving the conformal decomposition of the Hamiltonian constraint \eqref{eq:hamconsfinal}. 
The solution to \eqref{eq:hamconsfinal} is computed, after second order finite discretization, through a full approximation storage (FAS) multigrid algorithm with v-cycling and Newton-Gauss-Seidel relaxation, built into the PAMR/AMRD libraries. We ensure that initial data satisfies the generalized harmonic constraints. See Appendix~\ref{sec:initdata} for more details and the complete choice of initial data.

\subsection{Apparent Horizon Finder and Excision}
\label{sec:AH_exc}

Once the solution is obtained at a certain time $t$, we can search for the position $R(\theta,\phi)$ of an apparent horizon (AH). We use the following flow method in spherical coordinates $(\rho,\theta,\phi)$, obtained in the usual way from the Cartesian coordinates of the solution. We consider $n$ two-dimensional surfaces at constant, equally spaced, values of $\rho$ within a user-specified range included in (0,1), and we pick the one with smallest $L^2$-norm of the outward null expansion. Let $\rho_0$ be the $\rho$ coordinate on this surface. 
Starting from the initial guess $R(\theta,\phi)=\rho_0$, for any $(\theta,\phi)\in [0,\pi]\times [0,2\pi)$ we find the solution to the equation
\begin{equation}
\label{eq:floweq}
\frac{dR(\theta,\phi)}{d s}=-\Theta(\rho,\theta,\phi)|_{\rho=R(\theta,\phi)}\,,
\end{equation}
where $\Theta(\rho,\theta,\phi)|_{\rho=R(\theta,\phi)}$ is the outward null expansion of the two-dimensional surface given by $F(\rho,\theta,\phi)\equiv\rho-R(\theta,\phi)=0$. We iterate this process with starting point given by the solution $R(\theta,\phi)$ to \eqref{eq:floweq} found in the previous iteration. Assuming that the initial guess $\rho_0$ is not too distant from the position of the AH, $R(\theta,\phi)$ is expected to progressively approach the AH after each iteration. 
This process stops when either the $L^2$-norm of $\Theta(\rho,\theta,\phi)|_{\rho=R(\theta,\phi)}$ is below some specified tolerance, i.e., $R(\theta,\phi)$ is sufficiently close to the AH, or the user-specified maximum number of iterations has been reached, i.e., either there is no AH at time $t$ or this method was not able to find it. 

This AH finder is based on a $(\theta,\phi)$ grid with equal grid spacings $\Delta \theta=\Delta\phi=\Delta_{AH}$.\footnote{The grid on which the AH finder is executed is completely independent of the specifics of the Cartesian evolution grid that was described in Section~\ref{sec:numcauprob}.} 
The outward null expansion at a given AH finder grid point $(\theta,\phi)$ is obtained by first order interpolation in three dimensions from the values of the expansion at Cartesian grid points that surround $(\theta,\phi)$.
These values are calculated from the definition of outward null expansion once the spacetime metric at time $t$ is known.
We observe that a $N_\theta\times N_\phi=9\times 17$ resolution is enough to find the AH in the simulations considered in Section~\ref{sec:results} in less than $10^{4}$ iterations. Since \eqref{eq:floweq} is a parabolic equation, the ``time'' step $\Delta s$ must be at least of order $\Delta_{AH}^2$ for stability. When using $n=10$ initial trial surfaces and an initial range of $\rho$ values between 0.1 and 0.5, as we do in our simulations, we find that the AH finder works effectively if $\Delta s$ takes much smaller values. Specifically, we set $\Delta s=10^{-4}$.

When an AH is found, we excise Cartesian grid points in an ellipsoid included in the AH and centred at the centre of the AH, in order to avoid the formation of geometric singularities in the computational domain.\footnote{This method is effective in removing singularities if the following common assumptions are valid on the spacetimes that we consider: (i) weak cosmic censorship is not violated, i.e., geometric singularities are contained inside a black hole event horizon; (ii) the AH at any time $t$ is contained in $t$-constant slices of the event horizon; (iii) the AH at any $t$ provides a sufficiently accurate approximation for $t$-constant slices of the event horizon.}
More specifically, the excision ellipsoid has Cartesian semi-axes, $a_x^{ex},a_y^{ex},a_z^{ex}$, determined by $a_x^{ex}=x_{AH}(1-\delta_{ex})$, where $x_{AH}$ is the $x$-coordinate value of the intersection between the AH and the $x$-axis, and similarly for $a_y^{ex}$ and $a_z^{ex}$. We set the excision buffer to $\delta_{ex}=0.4$. 
In our simulations, we assume that the characteristics of the equations of motion in the AH region flow towards the origin, although we do not compute the characteristics explicitly. As a consequence, the solution at points inside the AH only evolves to affect points, at later times, that are further inside the AH. In other words, the information needed to solve the equations of motion on and outside the excision surface at a certain time is entirely contained in the numerical domain at previous times. This allows us to solve the equations of motion at the excision surface by employing one-sided stencils that do not reference points inside the excised region, with no need to impose conditions at the excision boundary. By construction, the excised surface is the same for all three time levels involved in the Newton-Gauss-Seidel relaxation for evolution variables at time $t$. Therefore, we only need to use the one-sided version of the spatial stencils. 

It commonly occurs that the excised surface moves during evolution and previously excised points become unexcised. In this case, we initialize the value of newly unexcised points closest to the previous surface using fourth order extrapolated values from adjacent exterior points along each Cartesian direction. We do so for any variable and at all three time levels of the hierarchy.
Finally, Kreiss-Oliger dissipation~\cite{kreiss1973methods} is essential to damp unphysical high-frequency noise that arise at excision grid boundaries; we use a typical dissipation parameter of $\epsilon_{KO}=0.35$.

\section{Results}\label{sec:results}

As a proof-of-principle, we evolve initial data that undergoes gravitational collapses within one light-crossing time, and follow the subsequent ring-down to the Schwarzschild-AdS solution.
The geometry of the initial slice is sourced by a massless real scalar field with a Gaussian profile, distorted along each Cartesian direction and centred at $x=y=z=0$:
\begin{eqnarray}
\label{eq:scaGaupro}
\bar{\varphi}\big|_{t=0}&=&A e^{-(\tilde{r}(x,y,z)/\Delta)^2},\\
\tilde{r}(x,y,z)&\equiv&\sqrt{ x^2(1-e_x^2)+ y^2(1-e_y^2)+ z^2(1-e_z^2)}. \nonumber
\end{eqnarray}
The amplitude of the profile is $A=0.55$ and the eccentricities are $e_x=0.3, e_y=0.2, e_z=0.25$, so that the most prominent distortion is on the $(x,y)$-plane. The width of the Gaussian is $\Delta=0.2$. 
We choose the initial slice to be a moment of time symmetry, and the details of the time-symmetric initial data sourced by this matter field are collected in Appendix~\ref{sec:initdata}. 
As we see in that appendix, the momentum constraint is trivially satisfied for this type of data, so only the Hamiltonian constraint has to be solved. We evolve this initial data up to $t=31$ in units of the characteristic length scale $L=1$ (approximately 20 light crossing times), well after the end of gravitational collapse and the resulting black hole formation. 
The initial data has zero total angular momentum, 
and angular momentum conservation \cite{Fischetti:2012rd} ensures that this is zero at all times. 
Therefore, we can expect the black hole to settle down to the Schwarzschild-AdS solution. However, for generic initial data with non-vanishing total angular momentum, this may not be the final state: Ref. \cite{Holzegel:2011uu} conjectured that Schwarzschild-AdS, or more generally Kerr-AdS, may suffer from a non-linear instability for generic perturbations.
We will leave this interesting problem for future work.

\subsection{Collapse and ringdown}\label{sec:rescolring}

We describe here the evolution in the bulk: this consists of an initial short phase, in which the scalar field collapses and forms a black hole, and a long ringdown stage, in which the spacetime settles down to Schwarzschild-AdS. 

\begin{figure*}[t!]
        \centering
        \includegraphics[width=5.2in,clip=true]{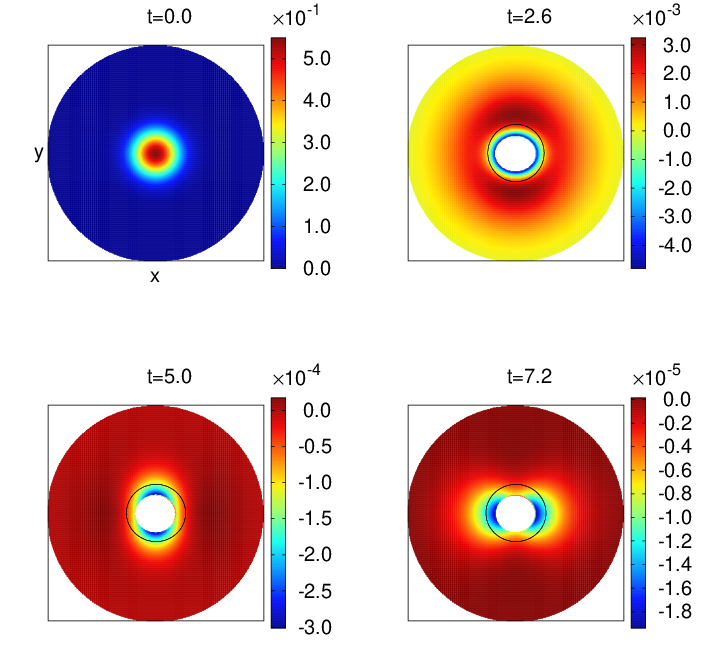}
\parbox{5.0in}{\caption{Snapshots of the scalar field profile $\bar{\varphi}$ on the $z=0$ slice in $(x,y)$ coordinates. In each plot, $x$ and $y$ are the horizontal and vertical axes, respectively, and the black square denotes the boundary of the numerical grid, i.e., $x=\pm 1$ and $y=\pm 1$. The external boundary of the coloured part is the AdS boundary. The black ellipse denotes the approximate position of the AH. This is obtained as the $z=0$ slice of the ellipsoid with Cartesian semi-axes, $x_{AH}$, $y_{AH}$, $z_{AH}$, where $x_{AH}$ is the $x$-coordinate value of the intersection between the AH and the $x$-axis, and similarly for $y_{AH}$ and $z_{AH}$.
The internal boundary of the coloured region is the excision surface: we excise points inside an ellipsoid whose semi-axes, $a_x^{ex}$, $a_y^{ex}$, $a_z^{ex}$, are given by $a_x^{ex}=x_{AH}(1-\delta_{ex})$, and similarly for $a_y^{ex}$ and $a_z^{ex}$. We use the value $\delta_{ex}=0.4$ for the excision buffer. Highest resolution: $N_x=N_y=N_z=325$.
        }\label{fig:snapshotsscalarfield}}
\end{figure*}

Figure \ref{fig:snapshotsscalarfield} shows the profile of the scalar field variable, $\bar{\varphi}$, at four representative times on the equatorial plane $z=0$ for the highest resolution grid, with $N_x=N_y=N_z=325$ grid points along each Cartesian direction. Notice that in all of these snapshots $\bar{\varphi}=0$ at the AdS boundary, as required by the Dirichlet boundary conditions. 
At $t=0$, the asymmetry of the initial Gaussian profile is too small to be visible. 
At the beginning of evolution, we see that the scalar field lump starts propagating away from the origin, and a portion of it soon forms an AH.
This occurs at $t=0.331$ in the highest resolution simulation. The rest of the scalar field remains outside the black hole, where it keeps bouncing back and forth the AdS boundary and is gradually absorbed.
The asymmetry on the $(x,y)$-plane is clearly visible at $t=2.6$, where the scalar field is stretched along the $x$-direction and squeezed along the $y$-direction. The elongation changes its direction multiple times during the evolution, as shown in the next two plots: it is along the $y$-axis at $t=5.0$ and again along the $x$-axis at $t=7.2$. 
At later times, $t\simeq 9$, the value of the scalar field becomes consistent with zero up to solution error\footnote{We estimate the solution error by comparing $\bar{\varphi}$ at different resolutions.} and the spacetime settles down to a Schwarzschild-AdS black hole spacetime with mass $M=0.403$.

\begin{figure*}[t!]
        \centering
        \includegraphics[width=5.02in,clip=true]{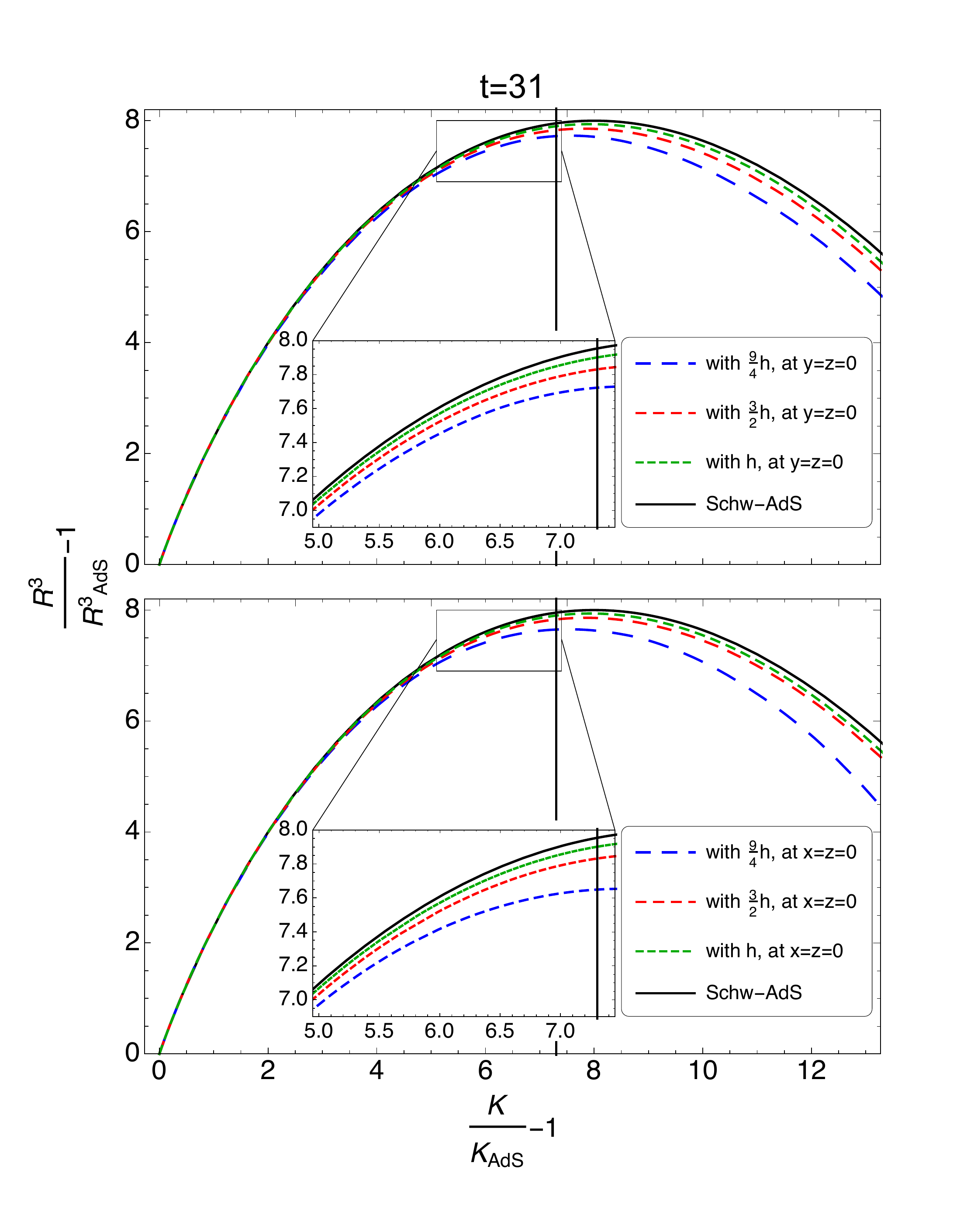}
\parbox{5.0in}{\caption{
Riemann cube scalar relative to AdS$_4$, $(R^3/R^3_{\text{AdS}})-1$, as a function of Kretschmann scalar relative to AdS$_4$, $(K/K_{\text{AdS}})-1$.
In each panel, the black curve denotes the result for a slice of Schwarzschild-AdS with mass given by $M_h=0.403$ (in units of the characteristic length scale $L=1$), i.e., the value of $M$ (see eq. \eqref{eq:AdSmasscalc}) for the highest resolution run with grid spacing $h$. The black vertical line denotes the value of $(K/K_{\text{AdS}})-1$ at the horizon of the Schwarzschild-AdS black hole. The relative Kretschmann increases as we move closer to the origin of the spacetime.
Top panel: the coloured lines denote the Riemann-Kretschmann dependence obtained from grid points on the $x$-axis (i.e., $y=z=0$) of the numerical solution at $t=31$. 
Bottom panel: the coloured lines denote the Riemann-Kretschmann dependence obtained from grid points on the $y$-axis (i.e., $x=z=0$) of the numerical solution at $t=31$.
        }\label{fig:relRiemanncube-relKretschmann-comparison-SchwAdS}}
\end{figure*}

The late-time solution is close to Schwarzschild-AdS, which can be seen explicitly in Figure \ref{fig:relRiemanncube-relKretschmann-comparison-SchwAdS}.
Here, we compare the numerical solution at the last time slice, i.e., $t=31$, to a slice of the Schwarzschild-AdS metric with conserved mass obtained from our highest resolution run ($M=0.403$).
This comparison is achieved with the following procedure.
First, we compute the Riemann cube scalar $R^3=R_{\mu\nu\rho\sigma}R^{\rho\sigma\gamma\delta}{R_{\gamma\delta}}^{\mu\nu}$, and the Kretschmann scalar $K=R_{\mu\nu\rho\sigma}R^{\mu\nu\rho\sigma}$.
Second, we compute the corresponding values, $R^3_{\text{AdS}}$ and $K_{\text{AdS}}$, for pure AdS$_4$.
We then use all four quantities to represent the relative Riemann scalar $(R^3/R^3_{\text{AdS}})-1$ as a function of the relative Kretschmann scalar $(K/K_{\text{AdS}})-1$ for the Schwarzschild-AdS black hole with $M=0.403$.
The same Riemann-Kretschmann dependence is estimated for our numerical solution at different resolutions from the values of $(R^3/R^3_{\text{AdS}})-1$ and $(K/K_{\text{AdS}})-1$ at each grid point along the $x$-axis ($y=z=0$ coloured lines of top panel) and the $y$-axis ($x=z=0$ coloured lines of bottom panel).

The black vertical lines in Figure \ref{fig:relRiemanncube-relKretschmann-comparison-SchwAdS} denotes the value of $\frac{K}{K_{\text{AdS}}}-1$ at the horizon of the Schwarzschild-AdS black hole. 
Notice that $\frac{K}{K_{\text{AdS}}}-1=0$ at the AdS boundary by construction, so going to larger values of $\frac{K}{K_{\text{AdS}}}-1$ is equivalent to moving towards the centre of the grid, and closer to the singularity.
Therefore the black vertical lines give an indication of the position of the AH relative to the AdS boundary.
The two panels of Figure \ref{fig:relRiemanncube-relKretschmann-comparison-SchwAdS} indicate that, sufficiently close to the AdS boundary, the curvature invariants of the numerical solution are almost identical to Schwarzschild-AdS. For clarity, this is shown using only values of the Riemann cube and Kretschmann scalars along the $x$ and $y$ axes, but we verified this for values from the entire grid. 
At any given resolution, the numerical curvature invariants start to differ from their Schwarzschild-AdS values as we get closer to the AH. This is expected since the gradients become larger as we approach the centre of the grid.
However, these differences converge away as resolution is increased.
Finally, although there is an asymmetry at any given resolution between the $x$ and $y$ axes even at this last time slice, this late-time asymmetry also converges away as resolution is increased.

\subsection{Boundary scalar field and stress-energy tensor}
\label{sec:resbouset}

\begin{figure*}[!t]
        \centering
        \includegraphics[width=5.0in,clip=true]{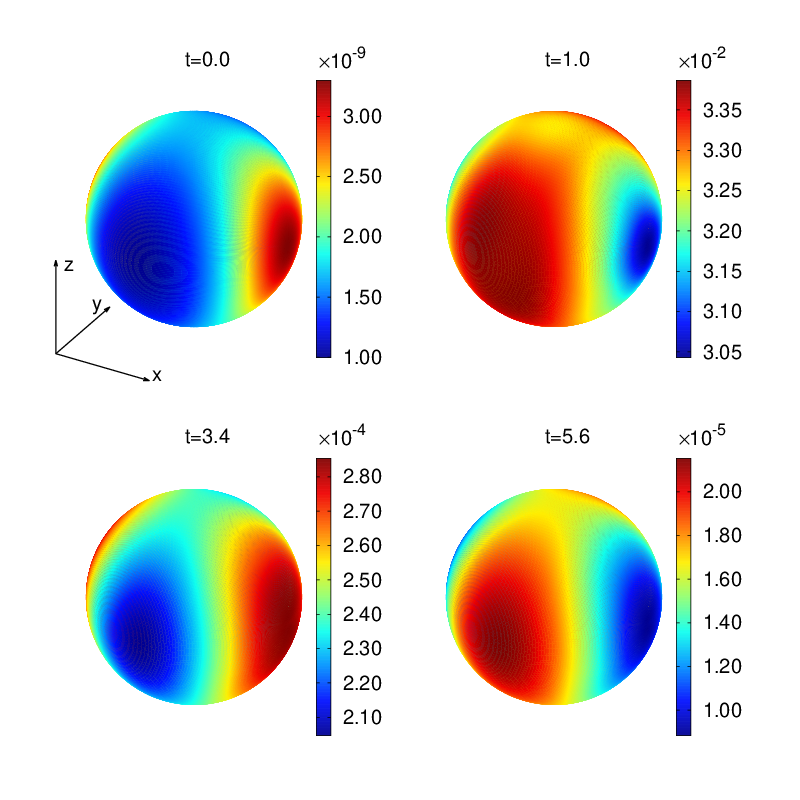}
\parbox{5.0in}{\caption{
Snapshots of the vacuum expectation value of the dual scalar field operator $\bar{\varphi}_{(1)}$. 
The first snapshot is obtained analytically from the initial scalar field profile. The remaining three are obtained by third order extrapolation and subsequent smoothening via a low-pass filter;
see Appendix~\ref{sec:extrapconvbdy}. Highest resolution: $N_x=N_y=N_z=325$.
        }\label{fig:snapshotsbdyphi}}
\end{figure*}

In this section we consider the evolution of the holographic quantities at the AdS boundary defined in Section~\ref{sec:bouset2}. 
These quantities are obtained via third order extrapolation from points in the interior, with the only exception of the $t=0$ plot of Figure~\ref{fig:snapshotsbdyphi}, which is computed analytically from the initial distorted Gaussian profile \eqref{eq:scaGaupro}. 
See Appendix~\ref{sec:extrapconvbdy} for a detailed explanation of the extrapolation scheme.

We start by noting that the numerical values for the total mass $M$ in AdS, obtained from equation \eqref{eq:AdSmasscalc}, are approximately constant during the evolution, as expected by mass conservation \cite{Fischetti:2012rd}. More precisely, a small drift of the total mass is observed numerically, however this becomes smaller as we increase the resolution and it is consistent with zero within our error estimate for boundary quantities that we will discuss shortly.

Figure \ref{fig:snapshotsbdyphi} shows four snapshots of the vacuum expectation value of the dual scalar field operator at the boundary, $\bar{\varphi}_{(1)}$, obtained from the near-boundary expansion of the bulk scalar field in \eqref{eqn:qexp}.
Unlike the $z=0$ slice snapshots of Figure~\ref{fig:snapshotsscalarfield}, these plots of the boundary $S^2$ encode the asymmetry in all three Cartesian directions in the bulk, as they appear on the boundary at $\rho = \sqrt{x^2+y^2+z^2}=1$. 
In fact, the asymmetry of the initial data is already visible at $t=0$, where the different values of eccentricities along the three Cartesian direction (largest along $x$ and smallest along $y$) are evident in this plot. At this time, the boundary scalar field is overall very small, which is expected since the initial $\bar{\varphi}$, given by \eqref{eq:scaGaupro}, is localized near $\rho=0$.
Notice from Figure \ref{fig:snapshotsbdyphi} that the asymmetry changes axes during evolution, but interestingly it is always strongest along $x$ and weakest along $y$ or vice-versa. Furthermore, a direct comparison with Figure~\ref{fig:snapshotsscalarfield} shows that the features present at a certain $t$ at the boundary take approximately $\pi/2\simeq1.6$ to reach the interior of the bulk, i.e., about a light-crossing time, as expected. At later times, mirroring the evolution in the bulk, $\bar{\varphi}_{(1)}$ decays exponentially in time as the bulk spacetime settles down to Schwarzschild-AdS.

\begin{figure*}[!t]
        \centering
        \includegraphics[width=5.0in,clip=true]{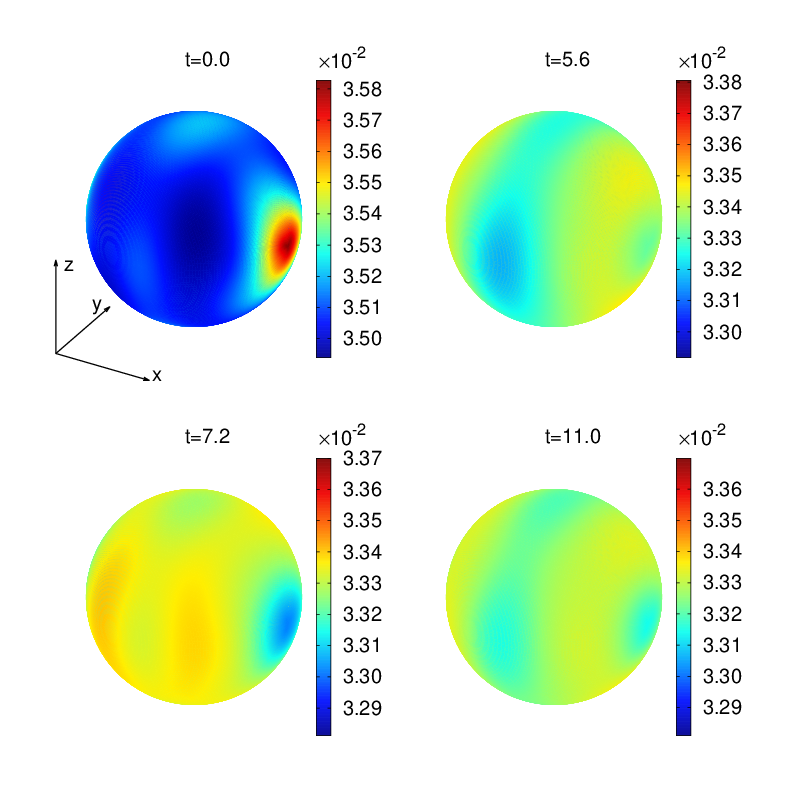}
\parbox{5.0in}{\caption{Snapshots of energy density $\epsilon$ of the dual boundary CFT, obtained by third order extrapolation and smoothened via a low-pass filter;
see Appendix~\ref{sec:extrapconvbdy}. The scale of each snapshot has fixed interval length centred at the mean value of $\epsilon$ at the corresponding evolution time to make the approach to a uniform configuration more visible. Highest resolution: $N_x=N_y=N_z=325$.
        }\label{fig:snapshotsenergydensity}}
\end{figure*}

Figure \ref{fig:snapshotsenergydensity} displays the energy density $\epsilon$ of the boundary CFT.
At $t=0$ this is strongly asymmetric along the $x$-direction, as expected from the shape of the initial scalar field profile \eqref{eq:scaGaupro}. 
After that, $\epsilon$ undergoes a phase of strong evolution with several changes of elongation axes, sampled at $t=5.6$ and terminating at approximately $t=7.2$. From that time onwards, $\epsilon$ settles down to a uniform configuration, as appropriate for the Schwarzschild-AdS black hole. Approach to uniformity is emphasized by using colour scales with fixed interval length, centred at the mean value of $\epsilon$ at the corresponding evolution time.

\begin{figure*}[!t]
        \centering
        \includegraphics[width=5.0in,clip=true]{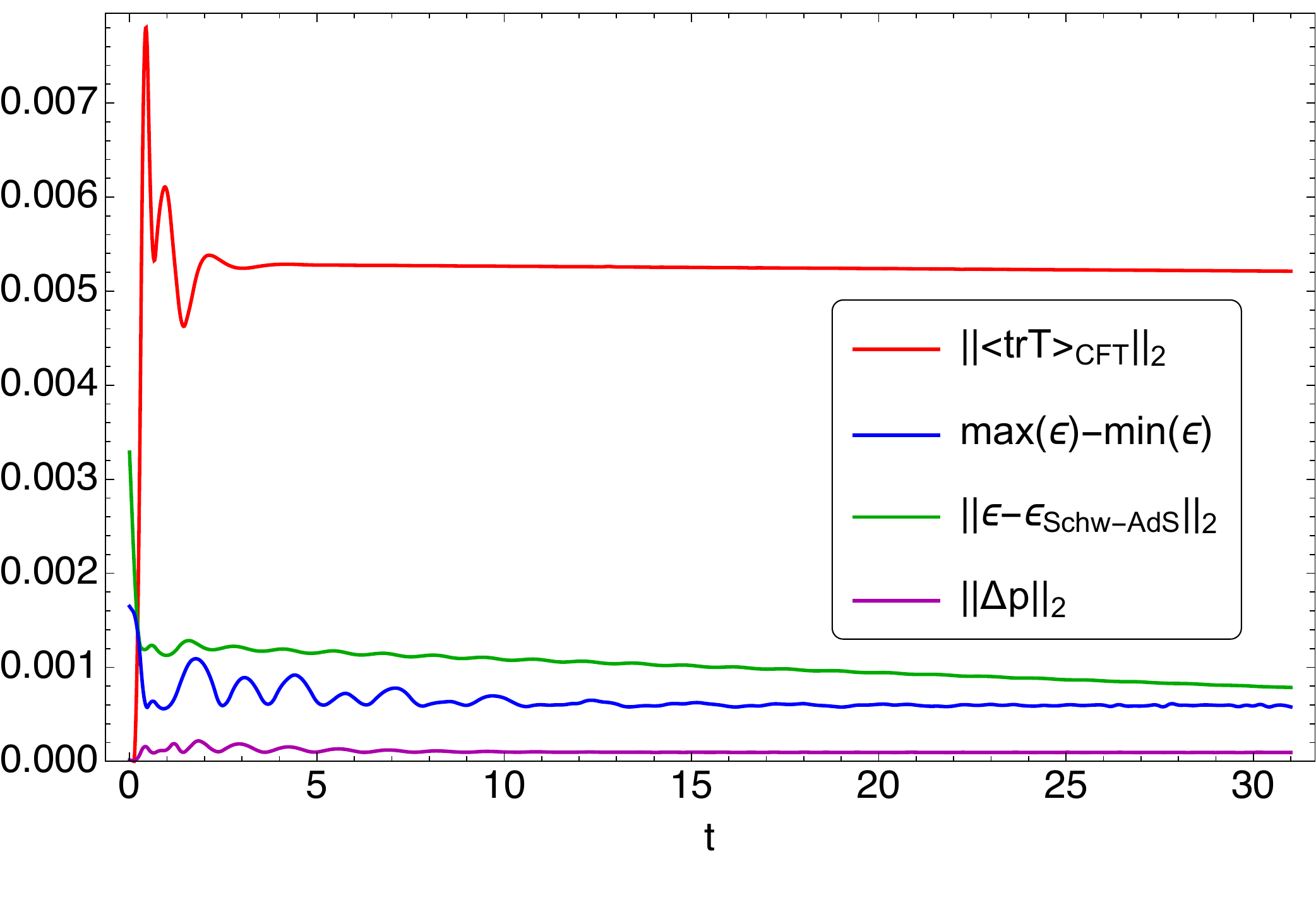}
\parbox{5.0in}{\caption{Comparison of boundary quantities with error estimate given by the deviation of the $L^2$-norm of $\langle \text{tr}T\rangle_{CFT}$ from its predicted zero value for the 2+1 CFT (red line). We consider the following boundary quantities: difference between maximum and minimum of boundary energy density $\epsilon$ (blue line), $L^2$-norm of difference between $\epsilon$ and the Schwarzschild-AdS value $\epsilon_{\text{Schw-AdS}}=\frac{M}{4\pi}$ (green line), with Schwarzschild mass $M=M_h=0.403$ (i.e., the value of $M$ for the resolution with grid spacing $h$), $L^2$-norm of boundary anistropy $\Delta p$ (magenta line). This plot is obtained from the data of the highest resolution run ($N_x=N_y=N_z=325$), but at any resolution these quantities exhibit the same hierarchy, although at different scales. Boundary quantities are computed by third order extrapolation.
        }\label{fig:fullplotfillregttraceanisotropyenergydensityminusschwmaxminusminbdyenergydensity.pdf}}
\end{figure*}

More information about the energy density of the boundary field theory can be deduced from Figure \ref{fig:fullplotfillregttraceanisotropyenergydensityminusschwmaxminusminbdyenergydensity.pdf}. 
The trace $\langle \text{tr}T\rangle_{CFT}$ vanishes for a conformal field theory in 2 +1 dimensions, which is the case for our $\mathbb{R} \times S^2$ boundary.
In Section~\ref{sec:bouset2}, we had spelled out how this trace, in our scheme, is tied to how well we are solving the Einstein field equations.
We thus use the $L^2$-norm of the numerical values of $\langle \text{tr}T\rangle_{CFT}$ (red line) as an error estimate for boundary quantities. We compare this error with the difference between maximum and minimum of $\epsilon$ (blue line), the $L^2$-norm of the difference between $\epsilon$ and its Schwarzschild-AdS value $\epsilon_{\text{Schw-AdS}}=\frac{M}{4\pi}$ (green line), with $M=M_h=0.403$, i.e., the highest resolution value of $M$, and the $L^2$-norm of $\Delta p$ (magenta line). We compute these quantities from the data of the highest resolution simulation, but at any resolution the hierarchy is the same, although it appears at different scales.
If we exclude very early times, we see that $\max(\epsilon)-\min(\epsilon)$ is consistent with zero, which confirms that the energy density becomes uniform in time.
We also see that $||\epsilon-\epsilon_{\text{Schw-AdS}}||_2$ is consistent with zero and decreasing in time, which shows that the energy density settles down to $\epsilon_{\text{Schw-AdS}}$, as expected.
Finally, $||\Delta p||_2$ is consistent with zero, as appropriate for the boundary anistropy of the Schwarzschild-AdS black hole.

\section{Discussion}\label{sec:Discussion}

We have presented the first proof-of-principle Cauchy evolution scheme with no symmetry assumptions that solves the Einstein-Klein-Gordon equations for asymptotically AdS spacetimes. 
Stability of this numerical scheme is achieved through the gauge choice \eqref{eqn:target_gauge_txyz} near the AdS boundary. 
We have used this scheme to obtain preliminary results using stationary initial data constructed from completely asymmetric Gaussian initial profiles of a massless scalar field.

We observe the collapse of the scalar field into a black hole and the subsequent ringdown to a Schwarzschild-AdS black hole spacetime, in both bulk and boundary quantities.
Deviations from Schwarzschild-AdS at late times are consistent with zero within estimates of the numerical error. 
At very late times, the spatial profiles of these small deviations appear to cascade towards higher harmonics.
Even though these deviations are consistent with our error estimates, they may nevertheless trigger a non-linear instability that can only be revealed by evolving for longer times and with higher spatial resolutions.\footnote{Schwarzschild-AdS has been shown to be stable under spherically symmetric deformations \cite{Holzegel:2011uu}.} 
It will be interesting to conduct a detailed analysis by decomposing the scalar field profile into spherical harmonics and showing that the radial part is non-vanishing near the boundary for a very long time.
We leave this for future studies.

In this work we limited ourselves to $D=4$ spacetime dimensions, but the calculation outlined in Section~\ref{sec:gauge_choice} would be almost identical if we were to study Cartesian evolution of asymptotically AdS spacetimes in any $D\geq4$ dimensions. In particular, the stable gauge found with this method would be the same up to a numerical factor. 
Interestingly, a comparison between \eqref{eqn:target_gauge_txyz} and the corresponding result in \cite{Bantilan:2017kok} (see eq. (S10) in that previous work) clearly suggests a trend for the expression of the stable gauge as we relax symmetries, and thus increase the number of spatial coordinates on which the solution depends.
If this trend were confirmed, repeating the calculation above would not be necessary when increasing the number of spatial degrees of freedom.
See \cite{Bantilan:2020pay} for an example in higher dimensions where this was done explicitly.
Furthermore, the scheme presented here can be applied to cases with different types of matter fields, different types of global coordinates, and to coordinates on the Poincar\'{e} patch.
For instance, in Appendix~\ref{sec:sphevvarboucon} we followed the prescription of Section~\ref{sec:pre_sta} to obtain the stable gauge also in spherical coordinates.
In Appendix~\ref{sec:poincare}, the same procedure leads to a gauge that stabilizes evolution on a Poincar\'e patch of AdS$_4$.
In other words, this framework makes numerical Cauchy evolution in asymptotically AdS spacetimes possible in full generality, with no need to impose symmetries.

We expect to be able to tackle several interesting problems in asymptotically AdS spacetimes using this Cauchy evolution scheme.
We want to highlight two of the most important of these here.
The first is the study of gravitational collapse in AdS with no symmetry assumptions and with angular momentum.
The numerical study of gravitational collapse in AdS was done in $D\geq 4$ spacetime dimensions by \cite{Bizon:2011gg,Jalmuzna:2011qw} in spherical symmetry. 
In these papers it was shown that a class of small perturbations of amplitude $\epsilon$ undergoes gravitational collapse and forms a black hole on a time-scale $\mathcal{O}(\epsilon^{-2})$, due to a turbulent cascade of energies from large to small distances until a horizon forms.
Subsequently, \cite{Bantilan:2017kok} considered the same massless scalar field model in AdS$_5$ in a 2+1 setting, and it was observed that for a certain class of initial data, the subsequent evolution resulted in collapse that happens faster away from spherical symmetry.
On the other hand, the authors in \cite{Choptuik:2017cyd} used a particular metric ansatz in a 1+1 setting to consider the inclusion of angular momentum, and observed delayed collapse.
A promising direction is provided in \cite{Moschidis:2018ruk}, \cite{Moschidis:2018kcf} with a proof of the instability of AdS in spherical symmetry for the Einstein-massless Vlasov system.
The scheme described in this article makes it possible for numerical investigations to help settle this question, by incorporating all the relevant physics needed to study gravitational collapse in AdS in full generality.

The second important problem we wish to highlight is the study of the superradiant instability in AdS.
Superradiantly unstable (see \cite{Brito:2015oca} for a review of superradiance) initial data around a Kerr-AdS black hole spacetime was evolved in \cite{Chesler:2018txn}, without imposing symmetries, up to approximately 580 light-crossing times using the characteristic scheme presented in \cite{Chesler:2013lia}. This paper showed a transition of the Kerr-AdS black hole to a rotating black hole with one helical Killing field consistent with a black resonator~\cite{Dias:2015rxy}. Since the known black resonators are rapidly rotating black holes with an ergo-region, they are also unstable to superradiance \cite{Hawking:1999dp,Green:2015kur}. Hence a cascade to smaller and smaller resonators, potentially leading to a violation of the weak cosmic censorship conjecture, was suggested \cite{Niehoff:2015oga}. 
The authors of \cite{Chesler:2018txn} see a second transition at late times that could be the beginning of such a cascade but they do not continue the evolution further. 
Hence, the endpoint of the Kerr-AdS superradiant instability is still unknown.
To settle this question, it will be necessary to keep track of progressively smaller spatial scales during the course of the evolution, which typically go hand-in-hand with progressively richer dynamics.
On top of the computationally expensive nature of 3+1 simulations, it will be necessary to keep track of the evolution on sufficiently long time scales until the endpoint is reached, requiring commensurately large-scale computational resources.
We leave this important study for future work.

\section*{Acknowledgements}
HB and PF are supported by the European Research Council Grant No. ERC-2014-StG 639022-NewNGR. PF is also supported by a Royal Society University Research Fellowship (Grant No. UF140319 and URF\textbackslash R\textbackslash 201026) and by a Royal Society Enhancement Award (Grant No. RGF\textbackslash EA\textbackslash 180260). LR is supported by a QMUL PhD scholarship. We acknowledge the use of Athena at HPC Midlands+, which was funded by the EPSRC on 
grant EP/P020232/1, in this research, as part of the HPC Midlands+ consortium. This research also utilised Queen Mary's Apocrita HPC facility, supported by QMUL Research-IT \cite{ApocritaHPC}. 

\appendix
\setcounter{tocdepth}{1}
\numberwithin{equation}{section}
\section{Generalized Harmonic Formulation}
\label{sec:GHfor}

The generalized harmonic formulation of the Einstein equations is based on coordinates $x^\alpha$ that each satisfies a wave equation $\Box x^{\alpha}=H^\alpha$ with source functions $H^\alpha$.
As long as the constraints $C^\alpha \equiv H^\alpha-\Box x^\alpha=0$ are satisfied,
we can then write the trace-reversed Einstein equations in $D$ dimensions with cosmological constant $\Lambda$,
\begin{equation}
0=R_{\alpha\beta} - \frac{2\Lambda}{d-2} g_{\alpha\beta} - 8\pi\left( T_{\alpha\beta} - \frac{1}{d-2} {T^\gamma}_\gamma g_{\alpha\beta} \right),
\end{equation}
as
\begin{eqnarray}
0
&=& R_{\alpha\beta} - \nabla_{(\alpha} C_{\beta)} - \frac{2\Lambda}{d-2} g_{\alpha\beta} - 8\pi \bar{T}_{\alpha\beta} \nonumber \\
&=& R_{\alpha\beta} - \nabla_{(\alpha} H_{\beta)} + \nabla_{(\alpha} \Box{x}_{\beta)} - \frac{2\Lambda}{d-2} g_{\alpha\beta} - 8\pi \bar{T}_{\alpha\beta} \nonumber \\
&=& -\frac{1}{2} g^{\gamma\delta} g_{\alpha\beta,\gamma\delta} - g^{\gamma\delta}{}_{,(\alpha}g_{\beta)\gamma,\delta} - H_{(\alpha,\beta)} + H_\gamma \Gamma^\gamma{}_{\alpha\beta} \nonumber \\
&&- \Gamma^\gamma{}_{\alpha\delta}\Gamma^\delta{}_{\gamma\beta} - \frac{2\Lambda}{d-2} g_{\alpha\beta} - 8\pi \bar{T}_{\alpha\beta}\,, \label{eqn:Eeqs} 
\end{eqnarray}
where $\bar{T}_{\alpha\beta} = T_{\alpha\beta} - {T^\gamma}_\gamma g_{\alpha\beta}/(d-2)$, the choice of $H_\alpha = g_{\alpha\beta} H^\beta$ fixes the gauge, $\Gamma^\alpha{}_{\beta\gamma}$ are the Christoffel symbols associated with
the spacetime metric $g_{\alpha\beta}$, and $T_{\alpha\beta}$ is the matter stress-energy tensor. 
It can be proven that the constraints $C_\alpha=0$ are satisfied for all times $t\equiv x^0$, as long as $C_\alpha=0$ and $\partial_{t}C_\alpha=0$ at $t=0$. In an initial-boundary value problem, this still holds if we assume that the boundary conditions are consistent with $C_\alpha=0$ being satisfied on the boundary for all times.
However, for numerical initial data, the constraints and their derivatives with respect to $t$ vanish at $t=0$ typically only up to truncation error. Thus, to suppress constraint-violating solutions, we supplement \eqref{eqn:Eeqs} with constraint damping terms as introduced in~\cite{Gundlach:2005eh}, controlled by the parameters $\kappa$ and $P$.
We thus obtain the final form of our evolution equations:
\begin{eqnarray}
\label{eqn:efe_gh_modified}
&-& \frac{1}{2} g^{\gamma \delta} g_{\alpha\beta, \gamma \delta} - 
{g^{\gamma\delta}}_{,(\alpha} g_{\beta) \gamma, \delta} - H_{(\alpha, \beta)} + H_\gamma {\Gamma^\gamma}_{\alpha\beta} \nonumber \\
&-& {\Gamma^\gamma}_{\delta \alpha} {\Gamma^\delta}_{\gamma \beta} - \kappa \left( 2 n_{(\alpha} C_{\beta)} - (1+P) g_{\alpha\beta} n^\gamma 
C_\gamma \right) \nonumber \\
&=&  \frac{2}{d-2} \Lambda g_{\alpha\beta} + 8\pi \left( T_{\alpha\beta} - 
\frac{1}{d-2} {T^\gamma}_\gamma g_{\alpha\beta} \right),
\end{eqnarray}
where $n_\alpha=-\partial_\alpha t$ is the timelike, future-directed unit 1-form normal to slices of constant $t$.
Notice that the principal part of \eqref{eqn:efe_gh_modified}, $-\frac{1}{2} g^{\gamma \delta} \partial_\gamma \partial_\delta g_{\alpha\beta}$, is a wave operator acting on metric components. Thus, the well-posedness of the wave equation suggests that the initial-boundary value problem in generalized harmonic form is well-posed, if we make reasonable assumptions on the remaining components of the problem.
See, for example, \cite{Pretorius:2004jg,Bantilan:2012vu} for more details on this formulation. 
In our simulations we use the values $\kappa=-10$ and $P=-1$.\footnote{Ref. \cite{Bantilan:2012vu} mentions that it is important to use $P$ close to $-1$, while the value of $\kappa$ is not too important to achieve effective constraint damping.}

In this work, we are interested in the case where matter fields are given by a single massless real scalar field $\varphi$, hence the stress-energy tensor reads
\begin{equation}
\label{eq:KHmomtens}
T_{\alpha\beta}=\partial_\alpha \varphi \partial_\beta \varphi - g_{\alpha\beta} \frac{1}{2} g^{\gamma\delta} \partial_{\gamma} \varphi \partial_{\delta} \varphi\,.
\end{equation}
For completeness, we also write the Klein-Gordon equation \eqref{eqn:eoms2} for the scalar field $\varphi$ in terms of partial derivatives with respect to the chosen set of coordinates:
\begin{equation}\label{eqn:eoms2cart}
g^{\alpha\beta} \partial_{\alpha} \partial_{\beta} \varphi -g^{\alpha\beta} \Gamma^{\gamma}{}_{\beta\alpha}\partial_\gamma\varphi= 0\,.
\end{equation}

\section{Boundary Prescription for Spherical Coordinates}\label{sec:sphevvarboucon}

Although spherical coordinates $x^\alpha=(t,\rho,\theta,\phi)$ are not suitable for numerically evolving points near the origin (see discussion in Section~\ref{sec:numcauprob}), they are convenient to extract the physics of the CFT at the AdS boundary, since they are adapted to the boundary topology $\mathbb{R}\times S^2$.
In this section we apply the prescription outlined in Section~\ref{sec:pre_sta} to the case of asymptotically AdS spacetimes in $D=4$ spacetime dimensions in spherical coordinates. Similarly to the Cartesian case, we first define the spherical coordinate version of the evolution variables $(\bar{g}_{\alpha\beta},\bar{\varphi},\bar{H}_\alpha)$. We also write down the transformations between these variables and their Cartesian version, \eqref{eq:gbarcart}--\eqref{eq:soufunb}. Then, we obtain the stable gauge in spherical coordinates by following the steps introduced in Section~\ref{sec:gauge_choice}. We compare this with a different potentially stable gauge that can be inferred from the one used in~\cite{Bantilan:2012vu}. Finally, we show that tracelessness and conservation of the boundary stress-energy tensor $\langle T_{ab}\rangle_{CFT}$, defined in Section~\ref{sec:bouset2}, is a consequence of the lowest order of the Einstein equations in the near boundary expansion, provided that the leading order of the generalized harmonic constraints is satisfied.\footnote{In fact, tracelessness was already proved in Section~\ref{sec:bouset2} by converting Cartesian variables into spherical ones. We prove it again in this section employing only spherical coordinates.}

\subsection{Evolution Variables and Boundary conditions}

We remind the reader that new evolution variables are defined in order to apply the boundary conditions found in Section~\ref{subsec:asyAdS} as simple Dirichlet conditions at the AdS boundary $\rho=1$. In the same way as in the Cartesian coordinate case where we defined metric evolution variables $\bar{g}_{\mu\nu}$ in \eqref{eq:gbarcart}, the metric evolution variables in spherical coordinates $\bar{g}_{\alpha\beta}$ are defined by (i) considering the deviation from pure AdS tensor $h_{\alpha\beta}=g_{\alpha\beta}-\hat{g}_{\alpha\beta}$ in spherical coordinates, and (ii) stripping $h_{\alpha\beta}$ of as many factors of $(1-\rho^2)$ as needed so that they fall off linearly in $(1-\rho)$ near the AdS boundary.

The boundary conditions on $h_{\alpha\beta}$ \eqref{eq:sphbounconh} tell us that
\begin{eqnarray}\label{eq:gbarsph}
\bar{g}_{\rho\alpha}&=&\frac{h_{\rho\alpha} }{1-\rho^2}\,,\qquad \textrm{ if $\alpha\neq\rho$}\,, \nonumber\\
\bar{g}_{\alpha\beta}&=&h_{\alpha\beta}\,, \qquad\;\;\;\;\, \textrm{ otherwise}.
\end{eqnarray}
Despite the notation, we emphasize that $\bar{g}_{\alpha\beta}$ and $\bar{g}_{\mu\nu}$ are not in general components of the same tensor (as it should be clear from their definition), therefore the usual transformation between tensor components in different sets of coordinates cannot be applied. The correct transformation can be easily deduced from \eqref{eq:gbarsph} and \eqref{eq:gbarcart}, remembering that $h$ is indeed a tensor: 
\begin{eqnarray}\label{eq:cartosph}
\bar{g}_{\rho\alpha}&=&\frac{1}{(1-\rho^2)}\frac{\partial x^\mu}{\partial \rho}\frac{\partial x^\nu}{\partial x^\alpha}\bar{g}_{\mu\nu}\,,\quad \textrm{ if $\alpha\neq\rho$}\,, \nonumber\\
\bar{g}_{\alpha\beta}&=&\frac{\partial x^\mu}{\partial x^\alpha}\frac{\partial x^\nu}{\partial x^\beta}\bar{g}_{\mu\nu}\,,\qquad\quad \;\;\;\;\;\; \textrm{ otherwise}.
\end{eqnarray}
Similarly, the boundary conditions on the scalar field \eqref{eq:sphbounconphi} suggest that we use the evolution variable
\begin{equation}
\bar{\varphi}=\frac{\varphi }{(1-\rho^2)^2}\,,
\end{equation}
which the same as the one in Cartesian coordinates, as expected for a scalar field.
Finally, the boundary conditions \eqref{eq:sphbouncondsoufunc} on $H_\alpha$ suggest the use of the evolution variables
\begin{eqnarray}
 \bar{H}_\alpha&=&\frac{H_\alpha-\hat{H}_\alpha}{(1-\rho^2)^2 }\,, \qquad \textrm{ if $\alpha\neq\rho$\,,} \\ \nonumber
 \bar{H}_\rho&=&\frac{H_\rho-\hat{H}_\rho}{1-\rho^2 }\,,
 \end{eqnarray}
in spherical coordinates.
Neither $H_\alpha,\hat{H}_\alpha,\bar{H}_\alpha$ nor $H_\mu,\hat{H}_\mu,\bar{H}_\mu$ are components of the same tensor, so there is no simple transformation from one set to the other. The two triplets of quantities can only be obtained from the definition of source functions in terms of the full metric $g$ in the appropriate set of coordinates, e.g., equation \eqref{eq:sphbouncondsoufunc} in spherical coordinates.

In a numerical scheme in spherical coordinates employing the framework presented in this article, reflective Dirichlet boundary conditions can be easily imposed as
\begin{equation}
\label{eq:dirbc_sphcoords}
\bar{g}_{\alpha\beta}\big|_{\rho=1}=0\,,\quad \bar{\varphi}\big|_{\rho=1}=0\,,\quad \bar{H}_\alpha\big|_{\rho=1}=0\,.
 \end{equation}

\subsection{Gauge Choice for Stability}\label{sec:gau_choice_sphcoords}

Since the evolution variables in spherical coordinates, $(\bar{g}_{\alpha\beta},\bar{\varphi},\bar{H}_\alpha)$, are linear in $q=1-\rho$ by construction, we can borrow the near-boundary expansions \eqref{eqn:qexpg}--\eqref{eqn:qexpphi}. 
We now substitute these into the evolution equations \eqref{eqn:efe_gh_modified}, and we expand each component in powers of $q$. Rewriting the resulting equations in the wave-like form \eqref{eq:waveEFE}, we obtain
\begin{widetext}
\begin{eqnarray}\label{eqn:efett_3p1}
\tilde{\Box}\bar{g}_{(1)tt}&=&q^{-2} \left(2 \bar{H}_{(1) \rho }-3 \bar{g}_{(1) \rho \rho }\right)+O\left(q^{-1}\right),\\
\label{eqn:efetrho_3p1}
\tilde{\Box}\bar{g}_{(1)t\rho}&=&\frac{1}{2}q^{-2} (-\bar{g}_{(1)\theta \theta,t}+\bar{g}_{(1) \rho \rho ,t}-\bar{g}_{(1)
  \text{$tt$},t}+\csc ^2\theta \left(2 \bar{g}_{(1) \text{$t$$\phi $},\phi }-\bar{g}_{(1)
  \phi \phi ,t}\right) \nonumber\\
   &&+2 \bar{g}_{(1) \text{$t$$\theta $},\theta }-2 \bar{H}_{(1) \rho ,t}-3
   \cot \theta  \bar{g}_{(1) \text{$t$$\theta $}}-40 \bar{g}_{(1) \text{$t$$\rho $}}+20
   \bar{H}_{(1) t}) +\mathcal{O}(q^{-1}),\\
\label{eqn:efettheta_3p1}
\tilde{\Box}\bar{g}_{(1)t\theta}&=&\mathcal{O}(q^{-1}),\\
\label{eqn:efetphi_3p1}
\tilde{\Box}\bar{g}_{(1)t\phi}&=&\mathcal{O}(q^{-1}),\\
\label{eqn:eferhorho_3p1}
\tilde{\Box}\bar{g}_{(1)\rho\rho}&=&3q^{-2} \left(\csc ^2\theta \bar{g}_{(1) \phi \phi }+\bar{g}_{(1)\theta \theta}-2 \bar{g}_{(1)
   \rho \rho }-\bar{g}_{(1) \text{$tt$}}+2 \bar{H}_{(1) \rho }\right)+\mathcal{O}(q^{-1}),\\
\label{eqn:eferhotheta_3p1}
\tilde{\Box}\bar{g}_{(1)\rho\theta}&=&\frac{1}{2} q^{-2}(\bar{g}_{(1)\theta \theta,\theta }+\csc ^2\theta \left(-\bar{g}_{(1)
   \phi \phi ,\theta }+2 \bar{g}_{(1)\theta \phi,\phi }+5 \cot \theta  \bar{g}_{(1) \phi
   \phi }\right)+\bar{g}_{(1) \rho \rho ,\theta }\nonumber\\
   &&-2 \bar{g}_{(1) \text{$t$$\theta
   $},t}+\bar{g}_{(1) \text{$tt$},\theta }-2 \bar{H}_{(1) \rho ,\theta }-3 \cot \theta 
   \bar{g}_{(1)\theta \theta}-40 \bar{g}_{(1) \rho \theta }+20 \bar{H}_{(1)\theta})+\mathcal{O}(q^{-1}),\\
\label{eqn:eferhophi_3p1}
\tilde{\Box}\bar{g}_{(1)\rho\phi}&=&\frac{1}{2} q^{-2}(2 \bar{g}_{(1)\theta \phi,\theta }+\csc ^2\theta \bar{g}_{(1) \phi
   \phi ,\phi }-\bar{g}_{(1)\theta \theta,\phi }+\bar{g}_{(1) \rho \rho ,\phi }-2
   \bar{g}_{(1) \text{$t$$\phi $},t}\nonumber\\
   &&+\bar{g}_{(1) \text{$tt$},\phi }-2 \bar{H}_{(1) \rho ,\phi }-13
   \cot \theta  \bar{g}_{(1)\theta \phi}-40 \bar{g}_{(1) \rho \phi }+20 \bar{H}_{(1) \phi
   })+\mathcal{O}(q^{-1}),\\
\label{eqn:efethetatheta_3p1}
\tilde{\Box}\bar{g}_{(1)\theta\theta}&=&q^{-2}  \left(3 \bar{g}_{(1) \rho \rho }-2 \bar{H}_{(1) \rho }\right)+\mathcal{O}(q^{-1}),\\
\label{eqn:efethetaphi_3p1}
\tilde{\Box}\bar{g}_{(1)\theta\phi}&=&\mathcal{O}(q^{-1}),\\
\label{eqn:efephiphi_3p1}
\tilde{\Box}\bar{g}_{(1)\phi\phi}&=&q^{-2}  \sin^2\theta\left(3 \bar{g}_{(1) \rho \rho }-2 \bar{H}_{(1) \rho }\right)+\mathcal{O}(q^{-1}).
\end{eqnarray}
\end{widetext}

Doing the same for the generalized harmonic constraints $0=C_\alpha\equiv H_\alpha-\Box x_\alpha$, we have
\vspace{\baselineskip}
\begin{widetext}
\begin{eqnarray}
\label{eqn:ct_3p1}
C_t&=&\frac{1}{2} q^3 (-\bar{g}_{(1)\theta \theta,t}-\bar{g}_{(1) \rho \rho ,t}-\bar{g}_{(1)
   \text{$tt$},t}+\csc ^2\theta \left(2 \bar{g}_{(1) \text{$t$$\phi $},\phi }-\bar{g}_{(1)
   \phi \phi ,t}\right)\nonumber\\
   &&+2 \bar{g}_{(1) \text{$t$$\theta $},\theta }-16 \bar{g}_{(1)
   \text{$t$$\rho $}}+8 \bar{H}_{(1) t})+\mathcal{O}(q^4),\\
\label{eqn:crho_3p1}
C_\rho&=&q^2 \left(2 \bar{H}_{(1) \rho }-\frac{3}{2} \left(\csc^2\theta \left(-\bar{g}_{(1) \phi
   \phi }\right)-\bar{g}_{(1)\theta \theta}+\bar{g}_{(1) \rho \rho }+\bar{g}_{(1)
   \text{$tt$}}\right)\right)+\mathcal{O}(q^3),\\
\label{eqn:ctheta_3p1}
C_\theta&=&\frac{1}{2} q^3 (\bar{g}_{(1)\theta \theta,\theta }+\csc ^2\theta
   \left(-\bar{g}_{(1) \phi \phi ,\theta }+2 \bar{g}_{(1)\theta \phi,\phi }+2 \cot \theta
    \bar{g}_{(1) \phi \phi }\right)\nonumber\\
   &&-\bar{g}_{(1) \rho \rho ,\theta }-2 \bar{g}_{(1)
   \text{$t$$\theta $},t}+\bar{g}_{(1) \text{$tt$},\theta }-16 \bar{g}_{(1) \rho \theta }+8
   \bar{H}_{(1)\theta})+\mathcal{O}(q^4),\\
\label{eqn:cphi_3p1}
C_\phi&=&\frac{1}{2} q^3 (\bar{g}_{(1)\theta \theta,\theta }+\csc ^2\theta
   \left(-\bar{g}_{(1) \phi \phi ,\theta }+2 \bar{g}_{(1)\theta \phi,\phi }+2 \cot \theta
    \bar{g}_{(1) \phi \phi }\right)\nonumber\\
   &&-\bar{g}_{(1) \rho \rho ,\theta }-2 \bar{g}_{(1)
   \text{$t$$\theta $},t}+\bar{g}_{(1) \text{$tt$},\theta }-16 \bar{g}_{(1) \rho \theta }+8
   \bar{H}_{(1)\theta})+\mathcal{O}(q^4).
\end{eqnarray}
\end{widetext}

We now follow the three steps of Section~\ref{sec:gauge_choice} to obtain a stable gauge choice.
\begin{enumerate}
\item Solve the leading order of the near-boundary generalized harmonic constraints, \eqref{eqn:ct_3p1}--\eqref{eqn:cphi_3p1}, for $\bar{H}_{(1)\alpha}$. We obtain
\begin{eqnarray}\label{eqn:target_gauge_trhothetaphi_step1}
\bar{H}_{(1)t}&=&\frac{1}{8} (\bar{g}_{(1)\theta \theta,t} +\csc ^2\theta \left(\bar{g}_{(1) \phi \phi ,t}-2 \bar{g}_{(1) \text{$t$$\phi$},\phi }\right)\nonumber \\
   &&+\bar{g}_{(1) \rho \rho ,t}+\bar{g}_{(1) \text{$tt$},t}-2 \bar{g}_{(1) \text{$t$$\theta $},\theta }+16 \bar{g}_{(1) \text{$t$$\rho
   $}})\,,\nonumber \\
\bar{H}_{(1)\rho}&=&\frac{3}{4} \left(-\csc ^2\theta \bar{g}_{(1) \phi \phi }-\bar{g}_{(1)\theta \theta}+\bar{g}_{(1) \rho \rho }+\bar{g}_{(1) \text{$tt$}}\right),\nonumber\\
\bar{H}_{(1)\theta}&=&\frac{1}{8} (-\bar{g}_{(1)\theta \theta,\theta } \nonumber \\
    &&+\csc ^2\theta \left(\bar{g}_{(1)
   \phi \phi ,\theta }-2 \left(\bar{g}_{(1)\theta \phi,\phi }+\cot \theta  \bar{g}_{(1)
   \phi \phi }\right)\right)\nonumber\\
   &&+\bar{g}_{(1) \rho \rho ,\theta }+2 \bar{g}_{(1) \text{$t$$\theta
   $},t}-\bar{g}_{(1) \text{$tt$},\theta }+16 \bar{g}_{(1) \rho \theta })\,,\nonumber\\
\bar{H}_{(1)\phi}&=&\frac{1}{8} (-2 \bar{g}_{(1)\theta \phi,\theta } +\csc ^2\theta \left(-\bar{g}_{(1)
   \phi \phi ,\phi }\right)\nonumber \\
   &&+\bar{g}_{(1)\theta \theta,\phi }+\bar{g}_{(1) \rho \rho ,\phi
   }+2 \bar{g}_{(1) \text{$t$$\phi $},t}-\bar{g}_{(1) \text{$tt$},\phi }\nonumber\\
   &&+4 \cot \theta 
   \bar{g}_{(1)\theta \phi}+16 \bar{g}_{(1) \rho \phi }).
\end{eqnarray}

\item Plug \eqref{eqn:target_gauge_trhothetaphi_step1} into the $q^{-2}$ terms of \eqref{eqn:efett_3p1}--\eqref{eqn:efephiphi_3p1}. This gives the following independent equations:
\begin{eqnarray}
\label{eq:indeq1_3p1}
&&\bar{g}_{(1) \text{$tt$}}-\csc ^2\theta\bar{g}_{(1) \phi \phi }-\bar{g}_{(1)\theta \theta }-\bar{g}_{(1) \rho \rho }=0, \\
\label{eq:indeq2_3p1}
&&\hspace{-0.4cm} \csc ^2\theta \left(-\bar{g}_{(1) \phi \phi ,\theta }+\bar{g}_{(1)\theta \phi,\phi}+\cot \theta  \bar{g}_{(1) \phi \phi }\right)\nonumber \\
   &&\hspace{0.5cm} -\frac{2}{3} \bar{g}_{(1) \rho \rho,\theta } +\bar{g}_{(1) \text{$tt$},\theta }+\cot \theta  \bar{g}_{(1)\theta \theta}=0,\\
   \label{eq:indeq3_3p1}
&&\hspace{-0.4cm}\csc ^2\theta 
   \left(\bar{g}_{(1) \phi \phi ,t}-\bar{g}_{(1) \text{$t$$\phi $},\phi }\right)+\bar{g}_{(1)\theta \theta,t}\nonumber \\
   &&\hspace{2.0cm}+\frac{2}{3} \bar{g}_{(1) \rho \rho ,t}-\cot \theta\bar{g}_{(1) \text{$t$$\theta $}}=0,\\
      \label{eq:indeq4_3p1}
&&\hspace{-0.4cm}\bar{g}_{(1)\theta \phi,\theta }-\bar{g}_{(1)\theta \theta,\phi }-\frac{2}{3} \bar{g}_{(1)\rho \rho ,\phi }\nonumber \\
   &&\hspace{2.2cm}+\bar{g}_{(1) \text{$tt$},\phi }+\cot \theta \bar{g}_{(1)\theta \phi}=0.
\end{eqnarray}
We prove below that these equations ensure tracelessness and conservation of the boundary stress-energy tensor $\langle T_{ab}\rangle_{CFT}$, defined in Section~\ref{sec:bouset2}.

\item Use \eqref{eq:indeq1_3p1}--\eqref{eq:indeq4_3p1} to eliminate $\bar{g}_{(1)tt}$, $\bar{g}_{(1)t\theta,t}$, $\bar{g}_{(1)t\theta,\theta}$, $\bar{g}_{(1)t\phi,t}$ from \eqref{eqn:target_gauge_trhothetaphi_step1}.
In this way we obtain a stable gauge in spherical coordinates
\begin{eqnarray}\label{eqn:target_gauge_trhothetaphi}
\bar{H}_{(1)t}&=&\frac{1}{12} \left(\bar{g}_{(1) \rho \rho ,t}+3 \cot \theta  \bar{g}_{(1) \text{$t$$\theta
   $}}+24 \bar{g}_{(1) \text{$t$$\rho $}}\right), \nonumber\\
\bar{H}_{(1)\rho}&=&\frac{3}{2} \bar{g}_{(1) \rho \rho }\,,\nonumber\\
\bar{H}_{(1)\theta}&=&\frac{1}{12} (3 \cot \theta  \left(\bar{g}_{(1)\theta\theta}-\csc ^2\theta \bar{g}_{(1) \phi \phi }\right)\nonumber \\
   &&\hspace{0.5cm}+\bar{g}_{(1) \rho \rho ,\theta }+24 \bar{g}_{(1) \rho \theta }),\nonumber\\
\bar{H}_{(1)\phi}&=&\frac{1}{12}(\bar{g}_{(1) \rho \rho ,\phi }+9 \cot \theta  \bar{g}_{(1)\theta \phi}\nonumber \\
   &&\hspace{0.5cm}+24 \bar{g}_{(1) \rho \phi }).
\end{eqnarray}
\end{enumerate}
By looking at the gauge choice made in~\cite{Bantilan:2012vu} (see eq. (74)) to obtain stability in simulations of 5-dimensional asymptotically AdS spacetimes with an SO(3) symmetry, and choosing numerical factors consistent with \eqref{eqn:target_gauge_trhothetaphi}, we can infer the following potentially stable gauge for the 4-dimensional case with no symmetry assumptions:
\begin{eqnarray}
\label{eq:hbold_3p1}
\bar{H}_{(1)t}&=&2 \bar{g}_{(1)\text{$t$$\rho $}}\,, \nonumber\\
\bar{H}_{(1)\rho}&=&\frac{3}{2} \bar{g}_{(1) \rho \rho }\,,\nonumber\\
\bar{H}_{(1)\theta}&=&2 \bar{g}_{(1) \rho \theta }\,,\nonumber\\
\bar{H}_{(1)\phi}&=&2 \bar{g}_{(1) \rho \phi }\,.
\end{eqnarray}
Notice that by setting certain terms in \eqref{eqn:target_gauge_trhothetaphi} to zero, one recovers \eqref{eq:hbold_3p1}.
It will be interesting to confirm numerical stability of \eqref{eqn:target_gauge_trhothetaphi} with empirical studies.

\subsection{Tracelessness and Conservation of Boundary Stress Tensor}

We conclude this subsection by showing that tracelessness and conservation of $\langle T_{ab}\rangle_{CFT}$ follow from \eqref{eq:indeq1_3p1}--\eqref{eq:indeq4_3p1}, i.e., from the lowest order of the Einstein equations, provided that the leading order of the generalized harmonic constraints are satisfied. 

With the notation of Section~\ref{sec:bouset2}, let $x^a=(t,\theta,\phi)$ be the coordinates along the AdS boundary, $\lambda_{ab}dx^a dx^b=-dt^2+d\theta^2+\sin^2\theta d\phi^2$ be the metric of the AdS boundary, and $\mathcal{D}$ be the Levi-Civita connection of $\lambda_{ab}$, i.e., $\mathcal{D}$ is torsion-free and $\mathcal{D}_a\lambda_{bc}=0$. Then, $\langle \text{tr}T\rangle_{CFT}=\lambda^{ab}\langle T_{ab}\rangle_{CFT}$ is the trace of the boundary-stress tensor and $\mathcal{D}^a \langle T_{ab}\rangle_{CFT}=\lambda^{ac}\mathcal{D}_c \langle T_{ab}\rangle_{CFT}$ is its divergence. 
We want to prove that $\langle \text{tr}T\rangle_{CFT}=0$ and $\mathcal{D}^a \langle T_{ab}\rangle_{CFT}=0$. 
The expression of $\langle \text{tr}T\rangle_{CFT}$ in terms of the leading order of the metric variables in spherical coordinates was already written in \eqref{eq:tracecalc}. We repeat it here for completeness:
\begin{equation}
\label{eq:tracecalc2}
\langle \text{tr}T\rangle_{CFT}=\frac{3}{8\pi}\biggl(\bar{g}_{(1)tt}-\bar{g}_{(1)\rho\rho}-\bar{g}_{(1)\theta\theta}-\csc ^2\theta \bar{g}_{(1)\phi\phi}\biggr).
\end{equation}
The divergence of the boundary stress tensor is given by
\begin{eqnarray}
\label{eq:divergence_t}
\mathcal{D}^a \langle T_{at}\rangle_{CFT}&=&\frac{1}{16\pi}(-3 \csc ^2\theta  \bar{g}_{(1) \phi \phi ,t}-3 \bar{g}_{(1)\theta \theta ,t}\nonumber\\
&&\hspace{-1.0cm}-2 \bar{g}_{(1) \rho \rho ,t}+3 \bar{g}_{(1) \text{$t$$\theta $},\theta }+3 \csc^2\theta \bar{g}_{(1) \text{$t$$\phi $},\phi }\nonumber\\
&&\hspace{-1.0cm}+3 \cot \theta  \bar{g}_{(1)\text{$t$$\theta $}})\,, \\
\label{eq:divergence_theta}
\mathcal{D}^a \langle T_{a\theta}\rangle_{CFT}&=&\frac{1}{16\pi}(3 \csc ^2\theta  \bar{g}_{(1)\theta \phi ,\phi }-2 \bar{g}_{(1) \rho \rho ,\theta } \nonumber \\
&&\hspace{-1.0cm} -3 \csc^2\theta \bar{g}_{(1) \phi \phi ,\theta }-3 \bar{g}_{(1) \text{$t$$\theta $},t}+3\bar{g}_{(1) \text{$tt$},\theta } \nonumber\\
   &&\hspace{-1.0cm}+3 \cot \theta  \bar{g}_{(1)\theta \theta }+3 \cot   \theta \csc ^2\theta  \bar{g}_{(1) \phi \phi })\,,\\
\label{eq:divergence_phi}
\mathcal{D}^a \langle T_{a\phi}\rangle_{CFT}&=&\frac{1}{16\pi}(3 \bar{g}_{(1)\theta \phi ,\theta }-3 \bar{g}_{(1)\theta \theta ,\phi }-2 \bar{g}_{(1) \rho\rho ,\phi }\nonumber\\
   &&\hspace{-1.0cm}-3 \bar{g}_{(1) \text{$t$$\phi $},t}+3 \bar{g}_{(1) \text{$tt$},\phi }+3 \cot\theta \bar{g}_{(1)\theta \phi })\,.
\end{eqnarray}
We immediately see that $\langle \text{tr}T\rangle_{CFT}=0$ as a consequence of \eqref{eq:indeq1_3p1}.
Moreover, by solving the system of 6 equations given by the first derivatives of \eqref{eq:indeq1_3p1} with respect to $t,\theta,\phi$ and \eqref{eq:indeq2_3p1}, \eqref{eq:indeq3_3p1}, \eqref{eq:indeq4_3p1} for $\bar{g}_{(1) \text{$tt$},t},\bar{g}_{(1) \text{$tt$},\theta},\bar{g}_{(1) \text{$tt$},\phi },\bar{g}_{(1) \text{$t$$\theta $},t}$,$\bar{g}_{(1)\text{$t$$\theta $},\theta },\bar{g}_{(1) \text{$t$$\phi $},t}$, and substituting the solution into the right hand side of \eqref{eq:divergence_t}--\eqref{eq:divergence_phi}, we see that $\mathcal{D}^a \langle T_{ab}\rangle_{CFT}=0$.

\subsection{Boundary stress tensor from holographic renormalization}
\label{sec:HoloRen}

We can straightforwardly compute the boundary stress tensor from \eqref{eq:asyFG} using the holographic renormalization prescription of \cite{deHaro:2000vlm} in spherical coordinates $x^{\bar{a}}=(\bar{t},\bar{\theta},\bar{\phi})$ on the AdS boundary. We have
\begin{equation}
\label{eq:bdysetFGform}
\langle T_{\bar{a}\bar{b}}\rangle_{CFT}=\frac{3}{16\pi}g^{(3)}_{\bar{a}\bar{b}}\,,
\end{equation}
where $g^{(3)}_{\bar{a}\bar{b}}$ are the $z^3$ terms of the metric components in FG form, \eqref{eq:asyFG}.
The explicit components of the stress tensor in \eqref{eq:bdysetFGform} are given by
\begin{eqnarray}
\label{eq:set_explicit_2}
\langle T_{\bar{t}\bar{t}}\rangle_{CFT}&=&\frac{1}{16\pi}(3f_{tt}-f_{\rho\rho})\,, \nonumber \\
\langle T_{\bar{t}\bar{\theta}}\rangle_{CFT}&=&\frac{3}{16\pi}f_{t\theta}\,, \nonumber \\
\langle T_{\bar{t}\bar{\phi}}\rangle_{CFT}&=&\frac{3}{16\pi}f_{t\phi}\,, \nonumber \\
\langle T_{\bar{\theta}\bar{\theta}}\rangle_{CFT}&=&\frac{1}{16\pi} (3f_{\theta\theta}+f_{\rho\rho})\,, \nonumber \\
\langle T_{\bar{\theta}\bar{\phi}}\rangle_{CFT}&=&\frac{3}{16\pi}f_{\theta\phi}\,, \nonumber \\
\langle T_{\bar{\phi}\bar{\phi}}\rangle_{CFT}&=&\frac{\sin^2\bar{\theta}}{16\pi} \biggl(\frac{f_{\phi\phi}}{ \sin^2\bar{\theta}}+\frac{1}{3}f_{\rho\rho}\biggr).
\end{eqnarray}

On the other hand, in Section~\ref{sec:bouset2} we compute the boundary stress-tensor starting from the metric in global spherical coordinates and then using the prescription of \cite{Balasubramanian:1999re}. Of course, the expressions \eqref{eq:set_explicit} and \eqref{eq:set_explicit_2} are equivalent, as we now explain. To obtain \eqref{eq:set_explicit}, we have not imposed that the metric components satisfy the Einstein equations. On the other hand, \eqref{eq:bdysetFGform} gives the correct boundary stress-energy tensor if the bulk metric solves the Einstein equations, in agreement with the assumptions of the FG theorem. It is thus expected that \eqref{eq:set_explicit} and \eqref{eq:set_explicit_2} agree if we assume the validity of the lowest order of the Einstein equations in the form that takes into account the generalized harmonic constraints, i.e., \eqref{eq:indeq1_3p1}--\eqref{eq:indeq4_3p1}.
In fact, we only need \eqref{eq:indeq1_3p1}. For example, starting from \eqref{eq:set_explicit}, imposing \eqref{eq:indeq1_3p1} and using the fact that $\bar{t}=t,\bar{\theta}=\theta,\bar{\phi}=\phi$ at the boundary $\rho=1$ together with $\bar{g}_{(1)\alpha\beta}=f_{\alpha\beta}$,\footnote{Note that $\bar{t}=t,\bar{\theta}=\theta,\bar{\phi}=\phi$ at $\rho=1$ (i.e., $\bar{z}=0$) from \eqref{eqn:invertFGcoords}, while $\bar{g}_{(1)\alpha\beta}\equiv\frac{\partial \bar{g}_{\alpha\beta}}{\partial q}\bigr|_{q=0}=f_{\alpha\beta}$, where the second equality is obtained by comparing \eqref{eq:sphbounconh} with \eqref{eq:gbarsph} to write $\bar{g}_{\alpha\beta}$ in terms of $f_{\alpha\beta}$ and the corresponding $\rho$-dependent factors.} we find precisely the expressions \eqref{eq:set_explicit_2}.

\section{Boundary Prescription for the Poincar\'e Patch}\label{sec:poincare}

Here we follow the prescription of Section~\ref{sec:pre_sta} in the case of Poincar\'e AdS and display a choice of generalized harmonic source functions that stabilizes the evolution in this case.

The metric of the Poincar\'e patch of AdS$_4$, with AdS radius set to $L=1$, can be written as
\begin{equation}
\label{eq:AdSpoincare}
\hat{g} = \frac{1}{z^2} \left( -dt^2 + dz^2 + dx_1{}^2 + dx_2{}^2  \right).
\end{equation}
in terms of Poincar\'e coordinates $(t,z,x_1,x_2)$. To include the Poincar\'e horizon $z\rightarrow \infty$ in our computational domain, we compactify the bulk coordinate $z=(1-\rho^2)/\rho^2$ to have the Poincar\'e horizon at $\rho=0$ and the AdS boundary at $\rho=1$. This gives the following form for the metric of AdS$_4$:
\begin{equation}
\hat{g} = \frac{\rho^4}{(1-\rho^2)^2} \left( -dt^2 + (4/\rho^6)d\rho^2 + dx_1^2 + dx_2^2  \right).
\end{equation}

Let us now consider asymptotically AdS spacetimes.
Since \eqref{eq:AdSpoincare} is in the form given by the leading order of the FG expansion, \eqref{eqn:FGmetric}--\eqref{eqn:FGbdymetric}, we see that $(t,z,x_1,x_2)$ are FG coordinates.
We can thus read off the fall-offs of the metric components from the rest of the FG expansion.
The evolved fields consist of the spacetime metric $g_{\mu\nu}$, possibly a scalar field $\varphi$, and the generalized harmonic source functions $H_\mu$.
The fall-offs of the metric components $g_{\mu\nu}$ read the same as \eqref{eq:carbouncondh}, with $f_{\mu\nu}(t,x_1,x_2)$ coefficients.
The scalar field fall-off that preserves the metric asymptotics is given by \eqref{eq:carbouncondphi}, with $c(t,x_1,x_2)$ coefficient.
The fall-offs of the source functions can be inferred from the metric fall-offs, which are given by \eqref{eq:carbouncondsoufun}, with $f_{\mu}(t,x_1,x_2)$ coefficients.
As a result, the corresponding evolution variables in this Poincar\'e setting are given exactly by the same expressions as we had written in \eqref{eq:gbarcart}--\eqref{eq:soufunb}.

Using the same steps as in Section~\ref{sec:gauge_choice}, we obtain the following gauge:
\begin{eqnarray}
\label{eq:hbold_poincare}
\bar{H}_{(1)t}&=&\frac{3}{2} \bar{g}_{(1)\text{$t$$\rho $}}\,, \nonumber\\
\bar{H}_{(1)\rho}&=&\frac{3}{2} \bar{g}_{(1) \rho \rho }\,,\nonumber\\
\bar{H}_{(1)x_1}&=&\frac{3}{2} \bar{g}_{(1) \rho x_1 }\,,\nonumber\\
\bar{H}_{(1)x_2}&=&\frac{3}{2} \bar{g}_{(1) \rho x_2 }\,.
\end{eqnarray}
We have verified that this gauge leads to stable evolution in asymptotically AdS$_4$ spacetimes in Poincar\'e coordinates. 
We close by noting that \cite{Bantilan:2020pay} obtained a similar stable gauge to evolve dynamical black holes in the background of the AdS soliton.

\section{Initial Data}
\label{sec:initdata}

The Cauchy problem in GR requires the prescription of initial data on a spacelike hypersurface $\Sigma$ and a choice of gauge throughout the entire evolution. In an asymptotically AdS spacetime, in addition we have to specify boundary conditions at the boundary of AdS; we have dealt with boundary conditions in Section \ref{sec:pre_sta}. We pick Cartesian coordinates $x^\mu=(t,x,y,z)$ such that $t=0$ on $\Sigma$. The spatial Cartesian coordinates on $\Sigma$ are denoted by $x^i=(x,y,z)$, and the corresponding indices by $i,j,k,\dots$. 
With this notation, the data needed for the Cauchy evolution in the generalized harmonic scheme is composed of the initial data $\bar{\varphi}|_{t=0}$, $\bar{g}_{ij}|_{t=0}$, $\partial_t\bar{\varphi}|_{t=0} $, $\partial_t\bar{g}_{ij}|_{t=0}$ and the source functions $\bar{H}_\mu$ at all times. The gauge used in our numerical scheme at $t>0$ is discussed in Appendix~\ref{sec:GCbulk}. With regard to the gauge at $t=0$, we do not set $\bar{H}_\mu|_{t=0}$ explicitly, but we make an equivalent choice for $\bar{g}_{t\mu}|_{t=0}$, and $\partial_t\bar{g}_{t\mu}|_{t=0}$, and then compute $\bar{H}_\mu|_{t=0}$ from \eqref{eq:defsoufunsph}.
In summary, the complete set of initial data that we prescribe is $\bar{\varphi}|_{t=0}$, $\bar{g}_{\mu\nu}|_{t=0}$, $\partial_t\bar{\varphi}|_{t=0}$ and $\partial_t\bar{g}_{\mu\nu}|_{t=0}$. In this section we explain how this is done in our simulations, taking into account two crucial facts. Firstly, initial data cannot be chosen in a completely arbitrary way, but it must satisfy the constraints of GR. Secondly, the choice of the initial degrees of freedom must be consistent with the desired gauge \eqref{eqn:target_gauge_txyz} near the AdS boundary. 

\subsection{Constraints}
\label{sec:constr}

Here we review the constraints in GR and how they are solved in our numerical scheme.
We start by defining the relevant quantities on the initial spacelike hypersurface $\Sigma$. 

The timelike, future-directed unit 1-form normal to $\Sigma$ is given by
\begin{equation}
\label{eq:uninormal}
n_\mu=-\alpha (dt)_\mu\,,
\end{equation}
where $\alpha=1/\sqrt{-g^{\mu\nu}(dt)_\mu (dt)_\nu}$ is the lapse function. 
The projection operator onto $\Sigma$ is defined by
\begin{equation}
\gamma^\mu_\nu=\delta^\mu_\nu+n^\mu n_\nu.
\end{equation}
(Notice that $\gamma^\mu_\nu$ is idempotent, i.e., $\gamma^\mu_\rho \gamma^\rho_\nu=\gamma^\mu_\nu$, as appropriate for a projector.)
This operator can be applied to any tensor at a point $p\in\Sigma$ to obtain the part of that tensor tangent to $\Sigma$. For instance, given a vector $X$ at a point $p\in\Sigma$, $X_{||}^\mu=\gamma^\mu_\nu X^\nu$ is the part of $X$ tangent to $\Sigma$, i.e., $X_{||}^\mu n_\mu=0$. 
Let us now consider a tensor defined on the tangent space of the spacetime manifold $M$ at a point $p\in\Sigma$. If the tensor is invariant under projection onto $\Sigma$, then it can be identified with a tensor defined on the tangent space of $\Sigma$ at $p$, under a natural (i.e., basis-independent) isomorphism. For example, $\gamma_{\mu\nu}=g_{\mu\nu}+n_\mu n_\nu$ at points on $\Sigma$ can be identified with the Riemannian metric of $\Sigma$ defined as the pull-back\footnote{We refer to the pull-back with respect to the inclusion map that embeds $\Sigma$ in $M$.} on $\Sigma$ of the spacetime metric $g_{\mu\nu}$, given by $\gamma_{ij}$ in spatial Cartesian coordinates. See \cite{Hawking:1973uf} for more details. Indices of tensors invariant under projection onto $\Sigma$ can be raised and lowered by $\gamma_{\mu\nu}$ or $g_{\mu\nu}$, equivalently. Indices $i,j,k,\dots$ of tensors on the tangent space of $\Sigma$ can be raised and lowered by $\gamma_{ij}$.

The projection of $\nabla_\mu n_\nu$ defines the extrinsic curvature of $\Sigma$:\footnote{We can make sense of covariant derivatives of $n_\mu$ by extending its definition on $\Sigma$ \eqref{eq:uninormal} to a 1-form field over a neighbourhood of $\Sigma$, which can be done in an arbitrary way without changing the value of $K_{\mu\nu}$ on $\Sigma$ given by \eqref{eq:extrcurv}.}
\begin{equation}
\label{eq:extrcurv}
K_{\mu\nu}=-\gamma^\rho_\mu \gamma^\sigma_\nu \nabla_\rho n_\sigma=-\frac{1}{2}\mathcal{L}_n\gamma_{\mu\nu}.
\end{equation}
The Lie derivative along the normal direction in the second equality suggests that a choice of $K_{\mu\nu}$ on $\Sigma$ is ``morally'' equivalent to a choice for the time-derivative of the metric components at $t=0$. $K_{\mu\nu}$ is identified with the tensor on the tangent space of $\Sigma$, given by $K_{ij}$.

As a final ingredient, the covariant derivative on $\Sigma$ of a tensor field invariant under projection onto $\Sigma$ is defined as the projection onto $\Sigma$ of the covariant derivative $\nabla$ of the tensor field, and we denote it by $D$. For instance, $D_\mu X_{||}^\nu=\gamma^\rho_\mu \gamma^\nu_\sigma \nabla_\rho X_{||}^\sigma$ and $D_\mu X_{||}^\nu$ is identified with the tensor on the tangent space of $\Sigma$ given by $D_i X_{||}^j=\gamma^\rho_i \gamma^j_\sigma \nabla_\rho X_{||}^\sigma$. $D$ is the Levi-Civita connection of $\gamma_{ij}$, i.e., it is torsion-free and $D_i\gamma_{jk}=0$.

We can now write the constraints that initial data on $\Sigma$ must satisfy. The ``normal-normal'' projection (i.e., contraction with $n^\mu n^\nu$) of the Einstein equations gives the Hamiltonian constraint
\begin{equation}
\label{eq:hamconstr}
^{(3)}R-K^{ij}K_{ij}+K^2-2\Lambda=16\pi\rho,
\end{equation}
where $^{(3)}R$ is the Ricci scalar associated with the connection $D$, $K=\gamma^{ij}K_{ij}$ and $\rho=T_{\mu\nu}n^\mu n^\nu$ is the matter energy density measured by an observer with 4-velocity $n^\mu$.
The ``tangent-normal'' projection (i.e., contraction with $\gamma^{\mu\nu} n^\rho$) of the Einstein equations gives the momentum constraint
\begin{equation}
\label{eq:momconstr}
D_j {K^j}_i-D_i K=8\pi j_i\,,
\end{equation}
where $j^i=-T_{\rho\sigma}n^\rho\gamma^{\sigma i}$ is the matter momentum density measured by an observer with 4-velocity $n^\mu$.

We now explain how these constraints are solved for massless real scalar matter, whose energy-momentum tensor is \eqref{eq:KHmomtens}, in the simplified case of time-symmetric data. Time symmetry in the scalar sector,
\begin{equation}
\partial_t \bar{\varphi}\big|_{t=0}=0,
\end{equation}
implies $j^i=0$. Time symmetry in the gravitational sector,
\begin{equation}
\partial_t \bar{g}_{ij}\big|_{t=0}=0,
\end{equation}
together with the initial gauge choice
\begin{equation}
\bar{g}_{ti}\big|_{t=0}=0,
\end{equation}
implies $K_{ij}=0$. Thus, we see that the momentum constraint is trivially satisfied.
The Hamiltonian constraint, instead, reduces to
\begin{equation}
\label{eq:redhamconstr}
^{(3)}R-2\Lambda=16\pi\rho.
\end{equation}
This can be solved through the conformal approach, initiated in \cite{Lichnerowicz:1994}, which assumes that the spatial metric $\gamma_{ij}$ is conformal to the spatial metric $\hat{\gamma}_{ij}$ of the $t=0$ slice of pure AdS in Cartesian coordinates:
\begin{equation}
\label{eq:confdec}
\gamma_{ij}=\zeta^4 \hat{\gamma}_{ij},
\end{equation}
where $\zeta$ is a smooth positive function on $\Sigma$, satisfying the AdS boundary condition $\zeta|_{\rho=1}=1$. Let $\hat{D}$ be the Levi-Civita connection of $\hat{\gamma}_{ij}$ and $^{(3)}\hat{R}$ the corresponding Ricci scalar. Using \eqref{eq:confdec} and its inverse, $\gamma^{ij}=\zeta^{-4} \hat{\gamma}^{ij}$, we obtain
\begin{equation}
\label{eq:confRicci}
^{(3)}R=\frac{1}{\zeta^4}\left(^{(3)}\hat{R}-\frac{8}{\zeta}\hat{\gamma}^{ij}\hat{D}_i \hat{D}_j \zeta\right).
\end{equation}
Plugging \eqref{eq:confRicci} into \eqref{eq:redhamconstr} gives
\begin{equation}
\label{eq:rehamconstr2}
^{(3)}\hat{R}\zeta-8\hat{\gamma}^{ij}\hat{D}_i \hat{D}_j \zeta-2\Lambda \zeta^5=16\pi\rho \zeta^5.
\end{equation}
$^{(3)}\hat{R}$ can be computed from the spatial part of the pure AdS metric \eqref{eqn:ads4_final}: $^{(3)}\hat{R}=-6/L^2=2\Lambda$. Thus, equation \eqref{eq:rehamconstr2} can be written as
\begin{equation}
\label{eq:rehamconstr3}
\hat{\gamma}^{ij}\hat{D}_i \hat{D}_j \zeta-\frac{1}{4}\Lambda\zeta+\frac{1}{4}(\Lambda+8\pi\rho)\zeta^5=0.
\end{equation}
Finally, the version of the Hamiltonian constraint that we are going to solve is obtained by writing the matter energy density $\rho$ in terms of $\zeta$. The time-symmetry requirement $\partial_t \varphi\big|_{t=0}=0$ gives
\begin{equation}
\label{eq:mattendens}
\rho=\frac{1}{2\zeta^4}\hat{\gamma}^{ij}\partial_i\varphi\partial_j\varphi,
\end{equation}
so the Hamiltonian constraint reads
\begin{equation}
\label{eq:hamconsfinal}
\hat{\gamma}^{ij}\hat{D}_i \hat{D}_j \zeta-\frac{1}{4}\Lambda\zeta+\frac{1}{4}(\Lambda\zeta^5+4\pi\zeta\hat{\gamma}^{ij}\partial_i\varphi\partial_j\varphi)=0.
\end{equation}
For any given choice of scalar field $\varphi$ on $\Sigma$, \eqref{eq:hamconsfinal} is an elliptic equation that can be solved for $\zeta$ with boundary condition $\zeta|_{\rho=1}=1$. In our simulations we pick the initial scalar field profile $\varphi|_{t=0}=\bar{\varphi}|_{t=0}(1-\rho^2)^2$ with $\bar{\varphi}|_{t=0}$ specified by \eqref{eq:scaGaupro}, and we solve \eqref{eq:hamconsfinal} with a multigrid algorithm, built into the PAMR/AMRD libraries.
The initial metric variables $\bar{g}_{ij}|_{t=0}$ are then easily reconstructed from \eqref{eq:confdec} and $\gamma_{ij}=g_{ij}|_{t=0}=\hat{g}_{ij}|_{t=0}+\bar{g}_{ij}|_{t=0}$, i.e.,
\begin{equation}
\bar{g}_{ij}\big|_{t=0}=\zeta^4\hat{\gamma}_{ij}-\hat{g}_{ij}\big|_{t=0}.
\end{equation}

\subsection{Consistency at the boundary}
\label{sec:consistbound}

In the previous section we explained how some components of the initial data for our simulations are obtained: (i) we impose time-symmetry, namely $\partial_t\bar{\varphi}|_{t=0}=0$ and $\partial_t\bar{g}_{ij}|_{t=0}=0$; (ii) we make the initial gauge choice $\bar{g}_{ti}|_{t=0}=0$; (iii) we choose the massless real scalar field profile $\bar{\varphi}|_{t=0}$ given by \eqref{eq:scaGaupro}; (iv) we determine $\bar{g}_{ij}|_{t=0}$ through the conformal decomposition of the Hamiltonian constraint. In this section we determine the remaining necessary components for Cauchy evolution based on the generalized harmonic scheme: $\bar{g}_{tt}|_{t=0}$ and $\partial_t\bar{g}_{t \mu}|_{t=0}$.

In doing so, the only restriction to consider is the one already obtained in step 2 of our gauge prescription in Section~\ref{sec:gauge_choice}: the Einstein equations in a gauge that satisfies the generalized harmonic constraints impose the condition $\bar{g}_{(1)tt}=\bar{g}_{(1)xx}+\bar{g}_{(1)yy}+\bar{g}_{(1)zz}$ near the boundary. This will hold at all times of the evolution and it must be imposed on initial data. Given that there is no requirement on the value of $\bar{g}_{tt}$ in the bulk, we make the simplest choice and set that to zero. In order to smoothly transition from the bulk value of $\bar{g}_{tt}$ to its required boundary value, we use the smooth transition function
  \begin{equation}
  \label{eq:transfunc}
    f(\rho) =
    \begin{cases*}
      1\,, & if $\rho\geq \rho_{b}$, \\
      1-R^3(\rho)\\
      \hspace{0.6cm}(6 R^2(\rho)-15 R(\rho)+10)\,, & if $\rho_{b} > \rho\geq\rho_{a}$, \\
      0\,,        & otherwise,
    \end{cases*}
  \end{equation}
where $R(\rho)=(\rho_{b}-\rho)/(\rho_{b}-\rho_{a})$ and $\rho_{a},\rho_{b}$ are the values between which the transition takes place, set to $\rho_{a}=0.5,\rho_{b}=0.9$ in our simulations.
Thus, our choice of $\bar{g}_{tt}|_{t=0}$ is
\begin{equation}
\bar{g}_{tt}\big|_{t=0}=f(\bar{g}_{xx}|_{t=0}+\bar{g}_{yy}|_{t=0}+\bar{g}_{zz}|_{t=0}).
\end{equation}

To conclude, the remaining initial variables can be chosen in a completely arbitrary way so we make the simplest choice everywhere on the grid:
\begin{equation}
\partial_t\bar{g}_{t \mu}|_{t=0}=0.
\end{equation}

\section{Complete Gauge Choice}
\label{sec:GCbulk}

In Section~\ref{sec:gauge_choice} we discussed the gauge choice of source functions that we impose near the boundary in order to obtain stable evolutions. Furthermore, the gauge at $t=0$, $\bar{H}_{\mu}|_{t=0}$, is determined from the initial data, detailed in Appendix~\ref{sec:initdata}, through the definition of source functions \eqref{eq:defsoufunsph} at $t=0$. All that remains is to make a gauge choice of $\bar{H}_\mu$ in the bulk, and smoothly join this with the target boundary values \eqref{eqn:target_gauge_txyz} on each spatial slice and with the initial values $\bar{H}_{\mu}|_{t=0}$ during evolution. In this section we describe how all this is implemented in our numerical scheme.

We start by choosing a zero value for $\bar{H}_\mu$ in the bulk, as this is the simplest choice. Therefore, the values of the source functions on each spatial slice, after the time transition from $t=0$, are given by
\begin{eqnarray}
\label{eqn:extend_gauge_txyz}
F_t&\equiv&\frac{3f_1}{2\sqrt{x^2+y^2+z^2}}(x \bar{g}_{tx}+y\bar{g}_{ty}+z\bar{g}_{tz})\,, \nonumber \\
F_x&\equiv&\frac{3f_1}{2\sqrt{x^2+y^2+z^2}}(x \bar{g}_{xx}+y\bar{g}_{xy}+z\bar{g}_{xz})\,, \nonumber \\
F_y&\equiv&\frac{3f_1}{2\sqrt{x^2+y^2+z^2}}(x \bar{g}_{xy}+y\bar{g}_{yy}+z\bar{g}_{yz})\,, \nonumber \\
F_z&\equiv&\frac{3f_1}{2\sqrt{x^2+y^2+z^2}}(x \bar{g}_{xz}+y\bar{g}_{yz}+z\bar{g}_{zz})\,,
\end{eqnarray}
where the spatial transition function $f_1(\rho)$ is defined as in \eqref{eq:transfunc} with transition occurring between $\rho_{1a}=0.05$ and $\rho_{1b}=0.95$.

Then, we define the time-transition function
\begin{equation}
g(t,\rho)=\left(\frac{t}{\xi_2 f_0(\rho)+\xi_1(1-f_0(\rho))}\right)^4\,,
\end{equation}
where $f_0(\rho)$ is defined as in \eqref{eq:transfunc} with transition interval between $\rho_{0a}=0.0$ and $\rho_{0b}=0.95$. Notice that $g(0,\rho)=0$, $g(t,\rho)\gg 1$ for $t\gg\xi_1,\xi_2$ and, in particular, $g(t,\rho)$ takes large values with characteristic time $\xi_1$ in the interior region $\rho\leq\rho_{0a}$ (i.e., where $f_0=0$) and characteristic time $\xi_2$ in the near-boundary region $\rho\geq\rho_{0b}$ (i.e., where $f_0=1$).

With these ingredients, we can finally write the complete gauge choice made in our simulations
\begin{equation}
\bar{H}_\mu=\bar{H}_{\mu}\big|_{t=0}\exp(-g)+F_\mu[1- \exp(-g)]\,.
\end{equation}
From the properties of $g(t,\rho)$, we see that $\bar{H}_\mu=\bar{H}_{\mu}|_{t=0}$ at $t=0$ and $\bar{H}_\mu=F_\mu$ for $t\gg\xi_1$ in the interior and $t\gg\xi_2$ near the boundary. Since the target gauge is crucial for stability and needs to be reached quickly, $\xi_2$ is typically set to a small value. On the other hand, it is not necessary, and perhaps even troublesome, to deal with a fast transition in the bulk, therefore $\xi_1$ takes a larger value. In our simulations, we set $\xi_1=0.1,\xi_2=0.0025$.

\section{Boundary Extrapolation}\label{sec:extrapconvbdy}

\begin{figure*}[t!]
        \centering
        \includegraphics[width=6.1in,clip=true]{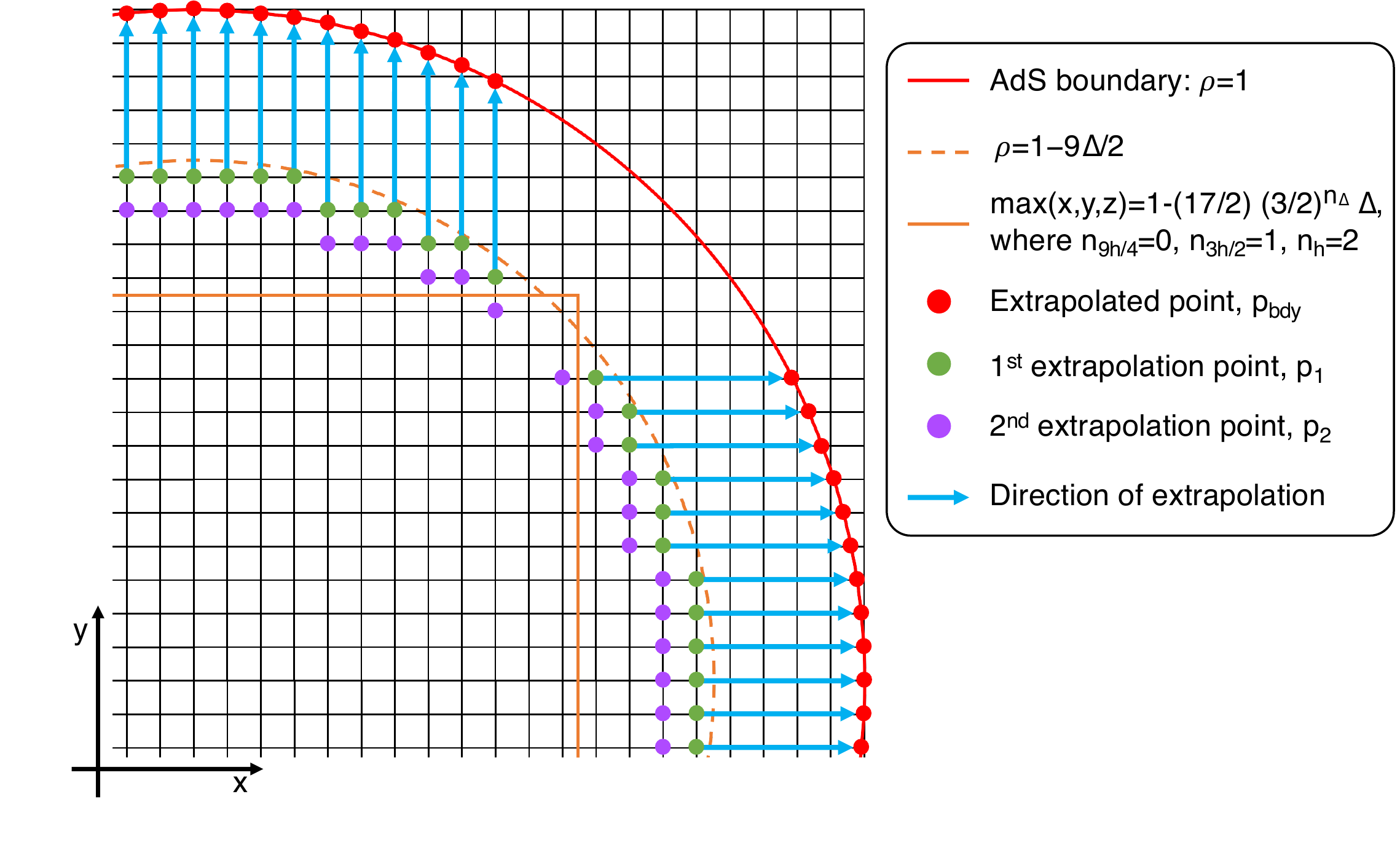}
\parbox{5.0in}{\caption{Visual description of first order extrapolation technique in the first quadrant of a $z=const.$ surface for a grid with spatial refinement $\Delta$.
        }\label{fig:lego_circle}}
\end{figure*}

As explained in Section~\ref{sec:bouset2}, since the AdS boundary generally does not lie on points of the Cartesian grid, we can only obtain the approximated value of any boundary quantity $f$ through extrapolation from the numerical values of $f$ on grid points near the boundary. In this section we describe how extrapolation is implemented in our scheme with the help of Figure~\ref{fig:lego_circle}. 

For simplicity, we consider first order extrapolation, i.e., extrapolation from two grid points. The following can be generalized to higher extrapolation orders in a straightforward way. In particular, third order extrapolation is used for the plots in Section~\ref{sec:resbouset}, since this improves the accuracy of the extrapolated numerical values.\footnote{This fact was tested by comparing values obtained with increasing extrapolation order and analytic values, in cases where the latter are known, e.g. boundary scalar field values at $t=0$.}

Given a Cartesian grid with spacing $\Delta$, let $f_\Delta$ denote the values of $f$ at bulk grid points and $f^{bdy}_{\Delta}$ denote the extrapolated values of $f$ at boundary points. 
We extrapolate the values $f^{bdy}_{\Delta}$ through the following procedure.
 \begin{enumerate}
 \item Restrict to the points with Cartesian coordinates $(x,y,z)$ satisfying $\rho(x,y,z)<1-9\Delta/2$ (inside the orange dashed line of Figure~\ref{fig:lego_circle}) and $\max(x,y,z)>1-\frac{17}{2}\left(\frac{3}{2}\right)^{n_\Delta}\Delta$ (outside the continuous orange line of Figure~\ref{fig:lego_circle}), where $n_\Delta$ denotes the degree of the three resolutions used for convergence, $n_{9h/4}=0,n_{3h/2}=1,n_{h}=2$ (notice that $\left(\frac{3}{2}\right)^{n_\Delta}\Delta$ is a constant for all three resolutions). We have empirically found that considering points outside of this region in the next steps leads to unphysical or non-converging values.
 \item For any point in the range defined at step 1, identify the coordinate with the largest absolute value, e.g., $x$, and its sign, say $x>0$. If two coordinates have the same absolute value, then we pick $x$ over $y$ and $z$, and $y$ over $z$. Each direction identified in this way is represented by a light blue arrow. Among all the points along the identified direction ($x$ in our example) and within the range of step 1, pick the closest point to the boundary. We denote this point by $p_1$ and its coordinates by $(x_1,y_1,z_1)$. For each direction identified as above, the corresponding $p_1$ point is represented as a green dot in Figure~\ref{fig:lego_circle}.
 \item Consider the nearest point to $p_1$ along the identified axis in the direction of the bulk (decreasing $x$ in the example). We denote this point by $p_2$ and its coordinates by $(x_2,y_2,z_2)$. For each $p_1$ point, the corresponding $p_2$ is represented as a purple dot in Figure~\ref{fig:lego_circle}. In our example $x_2=x_1-\Delta,y_2=y_1,z_2=z_1$.
 \item Use first order extrapolation on $f_\Delta(p_1),f_\Delta(p_{2})$ to determine the value of $f^{bdy}_{\Delta}(p_{bdy})$ where $p_{bdy}$ is the boundary point along the identified axis in the direction of the boundary. 
For each pair $p_1,p_2$, the corresponding $p_{bdy}$ is represented by a red dot in Figure~\ref{fig:lego_circle} and the AdS boundary is represented by a red line.
In our example, $p_{bdy}$ is the point with coordinates $(x_{bdy},y_{bdy},z_{bdy})=(\sqrt{1-y_1^2-z_1^2},y_1,z_1)$ and
 \begin{equation}
 \label{eq:firstordextrap}
 f^{bdy}_{\Delta}(p_{bdy})=\frac{x_{bdy}-x_2}{x_1-x_2}f_\Delta(p_1)+\frac{x_{bdy}-x_1}{x_2-x_1}f_\Delta(p_2).
 \end{equation}
 \item In order to avoid issues arising from singularities in the definition of spherical coordinates in terms of Cartesian coordinates, we do not extrapolate boundary points with $z_{bdy}=0$. 
Instead, we fill each of these points by copying the mean value of the closest boundary extrapolated points. This ensures continuity at the semi-circle $z_{bdy}=0, y_{bdy}\geq 0$, i.e., points with $\phi=0\sim 2\pi$.
 \end{enumerate}
 
Figure~\ref{fig:lego_circle} shows that the extrapolated values are not uniformly distributed on the boundary. 
We aim to improve this in the future by extrapolating the values at points on a uniform $(\theta,\phi)$ grid with given resolution on the $S^2$ at the boundary. For now, we fill the empty regions by linearly interpolating boundary values. The data obtained in this way displays high-frequency noise that does not allow for a clear visualisation of physical features. Therefore, we apply a low-pass filter to quantities to be shown on the boundary $S^2$. More precisely, we apply the filter on three copies of the boundary sphere joined along the semi-circle $z_{bdy}=0$, $y_{bdy}\geq 0$ and then we plot the smooth data of the central copy. After re-enforcing continuity at the semi-circle as explained in step 5 above, this strategy provides regular smooth data at the semi-circle if the original raw data is approximately periodic in $\phi$ with period $2\pi$, which is expected for data on a sphere.

Notice that, as \eqref{eq:firstordextrap} shows, second order convergence of boundary values $f^{bdy}_{\Delta}$ is a direct consequence of second order bulk convergence of $f_\Delta$, which is confirmed by Figure~\ref{fig:L2norm_iresallconvergence-crop} in our simulations. Despite this fact, some modifications must be made to our extrapolation scheme if we wish to perform explicit convergence tests on our boundary data. We now explain the reason for this and the necessary modifications. We assume the validity of the Richardson expansion \cite{doi:10.1098/rsta.1911.0009} for $f_{\Delta}$ at any grid point $p$,
\begin{equation}
\label{eq:Richexp}
f_\Delta(p)=f(p)+e(p)\Delta^2+\mathcal{O}(\Delta^3),
\end{equation}
where $f(p)$ is the true value of $f$ at $p$ and the rest of the right hand side is the solution error of $f_\Delta(p)$. The validity of this expansion is confirmed by bulk convergence of $f_\Delta$ to $f$.
Then, from \eqref{eq:firstordextrap}, we obtain the Richardson expansion for $f^{bdy}_{\Delta}$ at any extrapolated boundary point $p_{bdy}$:
 \begin{eqnarray}
 \label{eq:bdyRichexp}
 f^{bdy}_{\Delta}(p_{bdy})
&=&f(p_{bdy})+e_{extr}(p_{bdy},p_1,p_2)\nonumber \\
&&\hspace{0.8cm}+e_\Delta(p_1,p_2)\Delta^2 +\mathcal{O}(\Delta^3)\,,
 \end{eqnarray}
where the $f(p_{bdy})$ is the true value of $f$ at $p_{bdy}$, $e_{extr}(p_{bdy},p_1,p_2)$ is the error due to the extrapolation approximation. The remaining error terms come from the solution error in $f_\Delta$. The typical convergence test involves the computation of the convergence factor
\begin{equation}\label{eq:qconv}
Q(p_{bdy})=\frac{1}{\ln(3/2)}\ln\left( \frac{f_{9h/4}(p_{bdy})-f_{3h/2}(p_{bdy})}{f_{3h/2}(p_{bdy})-f_{h}(p_{bdy})} \right)
\end{equation}
at each boundary point $p_{bdy}$. We clearly see that $Q(p_{bdy})$ can be expected to asymptote to 2 as $\Delta\rightarrow0$, thus confirming second order convergence in the continuum limit, only if the points $p_1,p_2$ are the same for all 3 resolutions involved. Therefore, our extrapolation scheme must be modified to select pair of bulk points, $p_1$ and $p_2$, for extrapolation that are present in all three grids involved in the convergence test. 
In practice, we saw that boundary convergence follows the trend of bulk convergence only if, in addition to this modification, we restrict to $p_1$ points in the range mentioned in step 1 above. The reason for this should be investigated further. 

Finally, \eqref{eq:bdyRichexp} shows that this type of test does not prove convergence to the true value $f(p_{bdy})$, but rather to its approximation $f(p_{bdy})+e_{extr}(p_{bdy},p_1,p_2)$. For this reason, the convergence test \eqref{eq:qires} cannot be performed at the boundary for functions with vanishing true value (such as $\langle trT \rangle_{CFT}$), because their extrapolated value is not just the term linear in $\Delta^2$ but it also includes the extrapolation error $c_{extr}$. A more detailed analysis must be made to examine the explicit form $e_{extr}(p_{bdy},p_1,p_2)$ and be able to find the rate of convergence to $f(p_{bdy})$. In our study, we simply make the natural assumption that $e_{extr}(p_{bdy},p_1,p_2)$ decreases as we increase resolution, so $f^{bdy}_{\Delta}(p_{bdy})$ is a sufficiently accurate approximation of $f(p_{bdy})$ for sufficiently high resolution (i.e., sufficiently small $\Delta$).

\section{Convergence of the Independent Residual}\label{sec:convbulk}

To show that the solution is converging to a solution of the Einstein equations, we compute the independent residual that is obtained by taking the numerical solution, substituting it back into a discretized version of the Einstein equations.
At each grid point, we then take the maximum value over all components of the Einstein equations, which we denote by $\Phi_\Delta$. 
The independent residual should be purely numerical truncation error, so we can compute a convergence factor for it by using only two resolutions:
\begin{equation}\label{eq:qires}
Q_{EFE}(t,x,y,z)=\frac{1}{\ln(3/2)}\ln\left( \frac{\Phi_{3h/2}(t,x,y,z)}{\Phi_{h}(t,x,y,z)} \right).
\end{equation}
Again, with second-order accurate finite difference stencils and with a factor of 3/2 between successive resolutions, we expect $Q$ to approach $Q=2$ as $\Delta\rightarrow0$.
 \begin{figure*}[t!]
        \centering
        \includegraphics[width=5.0in,clip=true]{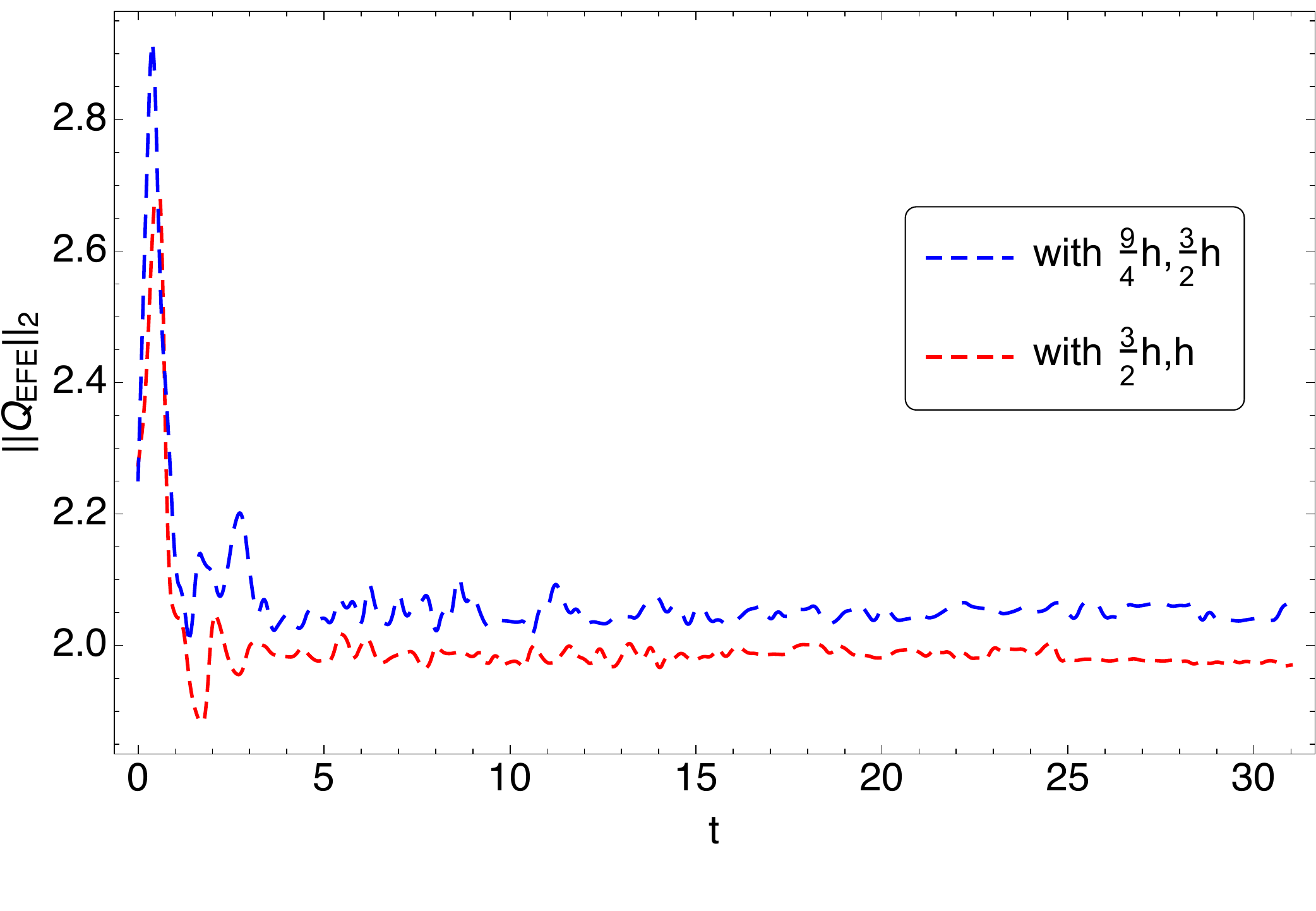}
\parbox{5.0in}{\caption{Time evolution for $L^2$-norm of convergence factor for independent residual of Einstein equations at different resolutions on the $z=0$ slice.
        }\label{fig:L2norm_iresallconvergence-crop}}
\end{figure*}
Figure~\ref{fig:L2norm_iresallconvergence-crop} displays the $L^2$-norm of the convergence factor \eqref{eq:qires} for two pairs of resolutions on the $z=0$ slice. It clearly shows second order convergence to a solution of the Einstein equations, after an initial transition phase.

\bibliographystyle{JHEP}
\bibliography{3p1}

\providecommand{\href}[2]{#2}\begingroup\raggedright\begin{thebibliography}{10}

\bibitem{Maldacena:1997re}


\bibitem{Gubser:1998bc}
S.~S. Gubser, I.~R. Klebanov and A.~M. Polyakov, \emph{{Gauge theory
  correlators from noncritical string theory}},
  \href{https://doi.org/10.1016/S0370-2693(98)00377-3}{\emph{Phys. Lett.}
  {\bfseries B428} (1998) 105}
  [\href{https://arxiv.org/abs/hep-th/9802109}{{\ttfamily hep-th/9802109}}].

\bibitem{Witten:1998qj}
E.~Witten, \emph{{Anti-de Sitter space and holography}},
  \href{https://doi.org/10.4310/ATMP.1998.v2.n2.a2}{\emph{Adv. Theor. Math.
  Phys.} {\bfseries 2} (1998) 253}
  [\href{https://arxiv.org/abs/hep-th/9802150}{{\ttfamily hep-th/9802150}}].

\bibitem{Baier:2007ix}
R.~Baier, P.~Romatschke, D.~T. Son, A.~O. Starinets and M.~A. Stephanov,
  \emph{{Relativistic viscous hydrodynamics, conformal invariance, and
  holography}},
  \href{https://doi.org/10.1088/1126-6708/2008/04/100}{\emph{JHEP} {\bfseries
  04} (2008) 100} [\href{https://arxiv.org/abs/0712.2451}{{\ttfamily
  0712.2451}}].

\bibitem{Bhattacharyya:2010owp}
S.~Bhattacharyya, \emph{{Nonlinear Fluid Dynamics From Gravity}}, Ph.D. thesis,
  Tata Inst., 2010.

\bibitem{Hubeny:2011hd}
V.~E. Hubeny, S.~Minwalla and M.~Rangamani, \emph{{The fluid/gravity
  correspondence}},  in \emph{{Theoretical Advanced Study Institute in
  Elementary Particle Physics}: {String theory and its Applications: From meV
  to the Planck Scale}}, pp.~348--383, 2012,
  \href{https://arxiv.org/abs/1107.5780}{{\ttfamily 1107.5780}}.

\bibitem{Chesler:2010bi}
P.~M. Chesler and L.~G. Yaffe, \emph{{Holography and colliding gravitational
  shock waves in asymptotically AdS$_{5}$ spacetime}},
  \href{https://doi.org/10.1103/PhysRevLett.106.021601}{\emph{Phys. Rev. Lett.}
  {\bfseries 106} (2011) 021601}
  [\href{https://arxiv.org/abs/1011.3562}{{\ttfamily 1011.3562}}].

\bibitem{Casalderrey-Solana:2013aba}
J.~Casalderrey-Solana, M.~P. Heller, D.~Mateos and W.~van~der Schee,
  \emph{{From full stopping to transparency in a holographic model of heavy ion
  collisions}},
  \href{https://doi.org/10.1103/PhysRevLett.111.181601}{\emph{Phys. Rev. Lett.}
  {\bfseries 111} (2013) 181601}
  [\href{https://arxiv.org/abs/1305.4919}{{\ttfamily 1305.4919}}].

\bibitem{Casalderrey-Solana:2013sxa}
J.~Casalderrey-Solana, M.~P. Heller, D.~Mateos and W.~van~der Schee,
  \emph{{Longitudinal Coherence in a Holographic Model of Asymmetric
  Collisions}},
  \href{https://doi.org/10.1103/PhysRevLett.112.221602}{\emph{Phys. Rev. Lett.}
  {\bfseries 112} (2014) 221602}
  [\href{https://arxiv.org/abs/1312.2956}{{\ttfamily 1312.2956}}].

\bibitem{Chesler:2015wra}
P.~M. Chesler and L.~G. Yaffe, \emph{{Holography and off-center collisions of
  localized shock waves}},
  \href{https://doi.org/10.1007/JHEP10(2015)070}{\emph{JHEP} {\bfseries 10}
  (2015) 070} [\href{https://arxiv.org/abs/1501.04644}{{\ttfamily
  1501.04644}}].

\bibitem{Gubser:2008px}
S.~S. Gubser, \emph{{Breaking an Abelian gauge symmetry near a black hole
  horizon}}, \href{https://doi.org/10.1103/PhysRevD.78.065034}{\emph{Phys. Rev.
  D} {\bfseries 78} (2008) 065034}
  [\href{https://arxiv.org/abs/0801.2977}{{\ttfamily 0801.2977}}].

\bibitem{Hartnoll:2008kx}
S.~A. Hartnoll, C.~P. Herzog and G.~T. Horowitz, \emph{{Holographic
  Superconductors}},
  \href{https://doi.org/10.1088/1126-6708/2008/12/015}{\emph{JHEP} {\bfseries
  12} (2008) 015} [\href{https://arxiv.org/abs/0810.1563}{{\ttfamily
  0810.1563}}].

\bibitem{Hartnoll:2008vx}
S.~A. Hartnoll, C.~P. Herzog and G.~T. Horowitz, \emph{{Building a Holographic
  Superconductor}},
  \href{https://doi.org/10.1103/PhysRevLett.101.031601}{\emph{Phys. Rev. Lett.}
  {\bfseries 101} (2008) 031601}
  [\href{https://arxiv.org/abs/0803.3295}{{\ttfamily 0803.3295}}].

\bibitem{CasalderreySolana:2011us}
J.~Casalderrey-Solana, H.~Liu, D.~Mateos, K.~Rajagopal and U.~A. Wiedemann,
  \emph{{Gauge/String Duality, Hot QCD and Heavy Ion Collisions}}. Cambridge
  University Press, 2014,
  \href{https://doi.org/10.1017/CBO9781139136747}{10.1017/CBO9781139136747},
  [\href{https://arxiv.org/abs/1101.0618}{{\ttfamily 1101.0618}}].

\bibitem{Chesler:2015lsa}
P.~M. Chesler and W.~van~der Schee, \emph{{Early thermalization, hydrodynamics
  and energy loss in AdS/CFT}},
  \href{https://doi.org/10.1142/S0218301315300118}{\emph{Int. J. Mod. Phys. E}
  {\bfseries 24} (2015) 1530011}
  [\href{https://arxiv.org/abs/1501.04952}{{\ttfamily 1501.04952}}].

\bibitem{Zaanen:2015oix}
J.~Zaanen, Y.-W. Sun, Y.~Liu and K.~Schalm, \emph{{Holographic Duality in
  Condensed Matter Physics}}. Cambridge Univ. Press, 2015.

\bibitem{Hartnoll:2016apf}
S.~A. Hartnoll, A.~Lucas and S.~Sachdev, \emph{{Holographic quantum matter}},
  \href{https://arxiv.org/abs/1612.07324}{{\ttfamily 1612.07324}}.

\bibitem{Gomez:1998uj}
R.~Gomez et~al., \emph{{Stable characteristic evolution of generic
  three-dimensional single black hole space-times}},
  \href{https://doi.org/10.1103/PhysRevLett.80.3915}{\emph{Phys. Rev. Lett.}
  {\bfseries 80} (1998) 3915}
  [\href{https://arxiv.org/abs/gr-qc/9801069}{{\ttfamily gr-qc/9801069}}].

\bibitem{Chesler:2018txn}
P.~M. Chesler and D.~A. Lowe, \emph{{Nonlinear Evolution of the AdS$_4$
  Superradiant Instability}},
  \href{https://doi.org/10.1103/PhysRevLett.122.181101}{\emph{Phys. Rev. Lett.}
  {\bfseries 122} (2019) 181101}
  [\href{https://arxiv.org/abs/1801.09711}{{\ttfamily 1801.09711}}].

\bibitem{Chesler:2013lia}
P.~M. Chesler and L.~G. Yaffe, \emph{{Numerical solution of gravitational
  dynamics in asymptotically anti-de Sitter spacetimes}},
  \href{https://doi.org/10.1007/JHEP07(2014)086}{\emph{JHEP} {\bfseries 07}
  (2014) 086} [\href{https://arxiv.org/abs/1309.1439}{{\ttfamily 1309.1439}}].

\bibitem{Bantilan:2020pay}
H.~Bantilan, P.~Figueras and D.~Mateos, \emph{{Real-time Dynamics of Plasma
  Balls from Holography}},
  \href{https://doi.org/10.1103/PhysRevLett.124.191601}{\emph{Phys. Rev. Lett.}
  {\bfseries 124} (2020) 191601}
  [\href{https://arxiv.org/abs/2001.05476}{{\ttfamily 2001.05476}}].

\bibitem{Bishop:1997ik}
N.~T. Bishop, R.~Gomez, L.~Lehner, M.~Maharaj and J.~Winicour, \emph{{High
  powered gravitational news}},
  \href{https://doi.org/10.1103/PhysRevD.56.6298}{\emph{Phys. Rev. D}
  {\bfseries 56} (1997) 6298}
  [\href{https://arxiv.org/abs/gr-qc/9708065}{{\ttfamily gr-qc/9708065}}].

\bibitem{Lehner:2001wq}
L.~Lehner, \emph{{Numerical relativity: A Review}},
  \href{https://doi.org/10.1088/0264-9381/18/17/202}{\emph{Class. Quant. Grav.}
  {\bfseries 18} (2001) R25}
  [\href{https://arxiv.org/abs/gr-qc/0106072}{{\ttfamily gr-qc/0106072}}].

\bibitem{Pretorius:2005gq}
F.~Pretorius, \emph{{Evolution of binary black hole spacetimes}},
  \href{https://doi.org/10.1103/PhysRevLett.95.121101}{\emph{Phys. Rev. Lett.}
  {\bfseries 95} (2005) 121101}
  [\href{https://arxiv.org/abs/gr-qc/0507014}{{\ttfamily gr-qc/0507014}}].

\bibitem{Campanelli:2005dd}
M.~Campanelli, C.~Lousto, P.~Marronetti and Y.~Zlochower, \emph{{Accurate
  evolutions of orbiting black-hole binaries without excision}},
  \href{https://doi.org/10.1103/PhysRevLett.96.111101}{\emph{Phys. Rev. Lett.}
  {\bfseries 96} (2006) 111101}
  [\href{https://arxiv.org/abs/gr-qc/0511048}{{\ttfamily gr-qc/0511048}}].

\bibitem{Baker:2005vv}
J.~G. Baker, J.~Centrella, D.-I. Choi, M.~Koppitz and J.~van Meter,
  \emph{{Gravitational wave extraction from an inspiraling configuration of
  merging black holes}},
  \href{https://doi.org/10.1103/PhysRevLett.96.111102}{\emph{Phys. Rev. Lett.}
  {\bfseries 96} (2006) 111102}
  [\href{https://arxiv.org/abs/gr-qc/0511103}{{\ttfamily gr-qc/0511103}}].

\bibitem{Bantilan:2012vu}
H.~Bantilan, F.~Pretorius and S.~S. Gubser, \emph{{Simulation of Asymptotically
  AdS5 Spacetimes with a Generalized Harmonic Evolution Scheme}},
  \href{https://doi.org/10.1103/PhysRevD.85.084038}{\emph{Phys. Rev.}
  {\bfseries D85} (2012) 084038}
  [\href{https://arxiv.org/abs/1201.2132}{{\ttfamily 1201.2132}}].

\bibitem{Bantilan:2017kok}
H.~Bantilan, P.~Figueras, M.~Kunesch and P.~Romatschke, \emph{{Nonspherically
  Symmetric Collapse in Asymptotically AdS Spacetimes}},
  \href{https://doi.org/10.1103/PhysRevLett.119.191103}{\emph{Phys. Rev. Lett.}
  {\bfseries 119} (2017) 191103}
  [\href{https://arxiv.org/abs/1706.04199}{{\ttfamily 1706.04199}}].

\bibitem{Pretorius:2004jg}
F.~Pretorius, \emph{{Numerical relativity using a generalized harmonic
  decomposition}},
  \href{https://doi.org/10.1088/0264-9381/22/2/014}{\emph{Class. Quant. Grav.}
  {\bfseries 22} (2005) 425}
  [\href{https://arxiv.org/abs/gr-qc/0407110}{{\ttfamily gr-qc/0407110}}].

\bibitem{Henneaux:1985tv}
M.~Henneaux and C.~Teitelboim, \emph{{Asymptotically anti-De Sitter Spaces}},
  \href{https://doi.org/10.1007/BF01205790}{\emph{Commun. Math. Phys.}
  {\bfseries 98} (1985) 391}.

\bibitem{Henneaux:2006hk}
M.~Henneaux, C.~Martinez, R.~Troncoso and J.~Zanelli, \emph{{Asymptotic
  behavior and Hamiltonian analysis of anti-de Sitter gravity coupled to scalar
  fields}}, \href{https://doi.org/10.1016/j.aop.2006.05.002}{\emph{Annals
  Phys.} {\bfseries 322} (2007) 824}
  [\href{https://arxiv.org/abs/hep-th/0603185}{{\ttfamily hep-th/0603185}}].

\bibitem{AST_1985__S131__95_0}
C.~Fefferman and C.~R. Graham, \emph{Conformal invariants},  in \emph{\'Elie
  Cartan et les math\'ematiques d'aujourd'hui - Lyon, 25-29 juin 1984},
  no.~S131 in Ast\'erisque, pp.~95--116.
\newblock Soci\'et\'e math\'ematique de France, 1985.

\bibitem{deHaro:2000vlm}
S.~de~Haro, S.~N. Solodukhin and K.~Skenderis, \emph{{Holographic
  reconstruction of space-time and renormalization in the AdS / CFT
  correspondence}}, \href{https://doi.org/10.1007/s002200100381}{\emph{Commun.
  Math. Phys.} {\bfseries 217} (2001) 595}
  [\href{https://arxiv.org/abs/hep-th/0002230}{{\ttfamily hep-th/0002230}}].

\bibitem{Balasubramanian:1999re}
V.~Balasubramanian and P.~Kraus, \emph{{A Stress tensor for Anti-de Sitter
  gravity}}, \href{https://doi.org/10.1007/s002200050764}{\emph{Commun. Math.
  Phys.} {\bfseries 208} (1999) 413}
  [\href{https://arxiv.org/abs/hep-th/9902121}{{\ttfamily hep-th/9902121}}].

\bibitem{Hawking:1973uf}
S.~Hawking and G.~Ellis, \emph{{The Large Scale Structure of Space-Time}},
  Cambridge Monographs on Mathematical Physics. Cambridge University Press, 2,
  2011,
  \href{https://doi.org/10.1017/CBO9780511524646}{10.1017/CBO9780511524646}.

\bibitem{PAMR}
\emph{{PAMR/AMRD}}, {\emph{http://laplace.physics.ubc.ca/Group/Software.html}
  }.

\bibitem{kreiss1973methods}
H.~Kreiss, H.~Kreiss, J.~Oliger and G.~A. R. P. J.~O. Committee, \emph{Methods
  for the approximate solution of time dependent problems}, GARP publications
  series. International Council of Scientific Unions, World Meteorological
  Organization, 1973.

\bibitem{Fischetti:2012rd}
D.~Marolf, W.~Kelly and S.~Fischetti, \emph{{Conserved Charges in
  Asymptotically (Locally) AdS Spacetimes}}, pp.~381--407.
\newblock 2014.
\newblock \href{https://arxiv.org/abs/1211.6347}{{\ttfamily 1211.6347}}.
\newblock 10.1007/978-3-642-41992-8\_19.

\bibitem{Holzegel:2011uu}
G.~Holzegel and J.~Smulevici, \emph{{Decay properties of Klein-Gordon fields on
  Kerr-AdS spacetimes}}, \href{https://doi.org/10.1002/cpa.21470}{\emph{Commun.
  Pure Appl. Math.} {\bfseries 66} (2013) 1751}
  [\href{https://arxiv.org/abs/1110.6794}{{\ttfamily 1110.6794}}].

\bibitem{Bizon:2011gg}
P.~Bizo\'n and A.~Rostworowski, \emph{{On weakly turbulent instability of
  anti-de Sitter space}},
  \href{https://doi.org/10.1103/PhysRevLett.107.031102}{\emph{Phys. Rev. Lett.}
  {\bfseries 107} (2011) 031102}
  [\href{https://arxiv.org/abs/1104.3702}{{\ttfamily 1104.3702}}].

\bibitem{Jalmuzna:2011qw}
J.~Ja{\l}mu\.zna, A.~Rostworowski and P.~Bizon, \emph{{A Comment on AdS
  collapse of a scalar field in higher dimensions}},
  \href{https://doi.org/10.1103/PhysRevD.84.085021}{\emph{Phys. Rev. D}
  {\bfseries 84} (2011) 085021}
  [\href{https://arxiv.org/abs/1108.4539}{{\ttfamily 1108.4539}}].

\bibitem{Choptuik:2017cyd}
M.~W. Choptuik, O.~J.~C. Dias, J.~E. Santos and B.~Way, \emph{{Collapse and
  Nonlinear Instability of AdS with Angular Momenta}},
  \href{https://arxiv.org/abs/1706.06101}{{\ttfamily 1706.06101}}.

\bibitem{Moschidis:2018ruk}
G.~Moschidis, \emph{{A proof of the instability of AdS for the
  Einstein--massless Vlasov system}},
  \href{https://arxiv.org/abs/1812.04268}{{\ttfamily 1812.04268}}.

\bibitem{Moschidis:2018kcf}
G.~Moschidis, \emph{{The characteristic initial-boundary value problem for the
  Einstein--massless Vlasov system in spherical symmetry}},
  \href{https://arxiv.org/abs/1812.04274}{{\ttfamily 1812.04274}}.

\bibitem{Brito:2015oca}
R.~Brito, V.~Cardoso and P.~Pani, \emph{{Superradiance}: {New Frontiers in
  Black Hole Physics}}, vol.~906. Springer, 2015,
  \href{https://doi.org/10.1007/978-3-319-19000-6}{10.1007/978-3-319-19000-6},
  [\href{https://arxiv.org/abs/1501.06570}{{\ttfamily 1501.06570}}].

\bibitem{Dias:2015rxy}
O.~J. Dias, J.~E. Santos and B.~Way, \emph{{Black holes with a single Killing
  vector field: black resonators}},
  \href{https://doi.org/10.1007/JHEP12(2015)171}{\emph{JHEP} {\bfseries 12}
  (2015) 171} [\href{https://arxiv.org/abs/1505.04793}{{\ttfamily
  1505.04793}}].

\bibitem{Hawking:1999dp}
S.~Hawking and H.~Reall, \emph{{Charged and rotating AdS black holes and their
  CFT duals}}, \href{https://doi.org/10.1103/PhysRevD.61.024014}{\emph{Phys.
  Rev. D} {\bfseries 61} (1999) 024014}
  [\href{https://arxiv.org/abs/hep-th/9908109}{{\ttfamily hep-th/9908109}}].

\bibitem{Green:2015kur}
S.~R. Green, S.~Hollands, A.~Ishibashi and R.~M. Wald, \emph{{Superradiant
  instabilities of asymptotically anti-de Sitter black holes}},
  \href{https://doi.org/10.1088/0264-9381/33/12/125022}{\emph{Class. Quant.
  Grav.} {\bfseries 33} (2016) 125022}
  [\href{https://arxiv.org/abs/1512.02644}{{\ttfamily 1512.02644}}].

\bibitem{Niehoff:2015oga}
B.~E. Niehoff, J.~E. Santos and B.~Way, \emph{{Towards a violation of cosmic
  censorship}},
  \href{https://doi.org/10.1088/0264-9381/33/18/185012}{\emph{Class. Quant.
  Grav.} {\bfseries 33} (2016) 185012}
  [\href{https://arxiv.org/abs/1510.00709}{{\ttfamily 1510.00709}}].

\bibitem{ApocritaHPC}
\emph{{Apocrita-High Performance Computing Cluster for Queen Mary University of
  London}}, .

\bibitem{Gundlach:2005eh}
C.~Gundlach, J.~M. Martin-Garcia, G.~Calabrese and I.~Hinder, \emph{{Constraint
  damping in the Z4 formulation and harmonic gauge}},
  \href{https://doi.org/10.1088/0264-9381/22/17/025}{\emph{Class. Quant. Grav.}
  {\bfseries 22} (2005) 3767}
  [\href{https://arxiv.org/abs/gr-qc/0504114}{{\ttfamily gr-qc/0504114}}].

\bibitem{Lichnerowicz:1994}
A.~Lichnerowicz, \emph{{L'int\'{e}gration des \'{e}quations de la gravitation
  relativiste et le probl\'{e}me des n corps}}, vol.~23, p. 37--63. Reprinted
  in A. Lichnerowicz : Choix d'{\oe}uvres math\'{e}matiques, Hermann, Paris
  (1982), p.4., 1994.

\bibitem{doi:10.1098/rsta.1911.0009}
L.~F. Richardson and R.~T. Glazebrook, \emph{Ix. the approximate arithmetical
  solution by finite differences of physical problems involving differential
  equations, with an application to the stresses in a masonry dam},
  \href{https://doi.org/10.1098/rsta.1911.0009}{\emph{Philosophical
  Transactions of the Royal Society of London. Series A, Containing Papers of a
  Mathematical or Physical Character} {\bfseries 210} (1911) 307}.

\end{thebibliography}\endgroup

\end{document}